\documentclass[12pt,twoside]{article}

\usepackage[
  left=2cm,right=2.5cm,top=3.1cm,bottom=3cm,headheight=2cm,headsep=1.2cm,footskip=1.4cm,twoside
]{geometry}
\usepackage[margin=1cm,font=small]{caption}
\usepackage{enumerate}
\usepackage{graphicx}
\usepackage{amsmath}
\usepackage{multirow}
\usepackage[section]{placeins}
\usepackage{booktabs}
\usepackage{array}
\usepackage{caption}
\usepackage{cite}

\numberwithin{equation}{section}

\newcommand{\babar}{\mbox{\ensuremath{{\displaystyle B}\!{\scriptstyle A}\!{\displaystyle B}\!{\scriptstyle {A\!R}}}\,\,}}

\begin{document}
\title{Constraints on exclusive branching fractions $\mathcal{B}_i(B^+\to X_c^il^+\nu)$ from moment measurements in
inclusive $B\to X_cl\nu$ decays
}

\date{\vspace{-5ex}}

\maketitle

\begin{raggedright}  

\hspace{1cm}{\it Florian U. Bernlochner \\
\hspace{1cm}University of Victoria, Victoria, British Columbia, Canada V8W 3P\\
}
\bigskip

\hspace{1cm}{\it Dustin Biedermann\\
\hspace{1cm}Humboldt-Universit\"at zu Berlin, 12489 Berlin, Germany\\
}
\bigskip

\hspace{1cm}{\it Heiko Lacker\\
\hspace{1cm}Humboldt-Universit\"at zu Berlin, 12489 Berlin, Germany\\
}

\bigskip

\hspace{1cm}{\it Thomas L\"uck \\
\hspace{1cm}University of Victoria, Victoria, British Columbia, Canada V8W 3P\\
}

\bigskip\bigskip
\end{raggedright}

\begin{abstract}
As an alternative to direct measurements, we extract exclusive branching fractions of 
semileptonic $B$-meson decays to charmed mesons, $\mathcal{B}_i(B^+\to X_c^il^+\nu)$
\footnote[1]{Charge conjugation is always implied.} 
with $X_c^i = D, D^*, D_0, D'_1, D_1, D_2, D', D'^{*}$ and non-resonant final states 
$(D^{(*)}\pi)_{nr}$, from a fit to electron energy, hadronic mass and 
combined hadronic mass-energy moments measured in inclusive $B\to X_cl\nu$ decays.
The fit is performed by constraining the sum of exclusive branching fractions to the 
measured value of $\mathcal{B} (B^+\to X_c l^+\nu)$, and with different
sets of additional constraining terms for the directly measured branching fractions.
There is no fit scenario in which a single branching fraction alone is enhanced to close the gap between
$\mathcal{B} (B^+\to X_c l^+\nu)$ and the sum of known branching fractions
$\mathcal{B}_i(B^+\to X_c^il^+\nu)$.
The fitted $\mathcal{B}(B^+\to \overline{D}^{*0}l^+\nu)$ is $5\%$ to $10\%$
larger than the direct measurement depending on whether or not
$\mathcal{B}(B^+\to \overline{D}^{*0} l^+\nu)$ is constrained to its direct
 measurement. $\mathcal{B}(B^+\to \overline{D}^0l^+\nu)$ values are in good 
agreement with the direct measurement unless $\mathcal{B}(B^+\to \overline{D}^{*0} l^+\nu)$ 
is constrained, which results in a higher $\mathcal{B}(B^+\to \overline{D}^0l^+\nu)$ value.
Within large uncertainties, $\mathcal{B}(B^+\to \overline{D}'^0_1l^+\nu)$ 
agrees with direct measurements. Depending on the fit scenario,
$\mathcal{B}(B^+\to \overline{D}^0_0l^+\nu)$ is consistent with or larger 
than its direct measurement.
The fit is not able to easily disentangle $B^+\to \overline{D}^0_1l^+\nu$ 
and $B^+\to \overline{D}^0_2l^+\nu$, and tends to increase the sum of these 
two branching fractions.
$\mathcal{B} (B^+\to (D^{(*)}\pi)_{nr}l^+\nu)$ with non-resonant 
$(D^{(*)}\pi)_{nr}$ final states is found to be $0.2-0.4\%$, consistent with 
$B^+\to D^{(*)}\pi l^+\nu$ and $B^+\to \overline{D}^{**}(D^{(*)}\pi) l^+\nu$ 
measurements. No indication is found for significant contributions from so far 
unmeasured $B^+\to \overline{D}'^{(*)0}l^+\nu$ decays assuming that the 
$\overline{D}'^{0}$ and $\overline{D}'^{*0}$ can be identified with the 
observed $D(2550)$, respectively, $D^{*}(2600)$ state.

\end{abstract} 

\section{INTRODUCTION}
\label{intro}

The Cabibbo-Kobayashi-Maskawa (CKM) quark mixing matrix~\cite{CKM} governs the weak 
coupling strength between up- and down-type quarks. The CKM-matrix element $|V_{cb}|$ 
can be extracted from semileptonic $B$-meson decays $B\to X_cl\nu$ with hadronic
final states $X_c$ containing mesons with charm whereby inclusive or exclusive final 
states can be used. For analyses of exclusive final states, such as $B\to D^{(*)}l\nu$ 
decays, good knowledge about the overall composition of the $X_c$ final states is 
crucial. Precise understanding of semileptonic $B\to X_cl\nu$ decays is also of 
utmost importance for the precision determination of $|V_{ub}|$ from $B\to X_u l\nu$ 
decays, since $B\to X_cl\nu$ decays represent the main source of background events 
in this kind of analyses.\\
Precise knowledge of $B\to X_cl\nu$ decays has also relevance for new physics searches.
A \babar\\ measurement of $\frac{\mathcal{B}(\overline{B}\to D^{(*)}\tau^- \overline{\nu})}{\mathcal{B}(\overline{B}\to D^{(*)}l^-\overline{\nu})}$, where $l\in\{e,\mu\}$, exceeds the Standard Model expectations by 
$3.4\,\sigma$~\cite{RMeasure}. In this analysis, an important systematic uncertainty 
originates from the detailed knowledge of the composition of $B\to D^{**}l\nu$ decays, 
where $D^{**}$ denotes the four 1P states of non-strange charmed mesons. In particular,
$D^{**}$ decays to $D^{(*)}\pi\pi$ states are seen to have a large impact on the measured 
ratio.

\subsection{BRANCHING FRACTION MEASUREMENTS}
\label{BrFrMeas}
Great efforts have been made in measuring the inclusive and exclusive branching 
fractions of $B\to X_cl\nu$ transitions. Exclusive branching fractions  
$\mathcal{B}(B\to X^i_cl\nu)$ have been determined for the hadronic 
final states $X_c^i\in\left\{D, D^*, D_0, D'_1, D_1, D_2, D^{(*)}\pi \right\}$.
Averages for these measurements and for the inclusive branching fraction
$\mathcal{B}(B\to X_c l\nu)$ are provided by the Heavy Flavor 
Averaging Group (HFAG)~\cite{HFAG}, which we use in this paper.
In Table~\ref{final}, we quote all branching fractions for semileptonic
decays of charged $B$-mesons, for which measurements exist and which are
used in our analysis. Thereby, we assume that the corresponding branching 
fractions for semileptonic decays of neutral $B$-mesons can be obtained by 
applying isospin invariance of strong interactions.
That is, the decay rates for semileptonic decays of charged and neutral 
$B$-mesons are set to be equal.
The $\mathcal{B}(B^{+}\to \overline{D}^{0} l^+\nu)$ and $\mathcal{B}(B^{+}\to \overline{D}^{*0} l^+\nu)$
values quoted in Table~\ref{final} are calculated from the HFAG (isospin) 
averages provided for $\mathcal{B}(B^{0}\to D^{(*)-} l^{+}\nu)$~\cite{HFAG} as 
\begin{equation}\label{isomodes}
\mathcal{B}(B^{+}\to \overline{D}^{(*)0} l^+\nu)=\tau_{+0}\mathcal{B}(B^{0}\to \overline{D}^{(*)-} l^+\nu),
\end{equation}
with
\begin{equation}
\tau_{+0}:=\frac{\tau_{B^+}}{\tau_{B^0}}=1.079\pm 0.007
\end{equation}
being the ratio of lifetimes $\tau_{B^+}$ and $\tau_{B^0}$ of charged and neutral 
$B$-mesons, respectively~\cite{HFAG}. 
Since for $\mathcal{B}(B\to X_c l\nu)$ quoted in Ref.~\cite{HFAG} 
charged and neutral $B$-meson decays were used, we calculate $\mathcal{B}(B^{+}\to X_c l^+\nu)$ 
according to
\begin{equation}
\mathcal{B}(B^{+}\to X_c l^+\nu)=\tau_{+0}\frac{f_{+0}+1}{1+f_{+0}\tau_{+0}}\mathcal{B}(B\to X_c l\nu),
\end{equation}
with
\begin{equation}
f_{+0}:=\frac{\mathcal{B}(\Upsilon(4S)\to B^+B^-)}{\mathcal{B}(\Upsilon(4S)\to B^0\overline{B}^0)}=1.065\pm0.026\;
\end{equation}
being the measured ratio of $\Upsilon(4S)$ branching fractions into charged and neutral 
$B$-meson pairs as quoted in Ref.~\cite{HFAG}.\\
In case of $X^i_c$ being one of the $D^{\ast\ast}$ mesons $D_0$, $D'_1$, $D_1$, or $D_2$,
only product branching-fractions
$\mathcal{B}(B^+\to {D}^{\ast\ast}(D^{(*)-}\pi^{+})l^+\nu)= \mathcal{B}(B^+\to {D}^{\ast\ast} l^+\nu) \times \mathcal{B}({D}^{\ast\ast} \to D^{(*)-}\pi^{+})$
are available~\cite{HFAG}. In these cases, we have to correct  
for the branching fraction $\mathcal{B}({D}^{\ast\ast}\to D^{(*)-}\pi^{+})$
to obtain $\mathcal{B}(B^+\to {D}^{\ast\ast}l^+\nu)$.
To do so we assume strong-isospin symmetry. As a result, in case
of a $D^{\ast\ast}$ two-body decay, we account for the decay mode of 
the $D^{\ast\ast}$ in which the created $u$ quark is interchanged by a $d$ quark 
by introducing a multiplicative factor of $\frac{3}{2}$.
Furthermore, we make the following assumptions: the $D_{0}$ can only 
decay into $D \pi$, the $D'_1$ only into $D^{*} \pi$, the $D_1$-meson only
into $D^*\pi$ and $D\pi\pi$, and the $D_2$-meson into $D\pi$ and $D^{*}\pi$.

In cases of $D_1$ decays we use the average of measurements for the ratio
(see Appendix~\ref{SuppD1}) 
\begin{equation}
\begin{split}
\frac{\mathcal{B}(B^+\to \overline{D}_1^0(D\pi\pi)\pi^+)}{\mathcal{B}(B^+\to \overline{D}_1^0(D^{*}\pi)\pi^+)}=0.53\pm 0.14,
\end{split}
\end{equation}
relying on Refs.~\cite{D1eins, D1zwei, LHCbFrac, ThomasDiss},
and obtain $\mathcal{B}(B^+\to \overline{D}^0_1l^+\nu)$ according to
\begin{equation}
\begin{split}
\mathcal{B}(B^+\to \overline{D}_1^0l^+\nu)=\left(1+\frac{\mathcal{B}(B^+\to \overline{D}_1^0(D\pi\pi)\pi^+)}{\mathcal{B}(B^+\to \overline{D}_1^0(D^{*}\pi)\pi^+)}\right)\times\mathcal{B}(B^+\to \overline{D}_1^0(D^*\pi)l^+\nu).
\end{split}
\end{equation}

For the $D_2$-meson we use the measured ratio~\cite{D2Frac}
\begin{equation}\label{D2ratio}
\frac{\mathcal{B}(D_2^0\to D^+\pi^-)}{\mathcal{B}(D_2^0\to D^{*+}\pi^-)}=1.56\pm0.16,
\end{equation}
and calculate $\mathcal{B}(B^+\to \overline{D}_2^0l^+\nu)$ in an analogous
way as $\mathcal{B}(B^+\to \overline{D}_1^0l^+\nu)$.\\
Measurements for $B^+\to D^{(*)}\pi l^+\nu$ have been performed as well. 
Using the HFAG averages of these branching fractions~\cite{HFAG} 
together with the $\mathcal{B}(B^+\to \overline{D}^{\ast\ast}(D^{(*)}\pi )l^+\nu)$
averages~\cite{HFAG} we determine the branching fraction for semileptonic 
decays into non-resonant ($nr$) final states $(D^{(*)}\pi)_{nr}$. 
For these cases, we calculate the isospin average according to 
equation~\ref{iso1} as described in Appendix~\ref{NrCalc}. \\
The values for $\mathcal{B}(B^+\to X^i_cl^+\nu)$, with $X_c^i$ being $D_0$, $D'_1$, 
$D_1$, $D_2$, and $(D^{(*)} \pi)_{nr}$, obtained in this way, are quoted in Table~\ref{final}.

\subsection{PUZZLES AND POSSIBLE SOLUTIONS}

Some serious problems arise from the quoted branching fractions:

\begin{itemize}
\item The most obvious puzzle and in the following denoted as "gap problem" results 
      from the fact that the sum of the directly measured exclusive branching fractions 
      does not saturate the measured inclusive branching fraction, i.e. 
\begin{equation}
\begin{split}
\mathcal{B}(B^+\to X_cl^+\nu) = (10.90 \pm 0.14)\% \neq \sum\limits_{i=D,D^*,D^{**}}\mathcal{B}_i(B^+\to
X_c^il^+\nu) = (9.2 \pm 0.2)\%.
\end{split}
\end{equation}
      Even if the branching fraction for decays into non-resonant 
      $(D^{(*)}\pi)_{nr}$, $\mathcal{B}(B^+\to(D^{(*)}\pi)_{nr}l^+\nu)$, 
      is taken into consideration as well, the gap can not be closed.
\item A more subtle problem and commonly referred to as the "$\frac{1}{2}$ vs. $\frac{3}{2}$ 
      puzzle"~\cite{Memorino} concerns the sector of $B\to D^{**}l\nu$ decays. 
      Theoretical deliberations~\cite{Memorino, puzzle, Memorino1} suggest 
      that the branching fraction of $B\to D_1/D_2l\nu$ decays should be 
      about one order of magnitude larger than $\mathcal{B}(B\to D_0/D'_1l\nu)$. 
      The measured values are in clear contradiction to this expectation.\\
      It should be noted though that a quark-model based calculation essentially 
      agrees with the measured values of all $B\to D^{**}l\nu$ transitions~\cite{QuModel}. 
      If correct this would be in contrast to the stated 
      "$\frac{1}{2}$ vs. $\frac{3}{2}$ puzzle"~\cite{Memorino, puzzle, Memorino1}.
\item Furthermore, the branching fraction of $B^+\to \overline{D}'^0_1l^+\nu$ decays 
      which is given in Table~\ref{final} is the result of a weighted average of 
      three measurements from DELPHI~\cite{D1PrimeMeasure1}, 
      Belle~\cite{D1PrimeMeasure2} and \babar~\cite{D1PrimeMeasure3}:
\begin{itemize}
\item $\mathcal{B}(B^+\to \overline{D}'^0_1(D^{*-}\pi^+)l^+\nu)=(0.74\pm0.17\pm0.18)\%$ (DELPHI),
\item $\mathcal{B}(B^+\to \overline{D}'^0_1(D^{*-}\pi^+)l^+\nu)=(-0.03\pm0.06\pm0.07)\%$ (Belle),
\item $\mathcal{B}(B^+\to \overline{D}'^0_1(D^{*-}\pi^+)l^+\nu)=(0.27\pm0.04\pm0.04)\%$ (\babar).
\end{itemize}
When averaging these three measurements one obtains a $\chi^2$ over degrees of freedom ($dof$) 
of $\chi^2/dof=\frac{18}{2}$ corresponding to a confidence level of $0.1\%$. Possibly, 
at least one measurement underestimates the uncertainty, thus the average might be biased 
and the uncertainty on the weighted average might be underestimated.
\end{itemize}
\begin{table}
\begin{center}
\begin{tabular}{ccc}
\hline\noalign{\smallskip}
Decay&Branching Fraction [\%]\\
\hline\noalign{\smallskip}
$B^+\to \overline{D}^0l^+\nu$& $2.30\pm0.10$\\
$B^+\to \overline{D}^{*0}l^+\nu$&$5.34\pm0.12$\\
$B^+\to \overline{D}_1^0l^+\nu$&$0.652\pm0.071$\\
$B^+\to \overline{D}_2^0l^+\nu$&$0.284\pm0.032$\\
$B^+\to \overline{D}_1'^0l^+\nu$&$0.195\pm0.060$\\
$B^+\to \overline{D}_0^0l^+\nu$&$0.435\pm0.075$\\
\hline\noalign{\smallskip}
$B^+\to (D^{(*)}\pi)_{nr} l^+\nu$&$0.17\pm0.14$\\
\hline\noalign{\smallskip}
$B^+\to X_cl^+\nu$&$10.90\pm0.14$\\
\hline\noalign{\smallskip}
\end{tabular}
\end{center}
\caption{Semileptonic branching fractions $\mathcal{B}(B^+\to X^{(i)}_cl^+\nu)$
taken or calculated from HFAG averages~\cite{HFAG} as described in the text.}
\label{final}
\end{table}
Possible experimental issues that might be the source for these puzzles are:
\begin{itemize}
\item Exclusive decay channels $B\to X_{c}^{i}l\nu$ into final states $X_{c}^{i}$ 
      not measured yet could contribute significantly to the 
      inclusive semileptonic decay rate. 
      Such transitions could be for example $B\to D'^{(*)}l\nu$, where $D'^{(*)}$ 
      might be the recently discovered resonances $D(2550)$ and $D^{*}(2600)$~\cite{D2S}. 
      In Ref.~\cite{BLT} a rough estimation suggested that a combined branching fraction 
      of about $1\%$ could be realized in nature for $B\to D'^{(*)}l\nu$ whereas in 
      Ref.~\cite{BroadProb} it was argued that such a large branching fraction would 
      be difficult to understand theoretically.
\item It is possible that not all $D^{**}$ decay channels were incorporated
      when determining $\mathcal{B}(B\to D_{1}/D_{2}l\nu)$ from the 
      measured product branching fractions. For example, there might be 
      $D_{1}\to D^*\pi \pi$, $D_{2}\to D^*\pi \pi$ (upper limits are given in Ref.~\cite{D1eins}), 
      $D_1/D_2\to D_0/D'_1\pi$ and $D_{2}\to D\eta$ decays with sizeable branching 
      fractions~\cite{FlorianDiss}. If true this would relax both the "gap problem" and the 
      "$\frac{1}{2}$ vs. $\frac{3}{2}$ puzzle" at the same time. 
\item Another possibility would be that the branching fraction of $B\to D^{(*)}l\nu$ decays is 
      experimentally underestimated, which would ease the "gap problem" but not the 
      "$\frac{1}{2}$ vs. $\frac{3}{2}$ puzzle". 
      However, $\mathcal{B}(B\to D^{(*)}l\nu)$ is measured with high precision.
      As a consequence, one would need to enlarge these branching fractions signifcantly more 
      than it is allowed by the quoted uncertainty in order to relax the "gap problem".\\
      One possible effect that could lead to a biased estimate of the $B\to D^{*}l\nu$ 
      branching fraction is an overestimate of the reconstruction efficiency of the 
      low-energy pion appearing in the $D^{*}$ decay to a $D$ and a $\pi$. It should be
      noted though that the experimental measurement that has a very strong weight in
      the average extracts $\mathcal{B}(B\to D^{*}l\nu)$ from a global fit
      to kinematical distributions without relying on the reconstruction of the
      low-energy pion from the $D^{*}$ decay~\cite{Aubert:2008yv}.\\
      If there is no experimental problem with the reconstruction of the low-energy pions
      or other issues relevant to the analyses, another explanation of underestimated 
      $B\to D^{(*)}l\nu$ branching fractions could be overestimated $D$ and/or $D^{*}$ 
      branching fractions. However, $D$-meson branching fractions are very well determined 
      by experiments running on the $\psi(3770)$ resonance, such as CLEO-c or BES-III. 
      Since the $\psi(3770)$ decays into $D\overline{D}$, absolute branching-fraction measurements 
      are possible by tagging one $D$-meson and measuring the decay of the other one into a 
      specific final state.\\ 
      For the $D^{*}$-meson, possible electromagnetic decays not measured yet are 
      $D^{*} \to D e^{+} e^{-}$ and $D^{*} \to D \gamma \gamma$. 
      These decays would have to compete at least with $D^{*} \to D \gamma$ in order to have a 
      sizeable effect on $\mathcal{B}(B\to D^{*}l\nu)$. This would come as a real 
      surprise since one would expect a rate suppression of these decays of the order the 
      fine-structure constant $\alpha \approx 1/137$ with respect to $D^{*} \to D \gamma$.
\item Reconstructing $B\to D'_1l\nu$ and $B\to D_0l\nu$ with $D_0 / D'_1 \to D^{*} \pi$
      is not an easy experimental task as the $D_0$ and the $D_1'$ are very broad resonances 
      and therefore hard to distinguish from non-resonant $(D^{(*)}\pi)_{nr}$ final 
      states. Therefore, the correct values for $\mathcal{B}(B\to D'_1l\nu)$ and 
      $\mathcal{B}(B\to D_0l\nu)$ could be indeed smaller than the HFAG averages, 
      which would relax the "$\frac{1}{2}$ vs. $\frac{3}{2}$ puzzle", but not the "gap problem".
\item Non-resonant decays $B\to (D^{(*)}\pi)_{nr}l\nu$ could fill the gap. 
      If this is true, this would suggest a serious problem in the $B\to D^{(*)}\pi l\nu$ and/or 
      $B\to D^{**}(D^{(*)}\pi)l\nu$ analysis since the $B\to D^{(*)}\pi l\nu$
      together with the $B\to D^{**}(D^{(*)}\pi)l\nu$ results leave only a
      small space for $B\to (D^{(*)}\pi)_{nr}l\nu$ decays. In addition, theoretical 
      expectations do not support a large branching fraction for non-resonant 
      $B\to (D^{(*)}\pi)_{nr}l\nu$ decays~\cite{BLT}. 
\item There might be contributions from yet to be discovered $B\to (D^{(*)}\pi\pi)_{nr}l\nu$ 
      or $B\to (D^{(*)}\eta)_{nr}l\nu$ decays, which would ease the "gap problem". 
      Such decays have not been observed yet and we did not investigate them in our analysis
      since our general findings do not prefer large contributions from high-mass states 
      like $B\to D'^{(*)} l\nu$ or from non-resonant decays $B\to (D^{(*)}\pi)_{nr}l\nu$ 
      so that we don't expect significant contributions from $B\to (D^{(*)}\pi\pi)_{nr}l\nu$ 
      or $B\to (D^{(*)}\eta)_{nr}l\nu$ decays either. Moreover, by adding too many 
      free parameters to the problem our analysis would loose in sensitivity. 
\end{itemize}

Kinematical distributions of the lepton energy $E_l$ and the hadronic invariant mass $m_{X_c}$ 
measured in inclusive $B\to X_cl\nu$ decays are sensitive to the composition of exclusive final 
states containing mesons with charm. Usually, moments of these kinematical distributions are 
used to extract non-perturbative parameters of a Heavy Quark Expansion (HQE)~\cite{HQE1,HQE2,HQE3,HQE4,HQE5,HQE6} with 
the aim to measure the CKM matrix element $\left|V_{cb}\right|$ (e.g. Ref.~\cite{BabarMom}) 
with highest precision. In this paper, we make use of such moment measurements to fit exclusive 
branching fractions $\mathcal{B}(B^+\to X_c^il^+\nu)$ with the aim to shed additional light 
on a solution to the puzzles described above. We investigate the contributions to the inclusive 
branching fraction from exclusive final states $X_c^i =D$, $D^*$, $D_0$, $D'_1$, $D_1$, $D_2$,
$(D^{(*)}\pi)_{nr}$, and $D'^{(*)}$. 
Hereby, we assume that $\overline{D}'^{0}$ and $\overline{D}'^{*0}$ can be identified 
with the observed $D(2550)$, respectively, $D^{*}(2600)$ state.
One should stress that a moment of a kinematical distribution for any specific exclusive decay 
$B\to X_c^{i}l\nu$, with $X_c^{i}$ being a resonant state such as $D$, $D^*$, $D_0$, $D'_1$, $D_1$, $D_2$ , 
or $D'^{(*)}$, does not depend on the branching fractions of such a resonance decaying into 
specific final states. Therefore, branching-fraction values found by the
fit being larger than the directly measured values may indicate that
the $X_c^{i}$ decay branching-fractions assumed are overestimated.

In Section~\ref{sec:1}, we describe the moments entering our analysis as fit inputs. 
Section~\ref{sec:2} provides information concerning the Monte-Carlo events used for 
the calculation of the moments for an exclusive decay. In Section~\ref{sec:3}, we 
outline the fit procedure and its validation, and we present the fit results in 
Section~\ref{sec:4}. In the last section we give a summary.

\section{MOMENTS IN SEMILEPTONIC DECAYS}
\label{sec:1}

For our analysis we use three different kinds of moments: moments of the electron-energy spectrum, 
of the hadronic mass spectrum and of the combined hadronic energy-mass spectrum, which were measured at the 
experiments \babar~\cite{BabarMom}, Belle~\cite{BelleLepMom,BelleMassMom}, CLEO~\cite{CleoMass}, 
and DELPHI~\cite{DelphiLepMom}. In Table~\ref{data}, we quote the moment measurements 
to which we fit the branching fractions.\\
In the following, moments which correspond to a single decay mode we refer to as "exclusive moments",
while when summing over exclusive decay modes we refer to the term "inclusive moments".\\
We calculate the theoretical prediction for these moments from Monte-Carlo (MC) simulated events 
using the following estimators, where the nomenclature is based on Ref.~\cite{BabarMom}:
\begin{itemize}
\item The estimator for the first electron-energy moment $M_1$ is given by\\
\begin{equation}
M_1\left(E_{cut_0}\right)=\left\langle E\right\rangle_{E_{cut_0}}=\frac{\sum\limits^{E_{i}>E_{cut_0}}_{i}{g_iE_{i}}}{\sum\limits^{E_i>E_{cut_0}}_{i}{g_i}}, 
\end{equation}
where $E_{cut_0}$ is the lower electron-energy cut-off above which the electron energies are 
included in the calculation of the moment and $E_i$ is the energy of the electron of the 
$i$-th event in the $B$-meson rest frame. To switch between different form-factor models 
in exclusive decays we introduce the event weights $g_i$.\\ 
For higher moments, the estimator is given by
\begin{equation}
\begin{split}
M_k\left(E_{cut_0}\right)=\left\langle\left(E-\left\langle E\right\rangle_{E_{cut_0}}\right)^k\right\rangle=\frac{\sum\limits^{E_i>E_{cut_0}}_{i}{g_i\left(E_i-\left<E\right>_{E_{cut_0}}\right)^k}}{\sum\limits^{E_i>E_{cut_0}}_{i}{g_i}},
\end{split}
\end{equation}
with $k>1$.\\
For later convenience, the exclusive and inclusive moments are arranged in vectors:
\begin{equation}
\begin{split}
\vec{M}=\left( {\begin{array}{*{20}c}
   {M_1\left(E_{cut_0}\right) }  \\
   {M_1\left(E_{cut_1}\right) }  \\
   \vdots   \\
   {M_2\left(E_{cut_0}\right) }  \\
   {M_2\left(E_{cut_1}\right) }  \\
   \vdots   \\
   {M_3\left(E_{cut_0}\right) }  \\
   {M_3\left(E_{cut_1}\right) }  \\
   \vdots   \\
\end{array}} \right)=\left( {\begin{array}{*{20}c}
   {\left\langle {E} \right\rangle_{E_{cut_0} } }  \\
   {\left\langle {E} \right\rangle_{E_{cut_1} } }  \\
   \vdots   \\
   {\left\langle {\left( {E  - \left\langle {E } \right\rangle } \right)^2 } \right\rangle_{E_{cut_0} } }  \\
   {\left\langle {\left( {E  - \left\langle {E } \right\rangle } \right)^2 } \right\rangle_{E_{cut_1} } }  \\
   \vdots   \\
   {\left\langle {\left( {E  - \left\langle {E } \right\rangle } \right)^3 } \right\rangle_{E_{cut_0} } }  \\
   {\left\langle {\left( {E  - \left\langle {E } \right\rangle } \right)^3 } \right\rangle_{E_{cut_1} } }  \\
   \vdots   \\
\end{array}} \right).\\
\end{split}
\end{equation}
Here, $E_{cut_i}$ denotes again the corresponding lower electron-energy cut-off.\\
We define in addition the vector
\begin{equation}
\begin{split}
\left<\vec{E}\right>=\left( {\begin{array}{*{20}c}
   {\left<E\right>_{E_{cut_0}} }  \\
   {\left<E\right>_{E_{cut_1}} }  \\
   \vdots   \\
   {\left<E^2\right>_{E_{cut_0}} }  \\
   {\left<E^2\right>_{E_{cut_1}} }  \\
   \vdots   \\
   {\left<E^3\right>_{E_{cut_0}} }  \\
   {\left<E^3\right>_{E_{cut_1}} }  \\
   \vdots   \\
\end{array}} \right).\\
\end{split}
\end{equation}

\item The non-central moments of the hadronic mass spectrum in $B\to X_cl\nu$ decays are defined 
as the mean of powers of the invariant hadronic mass. Again they are measured as a function of a 
lower lepton ($e$ or $\mu$) momentum cut-off $p_{cut_0}$ in the $B$-meson rest frame.\\
The estimators of the mass moments are given by:
\begin{equation}
\left\langle m^k\right\rangle_{p_{cut_0}}=\frac{\sum\limits^{p_i>p_{cut_0}}_{i}{g_im^k_{X_i}}}{\sum\limits^{p_i>p_{cut_0}}_{i}{g_i}},
\end{equation}
with $m_{X_i}$ being the invariant hadronic mass of event $i$.\\
The estimator of the central mass moments are defined as
\begin{equation}
\left\langle m^2_{centr}\right\rangle_{p_{cut_0}}=\frac{\sum\limits^{p_i>p_{cut_0}}_{i}{g_i\left(m_{X_i}^2-\bar{M}_D^2\right)}}{\sum\limits^{p_i>p_{cut_0}}_{i}{g_i}},
\end{equation}
as well as
\begin{equation}
\left\langle m^4_{centr}\right\rangle_{p_{cut_0}}=\frac{\sum\limits^{p_i>p_{cut_0}}_{i}{g_i\left(m_{X_i}^2-\bar{M}_D^2\right)^2}}{\sum\limits^{p_i>p_{cut_0}}_{i}{g_i}}.
\end{equation}
Here, $i$ runs over all events for which $p_i>p_{cut_0}$, where $p_i$ is the lepton momentum 
in the semileptonic decay, $p_{cut_0}$ the cut-off momentum, $g_i$ the event weight 
and $\bar{M}_D=\left(m_D+3m_{D^*}\right)/4=1.973\rm\,{GeV/c^2}$, with $m_D$ and $m_{D^*}$ 
the masses of the $D$ meson and $D^*$ meson, respectively.\\
Again, these moments are written in form of a vector:
\begin{equation}
\begin{split}
\left\langle \vec{ m }\right\rangle=\left( {\begin{array}{*{20}c}
   {\left\langle {m} \right\rangle_{p_{cut_0} } }  \\
   {\left\langle {m} \right\rangle_{p_{cut_1} } }  \\
   \vdots   \\
   {\left\langle {m^2 } \right\rangle_{p_{cut_0} } }  \\
   {\left\langle {m^2 } \right\rangle_{p_{cut_1} } }  \\
   \vdots   \\
   {\left\langle {m^3} \right\rangle_{p_{cut_0} } }  \\
   {\left\langle {m^3} \right\rangle_{p_{cut_1} } }  \\ 
   \vdots   \\
\end{array}} \right),
\end{split}
\end{equation}
whereby the vector of central moments with respect to $\bar{M}^2_D$ is defined analogously
\begin{equation}
\begin{split}
\left\langle \vec{ m }\right\rangle_{centr}=\left( {\begin{array}{*{20}c}
   {\left\langle {m^2_{centr} } \right\rangle_{p_{cut_0}} }  \\
   {\left\langle {m^2_{centr} } \right\rangle_{p_{cut_1}} }  \\
   \vdots   \\
   {\left\langle {m^4_{centr}} \right\rangle_{p_{cut_0}} }  \\
   {\left\langle {m^4_{centr}} \right\rangle_{p_{cut_1}} }  \\ 
   \vdots   \\
\end{array}} \right).
\end{split}
\end{equation}

\item In Ref.~\cite{GambinoUraltsev} a measurement of combined mass-energy moments was proposed. 
These moments are for instance better controlled theoretically and therefore they may result in 
a more reliable extraction of higher-order non-pertubative HQE parameters. Hence, a more accurate 
determination of the Standard Model parameters $|V_{cb}|$, the charm quark mass $m_c$ and the 
bottom quark mass $m_b$ should be possible. The first three even combined mass-energy moments 
were measured by the \babar collaboration~\cite{BabarMom}. Here, we use the following estimators 
for the prediction of these moments:
\begin{equation}
\begin{split}
\left\langle n^k\right\rangle_{p_{cut_0}}=\frac{1}{\sum\limits^{p_i>p_{cut_0}}_i{g_i}}\sum\limits^{p_i>p_{cut_0}}_i{g_i\left( m_{X_i}^2c^4-2\tilde{\Lambda}E_{X_i}+\tilde{\Lambda}^2\right)^{k/2}},
\end{split}
\end{equation}
where $i$ runs over all events for which $p_i>p_{cut_0}$, $E_{X_i}$ denotes the hadronic energy 
and $m_{X_i}$ the invariant mass of the hadronic system $X_i$, $p_i$ is the momentum of the 
involved lepton measured in the $B$-meson rest frame, $p_{cut_0}$ is the lower momentum cut-off, 
$g_i$ is again the corresponding event weight, and $\tilde{\Lambda}=0.65\,\rm{GeV}$~\cite{GambinoUraltsev}.\\
These moments are also arranged in a vector:
\begin{equation}
\begin{split}
\left\langle \vec{ n }\right\rangle=\left( {\begin{array}{*{20}c}
   {\left\langle {n^2} \right\rangle_{p_{cut_0} } }  \\
   {\left\langle {n^2} \right\rangle_{p_{cut_1} } }  \\
   \vdots   \\
   {\left\langle {n^4 } \right\rangle_{p_{cut_0} } }  \\
   {\left\langle {n^4 } \right\rangle_{p_{cut_1} } }  \\
   \vdots   \\
   {\left\langle {n^6} \right\rangle_{p_{cut_0} } }  \\
   {\left\langle {n^6} \right\rangle_{p_{cut_1} } }  \\
   \vdots   \\
\end{array}} \right).\\
\end{split}
\end{equation}
\end{itemize}
In Table~\ref{data}, we quote the moments measured at a specific lower cut-off which are used 
in the fit procedure. Since the measurements were unfolded for efficiency and detector resolution 
effects they can be directly compared with theoretical calculations.\\ 
Moments with lower cut-offs $E_{cut}$ or $p_{cut}$ close to each other are highly correlated 
and can result in numerical problems such as non-positive-definiteness of the final covariance 
matrices. As a consequence, we select data from a subset of available lower cut-offs to avoid 
these problems.

\begin{table}
\begin{center}
\begin{tabular}{cccc}
\hline\noalign{\smallskip}
&Exp.&$E_{cut}[\rm{GeV}]$ or $p_{cut}[\rm{GeV/c}]$&Ref.\\
\hline\noalign{\smallskip}
 $M_1$&\babar&0.6, 0.8, 1.0, 1.2, 1.5&\cite{BabarMom}\\
 $M_2$&\babar&0.6, 0.8, 1.0, 1.2, 1.5&\cite{BabarMom}\\
 $M_3$&\babar&0.6, 0.8, 1.0, 1.2, 1.5&\cite{BabarMom}\\
 $\left\langle m^1\right\rangle$&\babar&1.1, 1.3, 1.5, 1.7, 1.9&\cite{BabarMom}\\
 $\left\langle m^2\right\rangle$&\babar&0.8, 1.2, 1.4, 1.6, 1.8&\cite{BabarMom}\\
 $\left\langle m^3\right\rangle$&\babar&0.9, 1.1, 1.5, 1.7, 1.9&\cite{BabarMom}\\
 $\left\langle m^4\right\rangle$&\babar&0.8, 1.0, 1.2, 1.6, 1.8&\cite{BabarMom}\\
 $\left\langle m^5\right\rangle$&\babar&0.9, 1.1, 1.3, 1.5, 1.9&\cite{BabarMom}\\
 $\left\langle m^6\right\rangle$&\babar&0.8, 1.0, 1.2, 1.4, 1.6&\cite{BabarMom}\\
 $\left\langle n^2\right\rangle$&\babar&0.8 - 1.9, in steps of 0.1&\cite{BabarMom}\\
 $\left\langle n^4\right\rangle$&\babar&0.8 - 1.9, in steps of 0.1&\cite{BabarMom}\\
 $\left\langle n^6\right\rangle$&\babar&0.8 - 1.9, in steps of 0.1&\cite{BabarMom}\\
 $M_1$&Belle&1.0, 1.4&\cite{BelleLepMom}\\
 $M_2$&Belle&0.6, 1.4&\cite{BelleLepMom}\\
 $M_3$&Belle&0.8, 1.2&\cite{BelleLepMom}\\
 $M_4$&Belle&0.6, 1.2&\cite{BelleLepMom}\\
 $\left\langle m^2\right\rangle$&Belle&0.7 - 1.9, in steps of 0.2&\cite{BelleMassMom}\\
 $\left\langle m^4\right\rangle$&Belle&0.7 - 1.9, in steps of 0.2&\cite{BelleMassMom}\\
$\left\langle m^2_{centr}\right\rangle$&CLEO&1.0, 1.5&\cite{CleoMass}\\
 $\left\langle m^4_{centr}\right\rangle$&CLEO&1.0, 1.5&\cite{CleoMass}\\
 $M_1$&DELPHI&0.0&\cite{DelphiLepMom}\\
 $M_2$&DELPHI&0.0&\cite{DelphiLepMom}\\
 $M_3$&DELPHI&0.0&\cite{DelphiLepMom}\\
 \hline\noalign{\smallskip}
\end{tabular}
\end{center}
\caption{Experimentally measured moments used to constrain exclusive semileptonic branching 
fractions in $B\to X_cl\nu$ decays.}
\label{data}
\end{table}

\section{MODELLING OF $B\to X_{c}^{i}l\nu$ DECAYS}\label{sec:2}
For the calculation of the exclusive moments we use for every mode $5\cdot10^6$ MC events 
generated with the \textit{EvtGen} event generator~\cite{EvtGen}. After the generation we 
use the \textit{XslFF reweighting package}~\cite{Reweigh} to reweight the events according 
to more up-to-date form-factor models:
\begin{itemize}
\item $B\to Dl\nu$ decays are modeled according to the Heavy Quark Effective Theory (HQET) 
model with the Caprini-Lellouch-Neubert (CLN) parametrization~\cite{CLN}.
\item $B\to D^*l\nu$ are modeled according to the HQET model with the CLN parametrization~\cite{CLN}.
\item $B\to D^{**}l\nu$ decays are modeled according to approximation B1 of the Leibovich-Ligeti-Stewart-Wise (LLSW) 
      model~\cite{LLSW}.
\item $B\to (D^{(*)}\pi)_{nr} l\nu$ decays are modeled according to the Goity-Roberts model~\cite{Goity}.
\item $B\to D'^{(*)}l\nu$ decays are modeled according to the Bernlochner-Ligeti-Turczyk (BLT) model~\cite{BLT}.
       Here, we assume that the $2S$ states $D'$ and $D'^{*}$ can be identified 
       with the observed resonances $D(2550)$ and $D^*(2600)$~\cite{D2S} 
       and accordingly assign the masses and widths to the measured values given 
       in Ref.~\cite{D2S}:
\begin{itemize}
\item $m(D'^0)=2.5394\,\rm{GeV/c^2}$, $\Gamma(D'^0)=130\,\rm{MeV}$
\item $m(D'^\pm)=2.5394\,\rm{GeV/c^2}$, $\Gamma(D'^\pm)=130\,\rm{MeV}$
\item $m(D'^{*0})=2.6087\,\rm{GeV/c^2}$, $\Gamma(D'^{*0})=93\,\rm{MeV}$
\item $m(D'^{*\pm})=2.6213\,\rm{GeV/c^2}$, $\Gamma(D'^{*\pm})=93\,\rm{MeV}$
\end{itemize}
assuming $m(D(2550)^0)=m(D(2550)^{\pm})$, $\Gamma(D(2550)^0)=\Gamma(D(2550)^{\pm})$.
\end{itemize}
The parameters for the CLN model are taken from Ref.~\cite{HFAG} and were obtained 
by a global fit to all available $\mathcal{B}(B\to D^*l\nu)$ measurements:
\begin{itemize}
\item $\rho^2_{A_1}(1)=1.207\pm 0.026$,
\item $R_1(1)=1.403\pm 0.033$,
\item $R_2(1)=0.854\pm 0.020$,
\end{itemize}
with associated correlation coefficients:
\begin{itemize}
\item $\rho_{\rho^2,R1}=0.566$
\item $\rho_{\rho^2,R2}=-0.807$
\item $\rho_{R1,R2}=-0.759$
\end{itemize}

The parameter for the LLSW model is set to (see Ref.~\cite{LLSW})
\begin{itemize}
\item $\hat{\tau}'=-1.5\pm 0.5$
\end{itemize}
with the estimated uncertainty taken from Ref.~\cite{LLSW}.\\
The BLT parameters are chosen to be equal to (see Ref.~\cite{BLT})
\begin{itemize}
\item $\beta_0=0.13$, $\beta_1=1.95\pm 0.05$, $\beta_2=-7.0\pm 1.3$
\item $\beta^*_0=0.1$, $\beta^*_1=2.4\pm 0.1$, $\beta^*_2=-6.65\pm 3.15$
\end{itemize}
with estimated uncertainties taken from Ref.~\cite{BLT}.\\
The non-resonant decays $B\to (D^{(*)}\pi)_{nr} l\nu$ (with $l\in\{e,\mu\}$) 
are a mixture of several channels which is experimentally unknown (see Appendix~\ref{NrCalc}). 
Since Ref.~\cite{Goity} suggests that the mixture is dominated by $B\to (D\pi)_{nr} l\nu$ 
transitions, we choose the following composition:
\begin{itemize}
\item $B^0\to (D^{*-}\pi^0)_{nr}l^-\overline{\nu} : 8.6\%$
\item $B^0\to (\overline{D}^{*0}\pi^{-})_{nr}l^-\overline{\nu} : 17.3\%$
\item $B^0\to (D^{-}\pi^{0})_{nr}l^-\overline{\nu} : 24.7\%$
\item $B^0\to (\overline{D}^{0}\pi^{-})_{nr}l^-\overline{\nu} : 49.4\%$
\end{itemize} 
and analogously for $\overline{B}^0$ decays.\\
Non-resonant $B^+$ decays are composed as follows:
\begin{itemize}
\item $B^+\to (D^{*-}\pi^+)_{nr}l^+\nu : 15.4\%$
\item $B^+\to (\overline{D}^{*0}\pi^{0})_{nr}l^+\nu : 7.7\%$
\item $B^+\to (D^{-}\pi^{+})_{nr}l^+\nu : 51.3\%$
\item $B^+\to (\overline{D}^{0}\pi^{0})_{nr}l^+\nu : 25.6\%$
\end{itemize} 
and analogously for $B^-$ decays.\\

\section{MOMENT FITTER}\label{sec:3}

We perform a $\chi^2-fit$ to the mentioned moments, taking the full covariance 
of each moment vector into account.
\subsection{$\chi^2$-FUNCTION}\label{sec:chi2}

To estimate the semileptonic branching fractions in $B\to X_c^il\nu$ decays from moment 
measurements we minimize the following $\chi^2$-function using the MINUIT package~\cite{Minuit}:
\begin{equation}\label{chi2}
\chi^2=\sum\limits_j{\left(\chi ^2_{\rm{n},j}+\chi^2_{\rm{M},j}+\chi^2_{\rm{m},j}\right)}+\chi_{\rm{sum}}^2+\chi^2_{\rm{constr}},
\end{equation}with
\begin{equation}\label{chi2moment}
\begin{split}
\chi ^2_{\rm{n},j}  &= \left( {\vec{\left\langle {n } \right\rangle}\left(\mathcal{B}^i\right)  - \vec{\left\langle {n } \right\rangle}_{exp}^j  } \right)^T C_{\rm{n},j}^{ - 1} \left( {\vec{\left\langle {n } \right\rangle}\left(\mathcal{B}^i\right)  - \vec{\left\langle {n} \right\rangle}_{exp}^j } \right),\\
\chi^2_{\rm{M},j}  &= \left( {\vec{ {M } }\left(\mathcal{B}^i\right)  - \vec{ {M } }_{exp}^j  } \right)^T C_{\rm{M},j}^{ - 1} \left( {\vec{{M } }\left(\mathcal{B}^i\right)  - \vec{ {M } }_{exp}^j  } \right),\\
\chi^2_{\rm{m},j}  &= \left( {\vec{\left\langle {m } \right\rangle}\left(\mathcal{B}^i\right)  - \vec{\left\langle {m} \right\rangle}_{exp}^j  } \right)^T C_{\rm{m},j}^{ - 1} \left( {\vec{\left\langle {m } \right\rangle}\left(\mathcal{B}^i\right)  - \vec{\left\langle {m } \right\rangle}_{exp}^j  } \right),
\end{split}
\end{equation}
where the index $j$ runs over the different moment measurements, $C_{\rm{n},j}$, $C_{\rm{M},j}$ 
and $C_{\rm{m},j}$ are the sums of the related experimental and theoretical covariance 
matrices and the different theoretical moments are functions of the current values 
of the set of the fitted $N$ branching fractions 
$\mathcal{B}^i:=\mathcal{B}^i_{fit}(B^+\to X_c^il^+\nu)$, $i=1,\hdots, N$. 
The experimentally measured moment vectors are denoted with a subscript "{\it exp}".\\
It is assumed that there is no correlation between the moment vectors 
$\left\langle \vec{n}\right\rangle$, $\vec{M}$ and $\left\langle \vec{m}\right\rangle$ for both, 
experimentally measured and theoretically predicted. In case of \babar, electron moments were 
measured by leptonically tagging the other $B$-meson decay by a leptonic tag whereas mass and combined 
mass-energy moments were measured by fully reconstructing the other $B$-meson decay. Therefore, 
we consider the electron moments measured by \babar as being uncorrelated with the other moments. 
In case of the Belle measurements, the electron moments have been measured on the recoil
of fully reconstructed $B$-meson decays. In this case, there are correlations with
the hadronic moments measured by Belle which, however, we neglect in the fit since
the uncertainties on the electron moments are smaller for the \babar measurements.\\
The experimental correlation 
between the combined mass-energy moments $\left\langle n^k\right\rangle$, which were only 
measured by~{\babar}, and the mass moments $\left\langle m^l\right\rangle$ measured by~{\babar} 
is {\it a-priori} not negligible. Unfortunately, Ref.~\cite{KloseDiss} provides only a subset 
of the correlation coefficients and therefore the full set of correlations for these moments 
can not be included in the fit. Since the complete correlation matrix is missing, we decided 
to omit the correlations between $\left\langle n^k\right\rangle$ and 
$\left\langle m^l\right\rangle$ for the~{\babar} measurements and studied how the fit result changes when assigning a
correlation matrix which uses the partial information quoted in Ref.~\cite{KloseDiss} 
(see Section~\ref{Cov}).\\
In case of the theoretical prediction, the electron, the combined mass-energy and mass moments
are statistically correlated. However, since the experimental covariances are dominant, we
neglect these correlations.\\
The sum of the fitted branching fractions is always constrained to the inclusive branching
fraction by adding
\begin{equation}\label{inclConst}
\chi_{sum}^2=\frac{\left(\mathcal{B}_{exp}(B^+\to X_cl^+\nu)-\sum\limits_i\mathcal{B}^i\right)^2}{\sigma^2_{exp}},
\end{equation}
where $\mathcal{B}_{exp}(B^+\to X_cl^+\nu)$ and its uncertainty $\sigma_{exp}$ are given in the last line of Table~\ref{final}.\\
The number of degrees of freedom of the fit can be increased by using additional constraints 
for those branching fractions which are experimentally known from direct measurements as listed 
in Table~\ref{final}. This is achieved by adding the optional term:
\begin{equation}
\chi^2_{constr}=\sum\limits_{j}{\frac{\left(\mathcal{B}^j_{exp}(B^+\to X_c^jl^+\nu)-\mathcal{B}^j\right)^2}{\sigma^2_{exp,j}}},
\end{equation}
where $\mathcal{B}^j_{exp}(B\to X_c^jl\nu)$ are the values of those particular 
semileptonic branching fractions that are externally constrained with $\sigma_{exp,j}$ 
being their corresponding experimental uncertainties.\\

From the exclusive moments the inclusive moments are predicted by the sum of exclusive moments 
with coefficients proportional to the corresponding exclusive branching fractions.\\
Since the moment measurements get contributions from charged and neutral $B$-meson decays, 
we decompose the predicted moments into contributions from both. In this composition,
we assume equal semileptonic decay rates of charged and neutral $B$-meson decays as
a consequence of strong-isospin symmetry resulting in
\begin{equation}\label{isomodes}
\mathcal{B}(B^+\to X_c^il^+\nu)=\tau_{+0}\mathcal{B}(B^0\to X_c^il^+\nu).
\end{equation}
For any inclusive moment $\left\langle Z^k\right\rangle$ (except for central moments) 
with a lower leptonic energy or momentum cut-off "{\it cut}", one can write then
\begin{equation}\label{MomentCal}
\begin{split}
\left\langle Z^k\right\rangle _{cut}=\frac{\sum\limits_t^{N_m/2} {\mathcal{B}_t\left(\frac{f_{+0}}{1+f_{+0}}\frac{N_{cut}^t}{N_t} \left\langle Z^k\right\rangle_{cut}^t+\frac{1}{\tau_{+0}}\frac{1}{1+f_{+0}}\frac{N_{cut}^{\tilde{t}}}{N_{\tilde{t}}} \left\langle Z^k\right\rangle_{cut}^{\tilde{t}}  \right)}+\mathcal{B}_{nr}\frac{N_{cut}^{nr}}{N_{nr}}\left\langle Z^k\right\rangle_{cut}^{nr} }{\sum\limits_t^{N_m/2} {\mathcal{B}_t\left(\frac{f_{+0}}{1+f_{+0}}\frac{N_{cut}^t}{N_t}+ \frac{1}{\tau_{+0}}\frac{1}{1+f_{+0}}\frac{N_{cut}^{\tilde{t}}}{N_{\tilde{t}}}\right)}+\mathcal{B}_{nr}\frac{N_{cut}^{nr}}{N_{nr}} }
\end{split}
\end{equation}
Here, $N_m$ denotes the number of exclusive decay modes of charged and neutral $B$-mesons 
included in the fit, $N_t$ corresponds to the number of events of decay mode $t$ of a charged 
$B$-meson, $N^t_{cut}$ is the number of events of decay mode $t$ passing the cut-off, $\tilde{t}$ 
denotes the decay mode of a neutral $B$-meson that is isospin-symmetrical to the decay mode $t$, 
and $\left\langle Z^k\right\rangle^t_{cut}$ is the exclusive moment of mode $t$ measured 
at a cut-off "{\it cut}". We add the contribution of the non-resonant decays differently compared to the other decay channels, since the moments of the non-resonant decays were already calculated from a mixture of charged and neutral $B$-meson decays.\\

The central moment vector $\vec{Z}_{centr}$ is computed by a linear transformation from 
their non-central equivalents $\vec{Z}$ as $\vec{Z}_{centr}=J\vec{Z}$ with $J$ being
the Jacobian of the transformation $\vec{Z} \to \vec{Z}_{centr}$.\\

The theoretical inclusive covariance matrices are calculated from the set of theoretical 
exclusive covariance matrices. The inclusive moment vector $\left\langle\vec{Z}\right\rangle$ 
can be written as
\begin{equation}
 {\left\langle {\vec{Z} } \right\rangle }  = \left( {\begin{array}{*{20}c}
   {\frac{{\sum\limits_t {b^t_{cut_1} \left\langle {Z } \right\rangle^t_{cut_1} } }}{{\sum\limits_t {b^t_{cut_1} } }}}  \\
   {\frac{{\sum\limits_t {b^t_{cut_2} \left\langle {Z } \right\rangle^t_{cut_2} } }}{{\sum\limits_t {b^t_{cut_2} } }}}  \\
    \vdots   \\
   {\frac{{\sum\limits_t {b^t_{cut_n} \left\langle {Z } \right\rangle^t_{cut_n} } }}{{\sum\limits_t {b^t_{cut_n} } }}}  \\
\end{array}} \right),
\end{equation}
where $t$ runs over all incorporated decay modes of charged and neutral $B$-mesons, 
$b^t_{cut_i}=\tilde{f}_t\mathcal{B}_t\frac{N^t_{cut_i}}{N_t}$, $cut_i$ denotes either 
a lower lepton energy or momentum cut-off and $\tilde{f}_t$ either equals 
$\frac{f_{+0}}{1+f_{+0}}$ if $t$ corresponds to a decay mode of a charged $B$-meson 
or $\frac{1}{\tau_{+0}\left(1+f_{+0}\right)}$ if $t$ corresponds to a decay mode of 
a neutral $B$-meson.\\
$\vec{Z}$ can be rewritten as a sum of products of exclusive moment vectors 
$\left<\vec{Z}\right>_t$ and transformation matrices $F_t$:
\begin{equation}
\begin{split}
\left\langle {\vec{Z} } \right\rangle &=\sum\limits_t{ diag\left(\frac{b^t_{cut_1}}{{\sum\limits_t {b^t_{cut_1} } }}, \frac{b^t_{cut_2}}{{\sum\limits_t {b^t_{cut_2} } }},  \cdots,  \frac{b^t_{cut_n}}{{\sum\limits_t {b^t_{cut_n} } }}\right){\left\langle {\vec{Z} } \right\rangle _t } }\\
&=\sum\limits_t { F_t  {\left\langle {\vec{Z} } \right\rangle _t } }.
\end{split}
\end{equation}
As a result, the covariance matrix of an inclusive moment vector is given by
\begin{equation}
\begin{split}
C_{\left\langle\vec{Z}\right\rangle}=\sum_{t}{F_tC_{\left\langle\vec{Z}\right\rangle_t}F^T_t}.
\end{split}
\end{equation}
We approximate the theoretical covariance matrix of a central moment vector by
\begin{equation}\label{JTrafo}
C_{\langle\vec{Z}_{centr}\rangle}=JC_{\langle\vec{Z}\rangle}J^{T},
\end{equation}
with the Jacobian $J$ of the transformation $\vec{Z}\to\vec{Z}_{centr}$.

\subsection{STATISTICAL AND SYSTEMATIC COVARIANCES}\label{Cov}
\begin{itemize}
\item The computation of the statistical covariances is described in detail in Appendix~\ref{StatCov}
\item Within our fit model one systematic arises from the choice of the form-factor parameters. 
      To obtain an estimate for these systematics, the calculation of theoretical moments is 
      performed with randomly varied form-factor parameters. 
      The variation is made by adding Gaussian random numbers to the nominal values of the 
      form-factor parameters, whereby we account for the measured correlations of the parameters 
      $\rho^2_{A_1}(1)$, $R_1(1)$ and $R_2(1)$ in the case of $B^+\to\overline{D}^{*0}l^+\nu$. 
      The central values are chosen to be zero and the Gaussian standard deviations are chosen 
      to be the parameter uncertainties as quoted in Section~\ref{sec:2}.\\
      The variations of all form factors are performed 100 times and afterwards the systematic 
      covariances of the obtained moment vectors are calculated for each mode, according to\\ 
      $C_{\left\langle\vec{Z}\right\rangle}=\left\langle\left(\left\langle\vec{Z}\right\rangle-\overline{\left\langle\vec{Z}\right\rangle}\right)\left(\left\langle\vec{Z}\right\rangle-\overline{\left\langle\vec{Z}\right\rangle}\right)^T\right\rangle$,
      where $\vec{Z}$ is the respective moment vector.
\item As mentioned in Section~\ref{sec:chi2} the full correlation matrix between mass and 
      combined mass-energy moments is missing. To investigate the influence of the omitted 
      correlations we assume that the correlation matrix between $\left\langle n^2\right\rangle$ 
      and $\left\langle m^2\right\rangle$, which is quoted in Ref.~\cite{KloseDiss}, holds  
      between all $\left\langle n^k\right\rangle$ and $\left\langle m^l\right\rangle$, which 
      should overestimate the correlations. When performing the fits with these correlations, 
      the fit results change typically only by about $10\%$ and in rare cases up to $40\%$ with 
      respect to the associated fit uncertainty. The uncertainties themselves change by about $10\%$. 
      Therefore, the qualitative picture does not change and we conclude that neglecting
      the correlations between mass and mass-energy moments are of minor importance.
\item The sensitivity of the fit with respect to the modeling of $B\to(D^{(*)}\pi)_{nr}l\nu$ 
      decays is checked by varying the mixture of $(D\pi)_{nr}$ and $(D^*\pi)_{nr}$ 
      final states: 
      For the extreme case that only $B\to (D\pi)_{nr} l\nu$ is simulated to predict 
      the moments for $B\to(D^{(*)}\pi)_{nr}l\nu$, none of the fit results for any 
      branching fraction changes more than $20\%$ with respect to the associated fit error. 
      For the case that only $B\to (D^{*}\pi)_{nr} l\nu$ is simulated to predict the 
      moments for \\$B\to(D^{(*)}\pi)_{nr}l\nu$, the situation becomes more involved. 
      In this case, there are some scenarios in which the results for $D^{\ast\ast}$ 
      and $(D^{*}\pi)_{nr}$ change more than $100\%$ with respect to the fit uncertainty. 
      While Ref.~\cite{Goity} suggests a clear dominance of $B\to (D\pi)_{nr}l\nu$ 
      decays over $B\to (D^*\pi)_{nr}l\nu$ decays, the experimental constraints do not
      exclude the contrary (see Appendix~\ref{NrCalc}). Therefore, we provide in Appendix~\ref{ResultsDstar} 
      the results for our considered fit scenarios in which $B\to(D^{(*)}\pi)_{nr}l\nu$ decays 
      are modeled exclusively with $B\to(D^{*}\pi)_{nr}l\nu$ decays.
\item The widths of the broad $D^{**}$ mesons are only known with an uncertainty of 
      about $30\%$. By varying their widths in the fit we study the influence on the fit result. 
      Most of the branching fractions are - compared to the corresponding fit uncertainty - 
      quite insensitive to this variation. Only ${\mathcal{B}}(B^+\to\overline{D}^{0}_{1}l^+\nu)$ 
      and ${\mathcal{B}}(B^+\to\overline{D}'^{0}_{1}l^+\nu)$ show some sensitivity. But even
      in these case their fit values do not change more than $O(50\%)$ compared to their associated 
      fit uncertainties. Hence, our qualitative findings are not modified by this effect.
\end{itemize}

\subsection{FIT VALIDATION}\label{FitVal}
No change of the fit results is observed when the initial values of the branching fractions 
$\mathcal{B}(B^+\to X^i_cl^+\nu)$ are chosen arbitrarily in the interval 
[0,$\mathcal{B}(B^+\to X_cl^+\nu)$].\\
To check for a potential bias of the fit results and the calculated uncertainties, 
the fit is tested with a "statistical ensemble of pseudo-experiments". 
This "pseudo-data" is obtained from a statistical variation of a particular mixture 
of the exclusive moments according to the experimental covariance matrix $C_{exp}$ as follows:\\
Nominal inclusive moment vectors $\left<\vec{Z}_i\right>_0$ ($i$ runs over the different used moment vectors) 
calculated according to the measured set of values of the branching fractions $\{\mathcal{B}^{nom}_i\}$ are chosen. 
Then, random vectors $\vec{x}=L\vec{z}$ are added, where $L$ is defined by $C_{exp}=LL^T$ and $\vec{z}$ is a 
standard normal distributed random vector. Since the theoretical exclusive moment vectors are 
also only known within statistical uncertainties due to the limited Monte-Carlo statistics, 
they are varied in an analogous way.\\
From the set of ensemble fits normalized residuals (as defined in the Appendix~\ref{Toy}) and p-value distributions are obtained. We performed the fit on 1000 pseudo data sets for each fit scenario in Section~\ref{sec:4}. A typical example of residuals and p-value distributions can be found in Appendix~\ref{Toy}.\\ 
Some small fit bias is observed. Compared to the fit uncertainty the bias for the 
individual branching fractions found is:
$5\%$ or smaller for $\mathcal{B}(B^+\to\overline{D}^0l^+\nu)$, $7\%$ or smaller for
$\mathcal{B}(B^+\to\overline{D}^{*0}l^+\nu)$, up to $25\%$ but typically of order $10\%$ 
for the narrow-width $D^{**}$, $6\%$ or smaller for the broad-width $D^{**}$, and
$5\%$ or smaller for $(D^{(*)}\pi)_{nr}$.
Depending on the decay also a small underestimation of the fit uncertainty is
observed. Compared to the true uncertainty the underestimation for the individual 
branching fractions found is: $7\%$ or smaller for $\mathcal{B}(B^+\to\overline{D}^0l^+\nu)$,
no significant underestimation for $\mathcal{B}(B^+\to\overline{D}^{*0}l^+\nu)$, $15\%$ or
less for the narrow-width $D^{**}$, $10\%$ or less for the broad-width $D^{**}$, and
$8\%$ or smaller for $(D^{(*)}\pi)_{nr}$.\\
The observed p-value distributions are not perfectly uniform, typically with a mean of 
$0.56\pm0.01$ and a RMS of $0.29\pm0.01$. This small deviation from a uniform distribution 
is caused by approximating the covariance of the theoretical inclusive central moments with 
Eq.~\ref{JTrafo}. If the central moments are not included in the fit, the p-value distribution 
gets uniform and the (small) intrinsic bias as well as the (slight) underestimation of the 
uncertainties observed in the residuals is significantly reduced.

\section{RESULTS}\label{sec:4}

\begin{table*}
\scriptsize
\centering
\begin{tabular}{|c|c|c|c|c|c|c|c|c|c|}
 \hline
\multirow{2}{*}{$X_c$} & \multicolumn{2}{c|}{Fit 1}& \multicolumn{2}{c|}{Fit 2}& \multicolumn{2}{c|}{Fit 3}& \multicolumn{2}{c|}{Fit 4}&Measured\\ \cline{2-9}
&U/C&$\mathcal{B}[\%]$&U/C&$\mathcal{B}[\%]$&U/C&$\mathcal{B}[\%]$&U/C&$\mathcal{B}[\%]$&$\mathcal{B}[\%]$\\ \hline \hline
$ \overline{D}^{0}$&x/-&$2.43 \pm 0.15$&x/-&$2.43 \pm 0.15$&x/-&$2.37 \pm 0.15$&x/-&$2.61 \pm 0.14$&$2.30 \pm 0.10 $\\ \hline
$ \overline{D}^{*0}$&x/-&$5.81 \pm 0.16$&x/-&$5.86 \pm 0.16$&x/-&$5.89 \pm 0.16$&x/x &$5.53 \pm 0.09$&$5.34 \pm 0.12 $\\ \hline
$ \overline{D}_1^{0}$&x/-&$2.13 \pm 0.67$&x/-&$1.33 \pm 0.33$&x/x &$0.67 \pm 0.07$&x/-&$1.34 \pm 0.33$&$0.65 \pm 0.07 $\\ \hline
$ \overline{D}_2^{0}$&x/-&$-0.52 \pm 0.59$&x/x &$0.28 \pm 0.03$&x/-&$0.61 \pm 0.29$&x/x &$0.28 \pm 0.03$&$0.28 \pm 0.03 $\\ \hline
$ \overline{D}'^{0}_1$&x/-&$0.21 \pm 0.34$&x/-&$0.02 \pm 0.31$&x/-&$0.19 \pm 0.34$&x/-&$0.19 \pm 0.30$&$0.20 \pm 0.06 $\\ \hline
$ \overline{D}_0^{0}$&x/-&$0.48 \pm 0.33$&x/-&$0.60 \pm 0.32$&x/-&$0.93 \pm 0.26$&x/-&$0.55 \pm 0.31$&$0.43 \pm 0.07 $\\ \hline
$ \overline{D}'^{0}$&-/-&-&-/-&-&-/-&-&-/-&-&-\\ \hline
$ \overline{D}'^{*0}$&-/-&-&-/-&-&-/-&-&-/-&-&-\\ \hline
$ (D^{(*)}\pi)_{nr}$&x/-&$0.36 \pm 0.16$&x/-&$0.37 \pm 0.16$&x/-&$0.25 \pm 0.15$&x/-&$0.27 \pm 0.16$&-\\ \hline
$\overline{D}^{0}_1/\overline{D}^{0}_2$& \multicolumn{2}{c|}{$1.61\pm0.33$}& \multicolumn{2}{c|}{$1.61\pm0.33$}& \multicolumn{2}{c|}{$1.28\pm0.29$}& \multicolumn{2}{c|}{$1.62\pm0.32$}&$0.94\pm0.08$\\ \hline
$\overline{D}^{0}_0/\overline{D}'^{0}_1$& \multicolumn{2}{c|}{$0.69\pm0.54$}& \multicolumn{2}{c|}{$0.62\pm0.53$}& \multicolumn{2}{c|}{$1.12\pm0.50$}& \multicolumn{2}{c|}{$0.73\pm0.52$}&$0.63\pm0.10$\\ \hline
$\overline{D}^{0}_0/\overline{D}'^{0}_1/(D^{(*)}\pi)_{nr}$& \multicolumn{2}{c|}{$1.06\pm0.40$}& \multicolumn{2}{c|}{$0.99\pm0.39$}& \multicolumn{2}{c|}{$1.37\pm0.37$}& \multicolumn{2}{c|}{$1.00\pm0.39$}&$0.63\pm0.10$\\ \hline
$\sum_iX_c^i$& \multicolumn{2}{c|}{$10.90\pm0.14$}& \multicolumn{2}{c|}{$10.90\pm0.14$}& \multicolumn{2}{c|}{$10.90\pm0.14$}& \multicolumn{2}{c|}{$10.77\pm0.13$}&$9.21\pm0.20$\\ \hline
$X_c$& \multicolumn{2}{c|}{}& \multicolumn{2}{c|}{}& \multicolumn{2}{c|}{}& \multicolumn{2}{c|}{}&$10.90 \pm 0.14$\\ \hline
\hline
$\chi^2/dof$& \multicolumn{2}{c|}{75/104 = 0.73}& \multicolumn{2}{c|}{77/105 = 0.74}& \multicolumn{2}{c|}{80/105 = 0.77}& \multicolumn{2}{c|}{84/106 = 0.80}&-\\ \hline
p-value& \multicolumn{2}{c|}{0.98}& \multicolumn{2}{c|}{0.98}& \multicolumn{2}{c|}{0.96}& \multicolumn{2}{c|}{0.94}&-\\ \hline
\end{tabular}
\caption{Results for fits with moments of semileptonic decays with hadronic final states containing 
$D$, $D^*$, any $D^{**}$ or $(D^{(*)}\pi)_{nr}$. Hereby, ``U/C'' stands for ``used/constrained'' 
and the ``x'' denotes ``yes'', whereas ``-'' denotes ``no'', respectively. 
This table is further discussed and described in the text of Section~\ref{sec:4}.}
\label{Results1} 
\end{table*}

\begin{table*}
\scriptsize
\centering
\begin{tabular}{|c|c|c|c|c|c|c|c|c|c|}
\hline
\multirow{2}{*}{$X_c$} & \multicolumn{2}{c|}{Fit 5}& \multicolumn{2}{c|}{Fit 6}& \multicolumn{2}{c|}{Fit 7}& \multicolumn{2}{c|}{Fit 8}&Measured\\ \cline{2-9}
&U/C&$\mathcal{B}[\%]$&U/C&$\mathcal{B}[\%]$&U/C&$\mathcal{B}[\%]$&U/C&$\mathcal{B}[\%]$&$\mathcal{B}[\%]$\\ \hline \hline
$ \overline{D}^{0}$&x/x &$2.41 \pm 0.08$&x/x &$2.44 \pm 0.08$&x/x &$2.42 \pm 0.08$&x/-&$2.61 \pm 0.14$&$2.30 \pm 0.10 $\\ \hline
$ \overline{D}^{*0}$&x/x &$5.59 \pm 0.09$&x/x &$5.60 \pm 0.09$&x/x &$5.63 \pm 0.09$&x/x &$5.51 \pm 0.10$&$5.34 \pm 0.12 $\\ \hline
$ \overline{D}_1^{0}$&x/-&$1.10 \pm 0.30$&x/-&$1.38 \pm 0.18$&x/x &$0.78 \pm 0.07$&x/-&$1.42 \pm 0.33$&$0.65 \pm 0.07 $\\ \hline
$ \overline{D}_2^{0}$&x/x &$0.28 \pm 0.03$&x/x &$0.28 \pm 0.03$&x/x &$0.30 \pm 0.03$&x/x &$0.28 \pm 0.03$&$0.28 \pm 0.03 $\\ \hline
$ \overline{D}'^{0}_1$&x/-&$0.32 \pm 0.29$&x/-&$0.19 \pm 0.27$&x/x &$0.22 \pm 0.06$&x/-&$0.54 \pm 0.39$&$0.20 \pm 0.06 $\\ \hline
$ \overline{D}_0^{0}$&x/-&$0.78 \pm 0.29$&x/x &$0.46 \pm 0.07$&x/x &$0.56 \pm 0.07$&x/-&$0.54 \pm 0.31$&$0.43 \pm 0.07 $\\ \hline
$ \overline{D}'^{0}$&-/-&-&-/-&-&x/-&$0.35 \pm 0.12$&x/-&$-0.32 \pm 0.22$&-\\ \hline
$ \overline{D}'^{*0}$&-/-&-&-/-&-&-/-&-&-/-&-&-\\ \hline
$ (D^{(*)}\pi)_{nr}$&x/-&$0.20 \pm 0.15$&x/-&$0.31 \pm 0.11$&x/-&$0.31 \pm 0.07$&x/-&$0.21 \pm 0.16$&-\\ \hline
$\overline{D}^{0}_1/\overline{D}^{0}_2$& \multicolumn{2}{c|}{$1.38\pm0.30$}& \multicolumn{2}{c|}{$1.66\pm0.17$}& \multicolumn{2}{c|}{$1.08\pm0.07$}& \multicolumn{2}{c|}{$1.70\pm0.33$}&$0.94\pm0.08$\\ \hline
$\overline{D}^{0}_0/\overline{D}'^{0}_1$& \multicolumn{2}{c|}{$1.10\pm0.49$}& \multicolumn{2}{c|}{$0.64\pm0.29$}& \multicolumn{2}{c|}{$0.78\pm0.09$}& \multicolumn{2}{c|}{$1.07\pm0.58$}&$0.63\pm0.10$\\ \hline
$\overline{D}^{0}_0/\overline{D}'^{0}_1/(D^{(*)}\pi)_{nr}$& \multicolumn{2}{c|}{$1.30\pm0.36$}& \multicolumn{2}{c|}{$0.96\pm0.20$}& \multicolumn{2}{c|}{$1.09\pm0.10$}& \multicolumn{2}{c|}{$1.28\pm0.44$}&$0.63\pm0.10$\\ \hline
$\sum_iX_c^i$& \multicolumn{2}{c|}{$10.68\pm0.12$}& \multicolumn{2}{c|}{$10.65\pm0.12$}& \multicolumn{2}{c|}{$10.56\pm0.12$}& \multicolumn{2}{c|}{$10.78\pm0.13$}&$9.21\pm0.20$\\ \hline
$X_c$& \multicolumn{2}{c|}{}& \multicolumn{2}{c|}{}& \multicolumn{2}{c|}{}& \multicolumn{2}{c|}{}&$10.90 \pm 0.14$\\ \hline
\hline
$\chi^2/dof$& \multicolumn{2}{c|}{88/107 = 0.82}& \multicolumn{2}{c|}{89/108 = 0.83}& \multicolumn{2}{c|}{110/109 = 1.01}& \multicolumn{2}{c|}{82/105 = 0.79}&-\\ \hline
p-value& \multicolumn{2}{c|}{0.91}& \multicolumn{2}{c|}{0.90}& \multicolumn{2}{c|}{0.45}& \multicolumn{2}{c|}{0.94}&-\\ \hline
\end{tabular}
\caption{Results for fits with moments of semileptonic decays with hadronic final states containing 
$D$, $D^*$, any $D^{**}$ or $(D^{(*)}\pi)_{nr}$. Hereby, ``U/C'' stands for ``used/constrained'' and the ``x'' 
denotes ``yes'', whereas ``-'' denotes ``no'', respectively. 
This table is further discussed and described in the text of Section~\ref{sec:4}.}
\label{Results2} 
\end{table*}

\begin{figure*}
 
  \hspace{-0.03\textwidth}
   \includegraphics[width=1.1\textwidth]{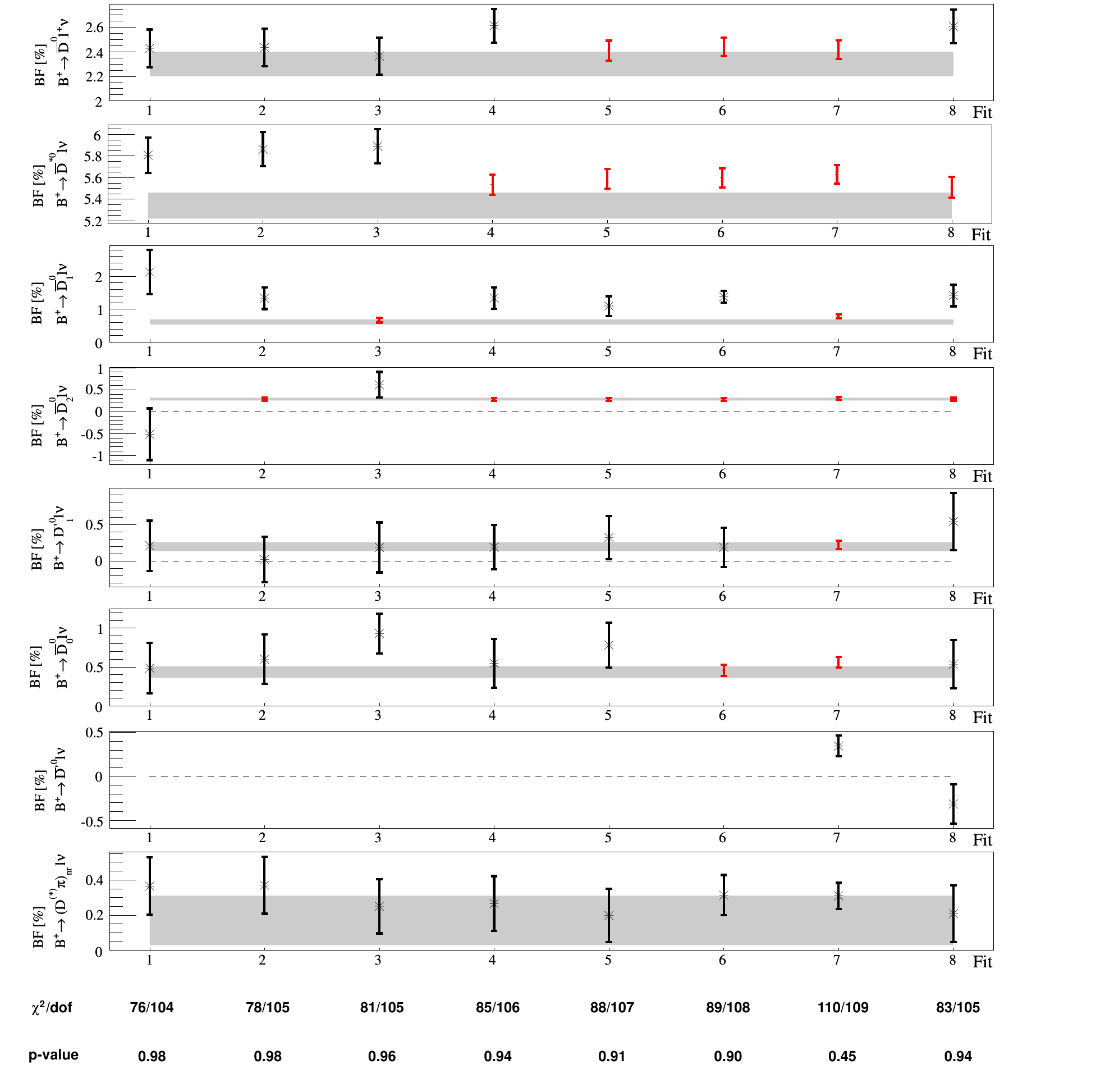}
\caption{Depiction of the fit results as quoted in Tables~\ref{Results1} and~\ref{Results2}. 
In each subplot the abscissa indicates a distinctive fit scenario labeled by a number, 
whereas the ordinate represents the branching fraction. The results for a constrained 
branching fraction are depicted as red points, whereas the results for a unconstrained 
branching fraction are depicted as a black star. The grey bands correspond to the particular 
one-sigma error band of the direct branching-fraction measurements. 
If the ordinate includes zero, a dashed black line visualizes the zero line.}    
\label{Plots}
\end{figure*}

The results of a representative set of fits are quoted in Tables~\ref{Results1} and~\ref{Results2} 
and are plotted in Fig.~\ref{Plots}. We provide further fit results in Appendix~\ref{AddRes} and the related 
plots in Appendix~\ref{AddPlots}. 
In the result tables, every fit result is presented as a column, consisting of an upper part, 
which is split into two subcolumns, and a lower part. The upper part provides in its left 
subcolumn the informations about the fit constellation. The ``x'' or ``-'' on the lefthand side
denotes whether the moments of this special decay mode are used (``U'') in the fit or not, 
whereas the ``x'' or ``-'' on the righthand side denotes whether this particular branching fraction 
was or was not constrained (``C'') by a $\chi^2_{constr}$ term (as described in Section~\ref{sec:3}). 
In addition to the individual branching fraction results, also the sums
$\mathcal{B}(B^+\to \overline{D}^0_1l^+\nu) + \mathcal{B}(B^+\to \overline{D}^0_{2}l^+\nu)$
and
$\mathcal{B}(B^+\to \overline{D}^0_0l^+\nu) + \mathcal{B}(B^+\to \overline{D}'^0_{1}l^+\nu)$ 
as well as 
$\mathcal{B}(B^+\to \overline{D}^0_0l^+\nu) + \mathcal{B}(B^+\to \overline{D}'^0_{1}l^+\nu)+\mathcal{B}(B^+\to (D^{*}\pi)_{nr}l^+\nu)$ are quoted in order to see whether the ``$\frac{1}{2}$ vs. $\frac{3}{2}$ puzzle''
is relaxed within a given fit scenario. Finally, the sum of all fitted branching 
fractions and its uncertainty taking into account the correlations between the 
fitted $\mathcal{B}_i(B^+\to X_c^il^+\nu)$ is compared with the
inclusive $\mathcal{B}(B^+\to X_c l^+\nu)$.
The lower part quotes the $\chi^2$ together with the number of degrees of freedom 
($dof$) and the corresponding p-value. In the last column, the directly measured 
values are quoted for comparison.
\\
The plot in Fig.~\ref{Plots} is composed of several subplots, where each row 
presents the result of a particular branching fraction. The abscissa labels
a certain fit scenario with a number. 
In addition, the error band (grey band) of the direct measurement and the 
zero line (dashed black line) are visualized. Those branching fractions that
are treated as a fit parameter can be deduced directly from the given plots, 
because only for branching fractions used in the fit results are plotted. 
If a branching fraction is constrained in the fit by a $\chi^2_{constr}$ term 
the fit result is plotted as a red point. Otherwise, if the branching fraction 
is free to vary, the fit result is visualized by a black star. At the bottom of 
the plots we quote the $\chi^2$ value, the number of degrees of freedom and 
the associated p-value for each fit scenario.\\
For one of the fits (fit 2 of Table~\ref{Results1}) we show in Appendix~\ref{MomDistr} 
how the fit result compares with the measured moments.\\ 

In general, none of the scenarios we have studied could be really excluded 
by the moment fit. However, it should be stressed that the initial $\chi^2/dof$ 
using the measured branching fractions as given in Table~\ref{final} 
($\mathcal{B}(B\to (D^{(*)}\pi)_{nr}l\nu)$ is set to zero) 
is found to be $\chi^2/dof=394/112=3.6$ (p-value: $5\cdot10^{-33}$) with a constraint applied 
for the sum of the branching fractions. 
Without this constraint we find for the initial $\chi^2/dof=248/111=2.2$ 
(p-value: $1.8\cdot10^{-12}$). That is, the inclusively measured moments are not well 
described by the exclusive moments calculated from the measured branching fractions.
As a consequence, solving simultaneously the gap problem and the poor description
of the inclusive moments requires enhancing certain branching fractions compared
to others or adding yet unmeasured exclusive semileptonic decays or both.\\
A general finding of our analysis is that not a single decay mode alone is capable
of filling the gap and better describing the inclusive moments, but that always 
a set of branching fractions is increased by the fit. \\
We did not apply a constraint in the fit that forces the branching fractions 
to be positive in order to avoid fit biases for branching fractions that are 
close to zero. For fit scenarios in which one of the branching fractions is 
found to be significantly negative we find that this branching fraction is 
highly anti-correlated to another one. In such cases, the fit rather constrains 
the sum of two particular branching fractions than both individually. For this 
reason, we consider fit scenarios in which one of such two branching fractions 
is constrained to its directly measured value as being more robust and more
meaningful.
\begin{itemize}
\item {\it Fit without any constraint on exclusive branching fractions}:\\
In the first fit scenario, we apply no additional constraint and include those
decays in the fit which are known to contribute to $B^+\to X_cl^+\nu$.\\
In this first fit scenario, $\mathcal{B}(B^+\to \overline{D}^0_2l^+\nu)$ is fitted 
to a negative value, whereas $\mathcal{B}(B^+\to \overline{D}^0_1l^+\nu)$ is extremely 
large. This is due to a large anti-correlation between these two branching fractions
as discussed above. The correlation matrix for the result vector
\begin{equation}
\begin{split}
 \vec{ \mathcal{B} }=\left( {\begin{array}{*{20}c}
   {\mathcal{B}(B^+\to \overline{D}^0l^+\nu)}  \\
   {\mathcal{B}(B^+\to \overline{D}^{*0}l^+\nu)}  \\
   {\mathcal{B}(B^+\to \overline{D}^0_1l^+\nu)}  \\
   {\mathcal{B}(B^+\to \overline{D}^0_2l^+\nu)}  \\
   {\mathcal{B}(B^+\to \overline{D}'^0_1l^+\nu)}  \\
   {\mathcal{B}(B^+\to \overline{D}^0_0l^+\nu)}  \\
   {\mathcal{B}(B^+\to (D^{(*)}\pi)_{nr} l^+\nu)}  \\
\end{array}} \right)\\
\end{split}
\end{equation}
of Fit 1 quoted in Table~\ref{Results1} is
\begin{equation}
\begin{split}
\left( {\begin{array}{*{20}c}
    1 & -0.54 & 0.19 & 0.03 & -0.16 & -0.45 & 0.13 \\
    -0.54& 1 & -0.21 & 0.24 & -0.33 &  0.13   & 0.32  \\    
    0.19 &  -0.21 & 1 & -0.87 & 0.01 & -0.63  & 0.34 \\
    0.03 & 0.24 & -0.87 & 1 & -0.40 &  0.26   & 0.03   \\
    -0.16& -0.33 & 0.01 & -0.40 & 1 &  0.28  & -0.79  \\
    -0.45 & 0.13 & -0.63 & 0.26 & 0.28 &  1  & -0.62 \\
     0.13 &  0.32 & 0.34 & 0.03 & -0.79 & -0.62 & 1 \\
\end{array}} \right).
\end{split}
\end{equation}
showing the large negative correlation coefficient of $-0.87$ between
${\mathcal{B}(B^+\to \overline{D}^0_1l^+\nu)}$ and\\ 
${\mathcal{B}(B^+\to \overline{D}^0_2l^+\nu)}$.
As a consequence, the fit is not able to clearly distinguish between these two modes 
and, therefore, it is more meaningful to always either constrain one of these branching 
fractions to the directly measured value or even fix one of them at zero and always 
consider only the combination of the two.\\

\item {\it Constraints on $\mathcal{B}(B^+\to \overline{D}^0_2l^+\nu)$ or $\mathcal{B}(B^+\to \overline{D}^0_1l^+\nu)$}:\\
In Fit 2, we constrain $\mathcal{B}(B^+\to \overline{D}^0_2l^+\nu)$ to its measured value.\\
Taking into account additional, similar fit scenarios as quoted in Appendix~\ref{AddRes}, we find 
a combined branching fraction into narrow $D^{**}$ mesons of about $1.3\%$ to $1.7\%$ with 
uncertainties varying between $0.18\%$ and $0.33\%$. 
These fit results are clearly above the sum of the directly measured branching fractions:\\ 
$\mathcal{B}(B^+\to \overline{D}^0_1l^+\nu)+\mathcal{B}(B^+\to \overline{D}^0_2l^+\nu)=\left(0.94\pm 0.08\right)\%$.\\
The increase in $\mathcal{B}(B^+\to \overline{D}^0_1l^+\nu)+\mathcal{B}(B^+\to \overline{D}^0_2l^+\nu)$ 
points to a possible solution of the ``$\frac{1}{2}$ vs. $\frac{3}{2}$ puzzle'':
$\mathcal{B}(B^+\to \overline{D}^0_0l^+\nu)$ and $\mathcal{B}(B^+\to \overline{D}'^0_1l^+\nu)$
are both in very good agreement with the directly measured values quoted in Table~\ref{final}. 
As a consequence, the ratio 
$(\mathcal{B}(B^+\to \overline{D}^0_1l^+\nu)+\mathcal{B}(B^+\to \overline{D}^0_2l^+\nu))/(\mathcal{B}(B^+\to \overline{D}^0_0l^+\nu)+\mathcal{B}(B^+\to \overline{D}'^0_1l^+\nu))$ 
is increased in all fit scenarios where $\mathcal{B}(B^+\to \overline{D}^0_2l^+\nu)$
is constrained by a $\chi^2_{\rm{constr}}$ term. \\
However, if $\mathcal{B}(B^+\to \overline{D}^0_1l^+\nu)$ is constrained instead of 
$\mathcal{B}(B^+\to \overline{D}^0_2l^+\nu)$, the findings are bit different (see fit scenario 3).
In this case, not only $\mathcal{B}(B^+\to \overline{D}^0_2l^+\nu)$ is increased, 
which is expected due to the large anticorrelation with $\mathcal{B}(B^+\to \overline{D}^0_1l^+\nu)$, 
but also $\mathcal{B}(B^+\to \overline{D}^0_0l^+\nu)$.
As a result, $(\mathcal{B}(B^+\to \overline{D}^0_1l^+\nu)+\mathcal{B}(B^+\to \overline{D}^0_2l^+\nu))/(\mathcal{B}(B^+\to \overline{D}^0_0l^+\nu)+\mathcal{B}(B^+\to \overline{D}'^0_1l^+\nu))$ is of order one and the
``$\frac{1}{2}$ vs. $\frac{3}{2}$ puzzle'' would be even more pronounced.
This is a general finding in all fits where $\mathcal{B}(B^+\to \overline{D}^0_1l^+\nu)$ 
is constrained instead of $\mathcal{B}(B^+\to \overline{D}^0_2l^+\nu)$.
It turns out that the moment fits in which $\mathcal{B}(B^+\to \overline{D}^0_2l^+\nu)$
are constrained instead of $\mathcal{B}(B^+\to \overline{D}^0_1l^+\nu)$ result in slightly 
better p-values as can be seen in Appendix~\ref{AddRes} although the difference between
the two fit scenarios is small. 
\\
We note that one could have used lower bounds for the branching fractions into states 
containing a narrow $D^{**}$ instead of constraining them directly to its measured value
since only product branching-fractions are directly measured. We have tested this option
and find that either the fit results do not change compared to the case when constraining
the branching fraction (in case of Fit 2) or reproduce another already covered fit scenario
(Fit 3 would reproduce the result of Fit 1 without any constraint; Fit 4, 5 and 6 
would reproduce the result of Fit 2 with usual constraints; Fit 7 and 8 would reproduce 
the result of a fit with only $\mathcal{B}(B\to \overline{D}^0_2l^+\nu)$ constrained). 
For this reason, we restrict our studies to scenarios in which we constrain 
$\mathcal{B}(B^+\to D^{\ast\ast}l^+\nu)$ to the values quoted in Table~\ref{final}.
\item {\it Increased $\mathcal{B}(B^+\to \overline{D}^{*0}l^+\nu)$ values}:\\
A general and puzzling observation is that the result for $\mathcal{B}(B^+\to \overline{D}^{*0}l^+\nu)$ 
in Fit 1, 2 and 3 deviates significantly by about $10\%$ from its directly measured value. 
Given the fact that for $B\to D^*l\nu$ one has the most precisely measured branching 
fraction of all exclusive $B\to X_c^il\nu$ decays this result comes as a surprise, 
in particular because several different measurement techniques were used at the 
$B$-factory experiments \babar and Belle that reported the most precise measurements. 
Therefore, in Fit 4 we apply a constraint on $\mathcal{B}(B^+\to \overline{D}^{*0}l^+\nu)$.
The $\mathcal{B}(B^+\to \overline{D}^{*0}l^+\nu)$ result is then lowered but still about $5\%$
above the directly measured value. 
Taking into account all additional results as quoted in the appendix we find the
following general picture: without the constraining term the $\mathcal{B}(B^+\to \overline{D}^{*0}l^+\nu)$
results vary between $5.81\%$ and $5.96\%$ with uncertainties of about $0.15\%$. 
With the constraint applied, the deviation is reduced but the fit results are in general 
still above the directly measured value of $\left(5.34\pm 0.12\right)\%$ and are in the 
range between $5.55\%$ and $5.65\%$ with an uncertainty of about $0.09\%$.\\
If the directly measured values do not suffer from an unknown systematic effect,
this finding might be caused by the fit model: If, for example, a yet unconsidered 
$B\to X_c^il\nu$ decay has a significant branching fraction, the fit result for 
the branching fractions under study might be biased to higher values.\\
Another possibility to explain our findings is that certain moment measurements 
drive the fit result for $B^+\to \overline{D}^{*0}l^+\nu$. To check this we perform
the fit by including only one class of moments at a time: either electron-energy moments, 
or combined hadronic mass-energy moments, or hadronic mass moments (see Appendix~\ref{DstarSys}).
Since the number of input points is drastically reduced we perform these test fits
by constraining $\mathcal{B}(B^+\to \overline{D}^{0}l^+\nu)$,
$\mathcal{B}(B^+\to \overline{D}_{0}^{0}l^+\nu)$, and either 
$\mathcal{B}(B^+\to \overline{D}_{2}^{0}l^+\nu)$ or 
$\mathcal{B}(B^+\to \overline{D}_{1}^{0}l^+\nu)$.
These studies also show how these three classes of moments influence the final
fit uncertainties. In terms of decreasing fit uncertainties we find the following
order: mass-energy moments, electron moments, mass moments.\\
We find that both, the combined mass-energy moments and in particular the mass moments, 
push the branching fraction for $B^+\to \overline{D}^{*0}l^+\nu$ decays to quite high values
(and in turn the branching fractions for $B^+\to \overline{D}_{1/2}^{0}l^+\nu$ decays to 
lower ones): 
$\mathcal{B}(B^+\to \overline{D}^{*0}l^+\nu)=(6.20\pm 0.18)\%$ (mass moments)
and
$\mathcal{B}(B^+\to \overline{D}^{*0}l^+\nu)=(5.91\pm 0.25)\%$  (mass-energy moments)
compared to
$\mathcal{B}(B^+\to \overline{D}^{*0}l^+\nu)=(5.55\pm 0.21)\%$ (electron moments).\\
While the result for the mass-energy moments is consistent with both, the mass moments and 
the electron moments, there is some discrepany between the mass moments and the electron moments.
We checked for the fit using only mass moments the consistency between the {\babar} and Belle 
measurements by removing the Belle measurements and find no significant shift in the fit 
results, which is expected since the measured mass moments of {\babar} and Belle are in good 
agreement (see Appendix~\ref{MomDistr}). This test also allows to check the consistency of the
fit results when either only using mass moments or combined mass-energy moments measured by 
{\babar} only: also these two fit results agree within uncertainties.\\
When inspecting how well the combined mass-energy moments can be described by the fit 
one observes for very high cut values in the lepton energy that the fit model undershoots 
the measured data points (see e. g. the distributions shown in Appendix~\ref{MomDistr}). Therefore, 
we study in Appendix~\ref{DstarSys} as well how the results change in case of the fit using only
combined mass-energy moments when removing from the list of inputs the two data points at the 
highest cut values. We find an improvement in the $\chi^{2}$-value of the fit, but no significant 
change in the branching fraction results.

\item {\it Results for $\mathcal{B}(B^+\to \overline{D}^0l^+\nu)$}:\\
Generally, we find that the unconstrained branching fraction 
of $B^+\to Dl\nu$ decays is fitted to values between $2.36\%$ and $2.49\%$ with 
corresponding uncertainties of about $0.15\%$. Most often it exceeds $2.40\%$
but is in agreement with the corresponding direct measurement, 
$\mathcal{B}(B^+\to \overline{D}^0l^+\nu)=\left(2.30\pm 0.10\right)\%$. 
If a constraint is applied the values lie also in that range but the fit 
uncertainties shrink to $0.08\%$. 
Once $\mathcal{B}(B^+\to \overline{D}^{*0}l^+\nu)$ 
is constrained to its directly measured value (as in Fit 4 and others) one finds 
that $\mathcal{B}(B^+\to \overline{D}^0l^+\nu)$ is pushed upwards, away 
from its directly measured value. Therefore, we choose in Fit 5 a constellation 
in which $\mathcal{B}(B^+\to \overline{D}^0l^+\nu)$ is constrained, too. \\
It should be also noted that fits in which both, $\mathcal{B}(B^+\to \overline{D}^0l^+\nu)$
and $\mathcal{B}(B^+\to \overline{D}^{*0}l^+\nu)$, are constrained, 
result in a sum of exclusive branching fractions that is lower by about two 
standard deviations than the inclusive branching fraction $\mathcal{B}(B^+\to X_c l^+\nu)$.
Hence, the moment fit prefers an enhancement of branching fractions for 
semileptonic $B$-meson decays into low-mass charmed mesons $D$ and/or $D^{*}$. 
We note, that the fit quality is slightly better when constraining 
$\mathcal{B}(B^+\to \overline{D}^0l^+\nu)$ instead of 
$\mathcal{B}(B^+\to \overline{D}^{*0}l^+\nu)$.

\item {\it The role of $B^+\to \overline{D}^0_0 l^+\nu$, $B^+\to \overline{D}'^0_1l^+\nu$, and $B^+\to (D^{(*)}\pi)_{nr}l^+\nu$}:\\
The fitted value for $\mathcal{B}(B^+\to \overline{D}'^0_1l^+\nu)$, without its constraint 
applied, varies between $-0.12\%$ and $0.63\%$ with uncertainties varying between 
$0.27\%$ and $0.51\%$. That is, the fit is not very sensitive to this mode but the 
results are in agreement with the directly measured value $\left(0.195\pm 0.06\right)\%$. 
The individual direct measurements for $\mathcal{B}(B^+\to \overline{D}'^0_1l^+\nu)$ 
are not in good agreement with each other. Therefore, we repeat the fits with 
a rescaled uncertainty of $0.18\%$ as a result of requiring $\chi^2/dof=1$ in
the weighted average for $\mathcal{B}(B^+\to \overline{D}'^0_1l^+\nu)$.
We find that this change does not influence the results of these fit scenarios 
significantly.\\
For $\mathcal{B}(B^+\to \overline{D}^0_0l^+\nu)$, the fit results vary in 
general between $0.44\%$ and $0.76\%$ and the uncertainties vary between $0.07\%$ and 
$0.33\%$, whereas the directly measured value is $\left(0.43\pm 0.07\right)\%$. 
As already mentioned above, $\mathcal{B}(B^+\to \overline{D}^0_0l^+\nu)$ 
is significantly higher than its directly measured value if a constraint on 
$\mathcal{B}(B^+\to \overline{D}^0_1l^+\nu)$ is applied.\\
In Fit 6, we present fit results when in addition to 
$\mathcal{B}(B^+\to \overline{D}^0l^+\nu)$ and
$\mathcal{B}(B^+\to \overline{D}^{*0}l^+\nu)$ also $\mathcal{B}(B^+\to \overline{D}^0_0l^+\nu)$
is constrained. To obtain a similar good description of the moments as in Fit 5
the fit shifts $\mathcal{B}(B^+\to (D^{(*)}\pi)_{nr}l^+\nu)$
upwards. This can be understood due to a large anticorrelation between
$\mathcal{B}(B^+\to \overline{D}^0_0l^+\nu)$ and $\mathcal{B}(B^+\to \overline{D}'^0_1l^+\nu)$ 
on one side and $\mathcal{B}(B^+\to (D^{(*)}\pi)_{nr}l^+\nu)$
on the other side (see e.g. the correlation matrix quoted for Fit 1 in this section).\\

As a very general result $\mathcal{B}(B^+\to (D^{(*)}\pi)_{nr}l^+\nu)$ 
is found to vary between $0.2\%$ and $0.4\%$ with uncertainties varying between $0.07\%$ and $0.16\%$ to be compared with the constraint of $\left(0.17\pm0.14\right)\%$ 
obtained from $B^+\to D^{(*)}\pi l\nu$ and $B^+\to D^{**}(D^{(*)}\pi)l^+\nu$ measurements. 
Thus, the inclusively measured moments suggest that $B^+\to (D^{(*)}\pi)_{nr}l^+\nu$ contribute 
to the inclusive $B^+\to X_cl^+\nu$ branching fraction with a value compatible with direct measurements, 
so that $B^+\to (D^{(*)}\pi)_{nr}l^+\nu$ decays are not able to solve the gap problem, 
in agreement with theoretical expectations (see Ref.~\cite{BLT}).\\

The dependence of the fit results on the modelling of $B^+\to (D^{(*)}\pi)_{nr}l^+\nu$ 
deserves some attention. When only allowing $(D^{*}\pi )_{nr}$ final states one finds significantly 
different results for $B^+\to (D^{(*)}\pi)_{nr}l^+\nu$ and $B^+\to \overline{D}'^0_1l^+\nu$ decays.
However, one has in these fits a very strong anticorrelation between $\mathcal{B}(B^+\to \overline{D}'^0_1l^+\nu)$ 
and $\mathcal{B}(B^+\to (D^{*}\pi)_{nr}l^+\nu)$ so that one needs to
consider rather the sum 
$\mathcal{B}(B^+\to \overline{D}'^0_1l^+\nu)+\mathcal{B}(B^+\to (D^{*}\pi)_{nr}l^+\nu)$ 
instead of the individual values. Keeping this in mind, the qualitative findings are similar 
to the ones observed with our default $B^+\to (D^{(*)}\pi)_{nr}l^+\nu$ modelling.

\item {\it No significant contribution from $B^+\to \overline{D}'^{(*)0} l^+\nu$}:\\
Besides the known decay modes discussed so far there might be $B^+\to \overline{D}'^{(*)0}l^+\nu$ 
transitions which contribute significantly to $B\to X_cl^+\nu$, {\it i.e.} with an order of 
magnitude of $1\%$. We assume that $\overline{D}'^0$ and $\overline{D}'^{*0}$ can 
be identified with the states $D(2550)$, respectively, $D^{*}(2600)$ found by \babar.
Since, as in the case of $B^+\to \overline{D}^0_1/\overline{D}^0_2l^+\nu$, 
very large anti-correlations between $\mathcal{B}(B^+\to \overline{D}'^0l^+\nu)$ and 
$\mathcal{B}(B^+\to \overline{D}'^{*0}l^+\nu)$ are found, {\it i.e}. of order $-0.90$,
it is sufficient to study fit scenarios in which either only $\mathcal{B}(B^+\to \overline{D}'^{0} l^+\nu)$
or only $\mathcal{B}(B^+\to \overline{D}'^{*0} l^+\nu)$ is added as a fit parameter.
Therefore, we add in Fit 7 and 8 the decay mode $B^+\to \overline{D}'^0l^+\nu$.
Additional fit scenarios including both modes are shown in Appendix~\ref{AddRes}.\\
In Fit 7, we constrain any mode except $B^+\to \overline{D}'^0l^+\nu$ and $B^+\to (D^{(*)}\pi)_{nr}l^+\nu$. 
It can be seen that the constrained branching fractions are still pushed upwards 
and that $\mathcal{B}(B\to \overline{D}'^0l^+\nu)$ is only of the order 
of $0.3\%$. Moreover, compared to the other fit scenarios discussed so far the p-value 
is significantly smaller. Therefore, $B^+\to \overline{D}'^{(*)0}l^+\nu$ does not seem to
be able to deliver the main contribution to solve the gap problem. 
In Fit 8, we provide another scenario with less constraints and there 
$\mathcal{B}(B\to \overline{D}'^0l^+\nu)$ becomes even negative.\\
From these observations and from the additional results in Appendix~\ref{AddRes}, 
we conclude: if there is any significant contribution from $B^+\to \overline{D}'^{(*)0}l^+\nu$ at all,
it is likely to be small and far from being sufficient to solve the gap problem.\\
\end{itemize}

\section{SUMMARY}\label{sec:5}
This paper is motivated by the various puzzles which occur in the sector of semileptonic 
$B\to X_cl\nu$ decays. Up to now the inclusive decay rate can not be saturated by the 
so far measured exclusive branching fractions ("gap problem"). In addition, theoretical 
predictions of the ratio of the branching fraction into states containing narrow $D^{**}$
and into states containing broad $D^{**}$-mesons are in conflict with the experimental 
data ("$\frac{1}{2}$ vs. $\frac{3}{2}$ puzzle"). Furthermore, the individual measurements 
of $\mathcal{B}(B^+\to \overline{D}'^0_1l^+\nu)$ do not agree very well among each other.\\
To find answers to the solution of these problems we extract the corresponding branching 
fractions of the exclusive modes from a fit to the moments of inclusive electron energy, 
hadronic mass and combined hadronic mass-energy spectra in which we constrain the sum
of exclusive branching fractions to the measured inclusive branching fraction $\mathcal{B}(B^+\to X_{c}l^+\nu)$.
We study the results when applying different sets of additional constraints coming
from the direct measurements of exclusive branching fractions.\\
Our main findings are:
\begin{itemize}
\item No single exclusive decay is able to solve the "gap problem" alone, and hence a 
      variety of fit scenarios is found to be able to describe the moments with a similar 
      good fit quality.\\ 
      For $\mathcal{B}(B^+\to D^{**}l^+\nu)$ decays, the fit uncertainties are much larger 
      than the ones from the direct branching-fraction measurements. 
      For $\mathcal{B}(B^+\to \overline{D}^{(*)0}l^+\nu)$ decays, the fit uncertainties 
      are slightly larger or of the same size as the directly measured values, and for 
      $B^+\to (D^{(*)}\pi)_{nr}l^+\nu$ decays they are of the same size or even smaller 
      than the value obtained from direct measurements. 
\item The individual classes of moments have different impacts on the final fit
      uncertainties. The fit uncertainties decrease in size when using either only 
      combined hadronic energy-mass moments, or only electron energy moments, or 
      only hadronic mass moments.
\item Semileptonic decays $B^+\to \overline{D}'^{(*)0}l^+\nu$ have been discussed in the 
      literature as possible candidates to solve the "gap problem". When $\overline{D}'^{0}$ 
      and $\overline{D}'^{*0}$ are identified with the observed $D(2550)$, respectively, 
      $D^{*}(2600)$ state, we find that $\mathcal{B}(B^+\to \overline{D}'^{(*)0}l^+\nu)$ is 
      small and is by far not able to saturate the inclusive semileptonic decay rate.
\item $B^+\to \overline{D}_1l^+\nu$ and $B^+\to \overline{D}_2l^+\nu$ are not easily 
      distinguished by the fit and hence the fit constrains rather the sum of these
      two branching fractions than their individual values. To avoid negative
      branching-fraction values one has to constrain at least one of these two
      branching fractions to its directly measured value. In general, the sum 
      $\mathcal{B}(B^+\to \overline{D}^0_1l^+\nu)+\mathcal{B}(B^+\to \overline{D}^0_2l^+\nu)$ 
      is found to be larger than its directly measured value.\\
      In cases in which
      $\mathcal{B}(B^+\to \overline{D}^0_2l^+\nu)$ is constrained, 
      $\mathcal{B}(B^+\to \overline{D}^0_1l^+\nu)+\mathcal{B}(B^+\to \overline{D}^0_2l^+\nu)$ 
      is significantly enhanced compared to 
      $\mathcal{B}(B^+\to \overline{D}^0_0l^+\nu)+\mathcal{B}(B^+\to \overline{D}'^0_1l^+\nu)$.
      If true, this would relax the "$\frac{1}{2}$ vs. $\frac{3}{2}$ puzzle" and
      would be possibly caused by neglecting yet unobserved $D_1$ and/or $D_2$ 
      decay modes $D_{1,2} \to Y$ when calculating $\mathcal{B}(B^+\to \overline{D}^0_{1,2}l^+\nu)$ 
      from the measured product branching-fractions \\$\mathcal{B}(B^+\to \overline{D}^0_{1,2}(Y)l^+\nu)$.\\
      On the contrary, if $\mathcal{B}(B^+\to \overline{D}^0_1l^+\nu)$ is constrained,
      $\mathcal{B}(B^+\to \overline{D}^0_1l^+\nu)+\mathcal{B}(B^+\to \overline{D}^0_2l^+\nu)$ 
      and $\mathcal{B}(B^+\to \overline{D}^0_0l^+\nu)+\mathcal{B}(B^+\to \overline{D}'^0_1l^+\nu)$
      are found to be of similar size because not only $\mathcal{B}(B^+\to \overline{D}^0_2l^+\nu)$
      is enhanced in the fit, but also $\mathcal{B}(B^+\to \overline{D}^0_0l^+\nu)$. 
      The latter fit constellation, which would even more pronounce the 
      "$\frac{1}{2}$ vs. $\frac{3}{2}$ puzzle", is slightly disfavoured compared 
      to the former although the differences in fit quality between the two fit 
      constellations are small.
\item $\mathcal{B}(B^+\to \overline{D}^0l^+\nu)$ is found to be slightly above 
      but in good agreement with its direct measurement unless a constraint is 
      applied on $\mathcal{B}(B^+\to \overline{D}^{*0}l^+\nu)$. 
      In this case, $\mathcal{B}(B^+\to \overline{D}^0l^+\nu)$ is significantly 
      shifted upwards.
\item Surprisingly, the most precisely measured branching fraction, 
      $\mathcal{B}(B^+\to \overline{D}^{*0}l^+\nu)$, is found to be $5\%$ to $10\%$ 
      larger than its directly measured value, depending on whether the branching 
      fraction is or is not constrained in the fit by its direct measurement. 
      The fit is able to describe the moments slightly better when 
      $\mathcal{B}(B^+\to \overline{D}^0l^+\nu)$ is constrained instead of 
      $\mathcal{B}(B^+\to \overline{D}^{*0}l^+\nu)$. 
\item The preference for higher $\mathcal{B}(B^+\to \overline{D}^{*0}l^+\nu)$ 
      values and in turn for smaller $\mathcal{B}(B^+\to \overline{D}^0_1l^+\nu)+\mathcal{B}(B^+\to \overline{D}^0_2l^+\nu)$
      values is mainly driven by the moments of the measured combined hadronic 
      mass-energy spectra and in particular by the mass-moment measurements. 
\item In general, $\mathcal{B}(B^+\to \overline{D}'^0_1l^+\nu)$ is found to 
      be small and in agreement with the HFAG average. 
      The fit errors for $\mathcal{B}(B^+\to \overline{D}'^0_1l^+\nu)$ are too 
      large though in order to draw a final conclusion about the inconsistency between 
      the direct measurements.
\item The fit results for semileptonic $B$-meson decays into non-resonant $(D^{(*)}\pi)_{nr}$
      final states, modeled by the Goity-Roberts model, are often slightly larger than, 
      but in good agreement with the value obtained from branching-fraction measurements 
      of $B\to D^{(*)}\pi l\nu$ and $B\to D^{**}(D^{(*)}\pi)l\nu$ decays.
\item The findings for $\mathcal{B}(B^+\to \overline{D}^0_0l^+\nu)$, 
      $\mathcal{B}(B^+\to \overline{D}'^0_1l^+\nu)$ and 
      $\mathcal{B}(B^+\to  (D^{(*)}\pi)_{nr}l^+\nu)$ show 
      a dependence how the $(D^{(*)}\pi)_{nr}$ part is modelled. As long as
      there is a substantial $(D \pi)_{nr}$ component
      all findings described above are unchanged. Once one goes to the
      extreme case that there are only $(D^{*}\pi)_{nr}$ but no 
      $(D \pi)_{nr}$ final states the fit produces enhanced 
      $\mathcal{B}(B^+\to (D^{*}\pi)_{nr}l^+\nu)$ values on one hand, and reduced 
      and even often negative $\mathcal{B}(B^+\to \overline{D}'^0_1l^+\nu)$ values 
      on the other hand. In these fits, there is a very large anticorrelation between 
      $\mathcal{B}(B^+\to  (D^{*}\pi)_{nr}l^+\nu)$ 
      and $\mathcal{B}(B^+\to \overline{D}'^0_1l^+\nu)$. As a result, the fit
      rather constrains their sum instead of their individual values. Seen from this
      point of view the other general findings agree qualitatively with the ones found
      in the fits with our default $B^+\to  (D^{*}\pi)_{nr}l^+\nu$ modelling.
\end{itemize}

\section*{Acknowledgments}
We thank Christoph Schwanda and Phillip Urquijo for their helpful remarks concerning the 
correlation matrices of Belle's moment measurements.

\newpage
\appendix

\section{CALCULATION OF STATISTICAL COVARIANCES}\label{StatCov}
Event samples for different cuts overlap and thus the corresponding moments are correlated.
All events of a sample of events $A$, which is associated with a lower momentum or 
energy cut-off $a$, are part of a sample of events $C$ being associated with a 
cut-off $c$ (without limitation of generality: $a>c$). \\
The moment $\left\langle {Z^k } \right\rangle_C$ corresponding to sample $C$ (with $N_C$ events) 
can be calculated with (the following calculation is based on Ref.~\cite{KloseDiss})
\begin{equation}
\left\langle {Z^k } \right\rangle_C=\frac{\sum\limits_{i=1}^{N_C}g_iZ_i^k}{\sum\limits_{i=1}^{N_C}g_i}.
\end{equation}
Note that a subscripted \textit{capital} letter indicates that the mean of the moment 
corresponds to a dedicated sample whereas small letters denote a particular cut-off.\\
Considering a third sample $B$, with $C = B\cup A$ and $A\cap B=\emptyset$, it follows
\begin{equation}
\begin{split}
\left\langle {Z^k } \right\rangle_C  &= \frac{{\sum\limits_{i = 1}^{N_A } {g_iZ_i^k }  + \sum\limits_{i = N_A  + 1}^{N_C } {g_iZ_i^k } }}{{\sum\limits_{i = 1}^{N_C } {g_i} }}=\frac{{\sum\limits_{i = 1}^{N_A } {g_iZ_i^k }\frac{\sum\limits_{i = 1}^{N_A } {g_i}}{\sum\limits_{i = 1}^{N_A } {g_i }}  + \sum\limits_{i = N_A  + 1}^{N_C } {g_iZ_i^k }\frac{\sum\limits_{i = N_A  + 1}^{N_C } {g_i }}{\sum\limits_{i = N_A  + 1}^{N_C } {g_i}} }}{{\sum\limits_{i = 1}^{N_C } {g_i} }}\\
&= \frac{{\left\langle {Z^k } \right\rangle _A\sum\limits_{i = 1}^{N_A } {g_i }   + \left\langle {Z^k } \right\rangle _B\sum\limits_{i = N_A  + 1}^{N_C } {g_i}  }}{{{\sum\limits_{i = 1}^{N_C } {g_i} } }}.\\
\end{split}
\end{equation}
Thus, the covariance of $\left\langle {Z^k } \right\rangle_C$ and $ \left\langle {Z^l } \right\rangle _A$ is
\begin{equation}
\begin{split}
C\left( {\left\langle {Z^k } \right\rangle _A ,\left\langle {Z^l } \right\rangle _C } \right) &= C\left( {\left\langle {Z^k } \right\rangle _A ,\frac{{\left\langle {Z^l } \right\rangle _A \sum\limits_{i = 1}^{N_A } {g_i }  + \left\langle {Z^l } \right\rangle _B\sum\limits_{i = N_A  + 1}^{N_C } {g_i}  }}{{\sum\limits_{i = 1}^{N_C } {g_i}}}} \right)\\
&= C\left( {\left\langle {Z^k } \right\rangle _A ,\left\langle {Z^l } \right\rangle _A \frac{{\sum\limits_{i = 1}^{N_A } {g_i } }}{{\sum\limits_{i = 1}^{N_C } {g_i}}}} \right) + C\left( {\left\langle {Z^k } \right\rangle _A ,\left\langle {Z^l } \right\rangle _B \frac{{\sum\limits_{i = N_A  +1}^{N_C } {g_i}}}{{\sum\limits_{i = 1}^{N_C } {g_i} }}} \right)\\
&= \frac{{\sum\limits_{i = 1}^{N_A } {g_i }}}{{\sum\limits_{i = 1}^{N_C } {g_i}}}C\left( {\left\langle {Z^k } \right\rangle _A ,\left\langle {Z^l } \right\rangle _A } \right)\\
\end{split}
\end{equation}
since $A\cap B=\emptyset$. Further, it is
\begin{equation}
\begin{split}
C\left( {\left\langle {Z^k } \right\rangle _A ,\left\langle {Z^l } \right\rangle _A } \right)= C\left(\frac{{\sum\limits_{i = 1}^{N_A } {g_i }  {Z^k }  }}{{\sum\limits_{i = 1}^{N_A } {g_i}}},\frac{{\sum\limits_{i = 1}^{N_A } {g_i } {Z^l } }}{{\sum\limits_{i = 1}^{N_A } {g_i}}} \right)=\frac{\sum\limits_{i = 1}^{N_A } {g_i^2 }}{\left(\sum\limits_{i = 1}^{N_A } {g_i }\right)^2}C\left(Z^k,Z^l\right)_A
\end{split}
\end{equation}
and therefore
\begin{equation}
\begin{split}
C\left( {\left\langle {Z^k } \right\rangle _A ,\left\langle {Z^l } \right\rangle _C } \right)=\frac{\sum\limits_{i = 1}^{N_A } {g_i^2 }}{\sum\limits_{i = 1}^{N_A } {g_i }\sum\limits_{i = 1}^{N_C } {g_i}}\left<\left(Z^k-\left<Z^k\right>_A\right)\left(Z^l-\left<Z^l\right>_A\right)\right>_A.
\end{split}
\end{equation}
Finally, this gives
\begin{equation}
\begin{split}
C\left( {\left\langle {Z^k } \right\rangle _A ,\left\langle {Z^l } \right\rangle _C } \right)=\frac{\sum\limits_{i = 1}^{N_A } {g_i^2 }}{\sum\limits_{i = 1}^{N_A } {g_i }\sum\limits_{i = 1}^{N_C } {g_i}}\left(\left<Z^{k+l}_A\right>-\left<Z^k\right>_A\left<Z^l\right>_A\right).
\end{split}
\end{equation}

\section{SUPPLEMENTARY INFORMATION ON THE $\mathcal{B}(B^+\to \overline{D}_1l^+\nu)$ CALCULATION}\label{SuppD1}
In Section~\ref{intro}, we use a weighted average for the ratio between $\mathcal{B}(B^+\to D_1^0(D^0\pi^+\pi^-)\pi^+)$ 
and $\mathcal{B}(B^+\to D_1^0(D^{*+}\pi^-)\pi^+)$.\\
One value for the ratio is obtained from the measurements
\begin{equation}
\begin{split}
\mathcal{B}(B^+\to D_1^0(D^0\pi^+\pi^-)\pi^+)=\left(1.85\pm0.29\pm0.35^{+0.0}_{-0.48}\right)\times 10^{-4},
\end{split}
\end{equation}
and
\begin{equation}
\begin{split}
\mathcal{B}(B^+\to D_1^0(D^{*+}\pi^-)\pi^+)=\left(6.8\pm0.7\pm1.3\pm0.3\right)\times 10^{-4},
\end{split}
\end{equation}
quoted in Ref.~\cite{D1eins} and Ref.~\cite{D1zwei}, respectively.\\
Combining them results in the ratio
\begin{equation}
\frac{\mathcal{B}(B^+\to D_1^0(D^0\pi^+\pi^-)\pi^+)}{\mathcal{B}(B^+\to D_1^0(D^{*+}\pi^-)\pi^+)}=0.28\pm 0.11.
\end{equation}
Furthermore, a LHCb measurement~\cite{LHCbFrac} finds
\begin{equation}
\frac{\mathcal{B}(B^+\to D_1^0(D^0\pi^+\pi^-)\pi^+)}{\mathcal{B}(B^+\to D_1^0(D^{*+}\pi^-)\pi^+)}=0.43\pm 0.14.
\end{equation}
From their weighted average, and noting that $D_1^0\to D^0\pi^+\pi^-$ 
contributes $\frac{3}{7}$ to the total $D_1^0\to D\pi\pi$ rate, and that 
$D_1^0\to D^{*+}\pi^-$ contributes $\frac{2}{3}$ to the total $D_1^0\to D^{*}\pi$ 
(if isospin-invariance is assumed)~\cite{ThomasDiss}, one obtains
\begin{equation}
\begin{split}
\frac{\mathcal{B}(B^+\to D_1^0(D\pi\pi)\pi^+)}{\mathcal{B}(B^+\to D_1^0(D^{*}\pi)\pi^+)}=0.53\pm 0.14.
\end{split}
\end{equation}

\section{CALCULATION OF $\mathcal{B}(B\to (D^{(*)}\pi)_{nr}l\nu)$ FROM $\mathcal{B}(B\to D^{(*)}\pi l\nu)$ AND $\mathcal{B}(B\to D^{**}l\nu)$ MEASUREMENTS}\label{NrCalc}
In Table~\ref{Nr}, we quote the averages for $\mathcal{B}(B\to D^{(*)}\pi l\nu)$ (inclusive), 
which we beforehand had corrected for unmeasured decay modes (e.g. to account for $B^+\to D^0\pi^0l^+\nu$ decays, 
$\mathcal{B}(B^+\to D^-\pi^+l^+\nu)$ was multiplied by a factor of $\frac{3}{2}$ to give 
$\mathcal{B}(B^+\to D\pi l^+\nu)$). We always assume isospin symmetry which implies the 
equality of the decay widths $\Gamma(B^+\to X_c^il^+\nu)=\Gamma(B^0\to X_c^il^+\nu)$ 
and therefore it is meaningful to take the isospin average according to
\begin{equation}\label{iso1}
\left\langle\mathcal{B}(B^+\to X_c^il^+\nu)\right\rangle_{iso}=\frac{\mathcal{B}(B^+\to X_c^il^+\nu)+\tau_{+0}\mathcal{B}(B^0\to X_c^il^+\nu)}{2}
\end{equation}
with $\tau_{+0}:=\frac{\tau_{B^+}}{\tau_{B^0}}$
being the lifetime ratio of charged and neutral $B$-mesons (see~\ref{BrFrMeas}).\\

\begin{table}
\begin{center}
\begin{tabular}{ccc}
\hline\noalign{\smallskip}
Decay&Branching Fraction [\%]\\
\hline\noalign{\smallskip}
$B^+\to D\pi l^+\nu$& $0.65\pm0.075$\\
$B^+\to D^{*}\pi l^+\nu$&$0.915\pm0.075$\\
$B^0\to \overline{D}\pi l^+\nu$&$0.645\pm0.09$\\
$B^0\to \overline{D}^{*}\pi l^+\nu$&$0.735\pm0.12$\\
\hline\noalign{\smallskip}
$\left\langle(B^+\to D\pi l^+\nu)\right\rangle_{iso}$& $0.67\pm0.06$\\
$\left\langle(B^+\to D^{*}\pi l^+\nu)\right\rangle_{iso}$&$0.85\pm0.08$\\
\hline\noalign{\smallskip}
\end{tabular}
\end{center}
\caption{Measured values for $\mathcal{B}(B\to D^{(*)}\pi l\nu)$ (inclusive) taken 
from Ref.~\cite{HFAG} and rescaled to account for unmeasured isospin-symmetric decay modes. The last two lines give isospin-averages.}
\label{Nr}
\end{table}
To compute $\mathcal{B}(B^+\to (D^{(*)}\pi)_{nr}l^+\nu)$ we have to 
substract the contributions of $B^+\to D^{**}(D^{(*)}\pi)l^+\nu$ decays. 
For this we use the assumptions quoted in Section~\ref{intro}, which gives
\begin{equation}
\begin{split}
\mathcal{B}(B^+\to (D\pi)_{nr} l^+\nu)&=\left(0.053\pm0.100\right)\%,\\
\mathcal{B}(B^+\to (D^*\pi )_{nr}l^+\nu)&=\left(0.117\pm0.104\right)\%,
\end{split}
\end{equation}
which results in a combined branching fraction of
\begin{equation}
\mathcal{B}(B^+\to(D^{(*)}\pi)_{nr} l^+\nu)=\left(0.17\pm0.14\right)\%.\\
\end{equation}

\FloatBarrier
\section{MOMENT DISTRIBUTIONS}\label{MomDistr}
\FloatBarrier
This appendix provides an example for the moment distributions. The theoretical distributions 
resulting from the fit are compared with the experimentally measured moments. The non-central 
theoretical moments are decomposed into the several exclusive contributions, which is implied 
by different colors. Since such a decomposition is not possible for central moments, the fit 
result is plain and referred to as "Theory".
\twocolumn
\begin{figure}
	
	\includegraphics[width=\columnwidth]{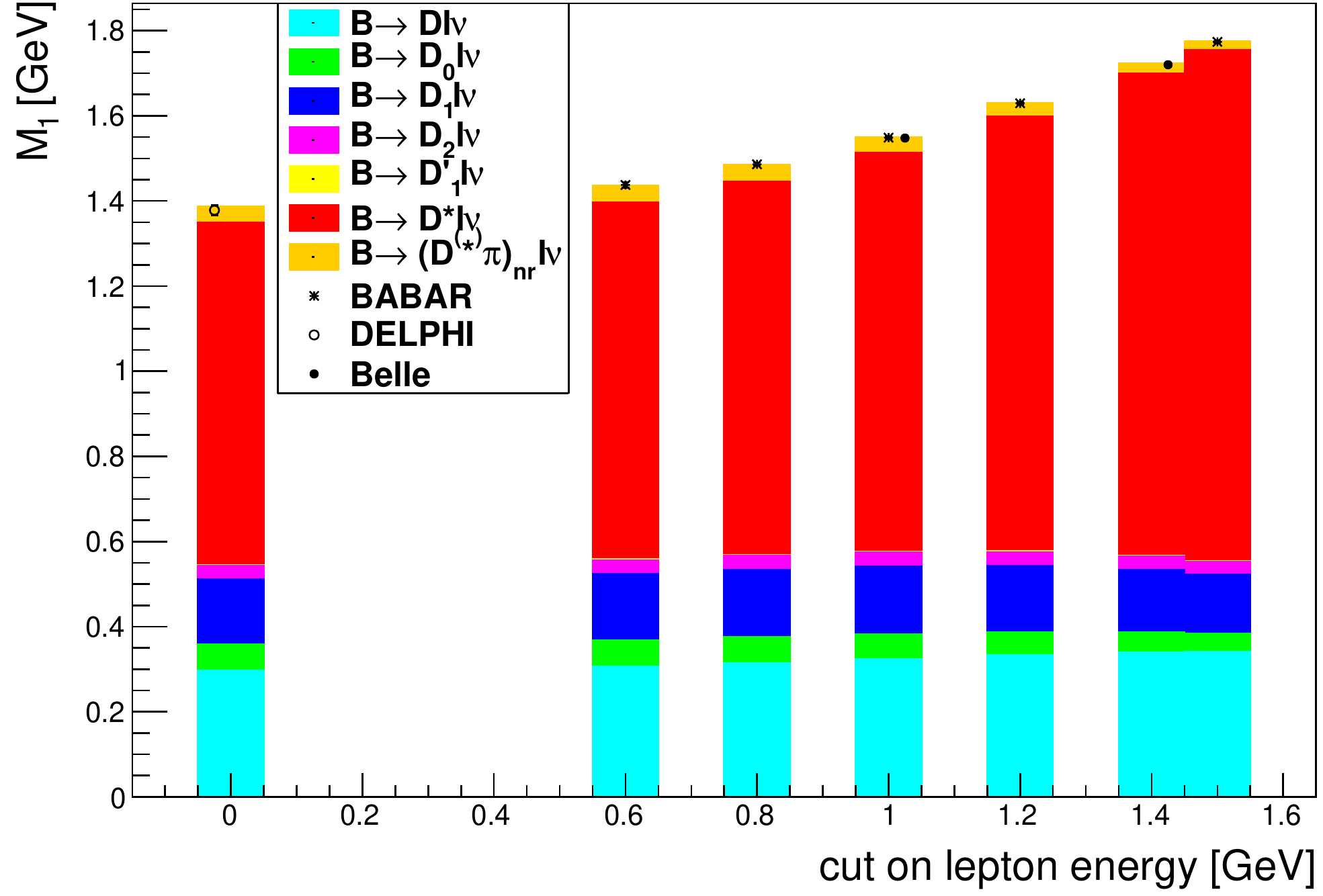}
\caption{First electron moment $M_{1}$: Result of Fit 2 in Table~\ref{Results1} and the corresponding experimental data used.}\label{M1}
\end{figure}
\begin{figure}
	
	\includegraphics[width=\columnwidth]{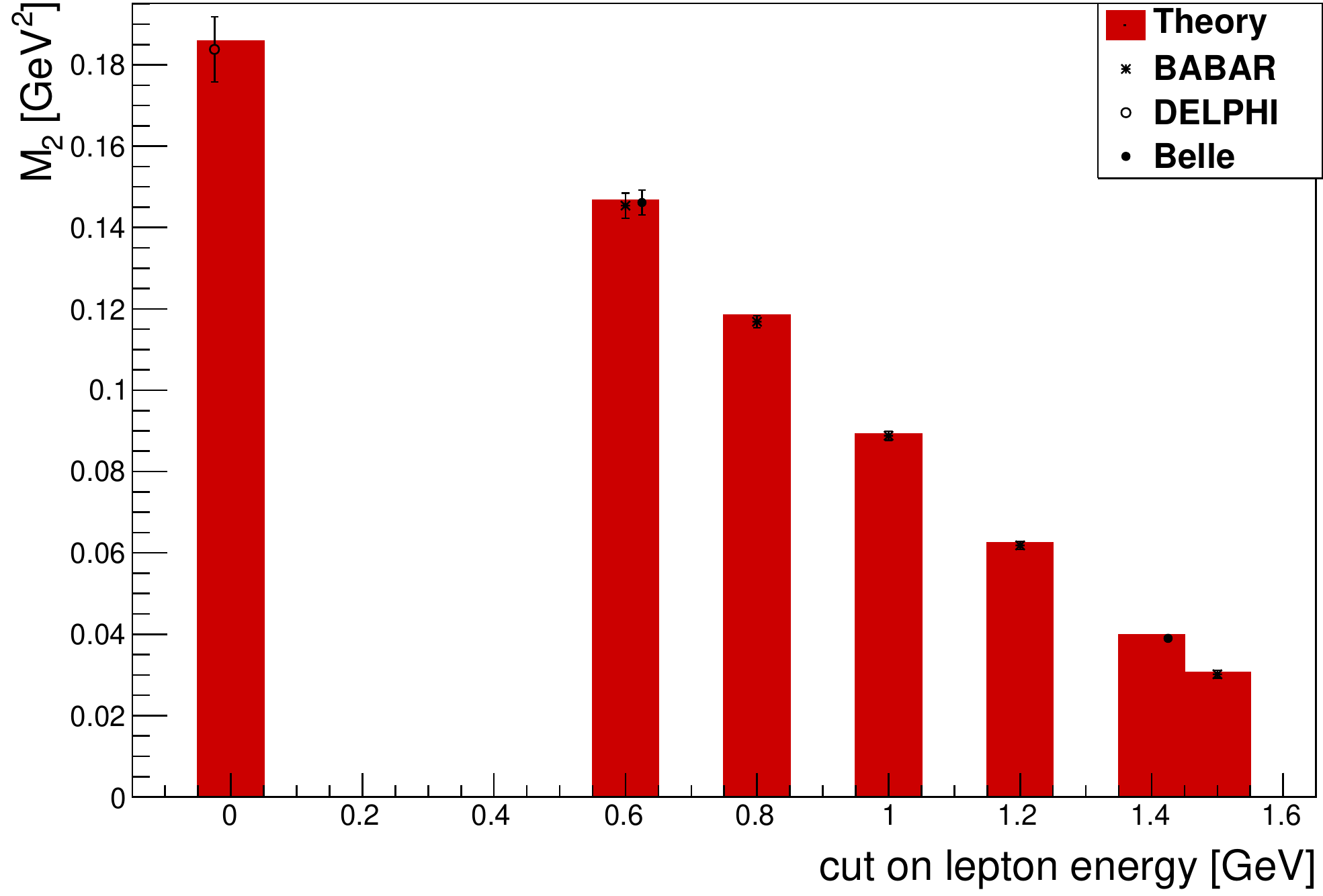}
\caption{Second electron moment $M_{2}$: Result of Fit 2 in Table~\ref{Results1} and the corresponding experimental data used.}
\end{figure}
\begin{figure}
	
	\includegraphics[width=\columnwidth]{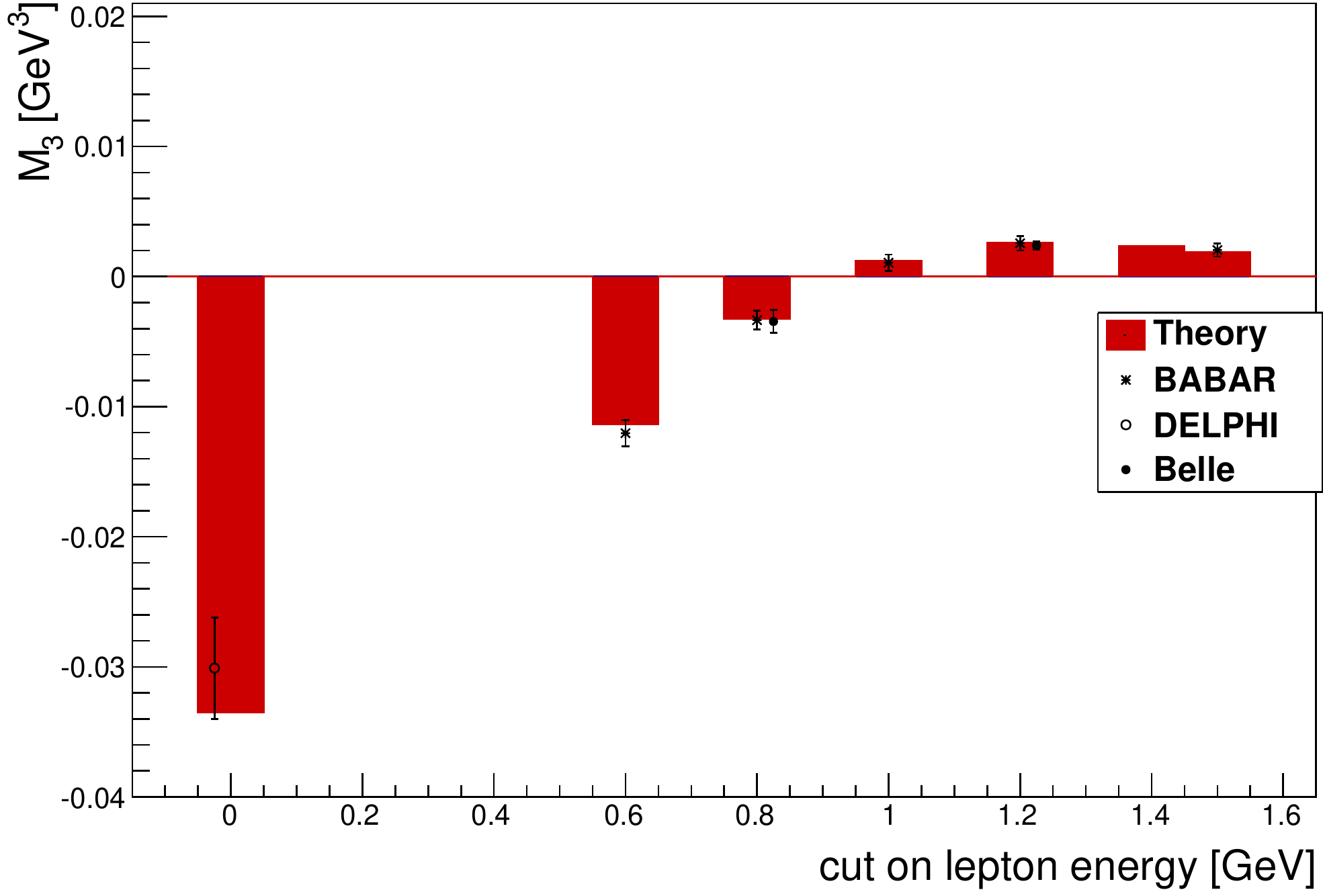}
\caption{Third electron moment $M_{3}$: Result of Fit 2 in Table~\ref{Results1} and the corresponding experimental data used.}
\end{figure}
\begin{figure}
	
	\includegraphics[width=\columnwidth]{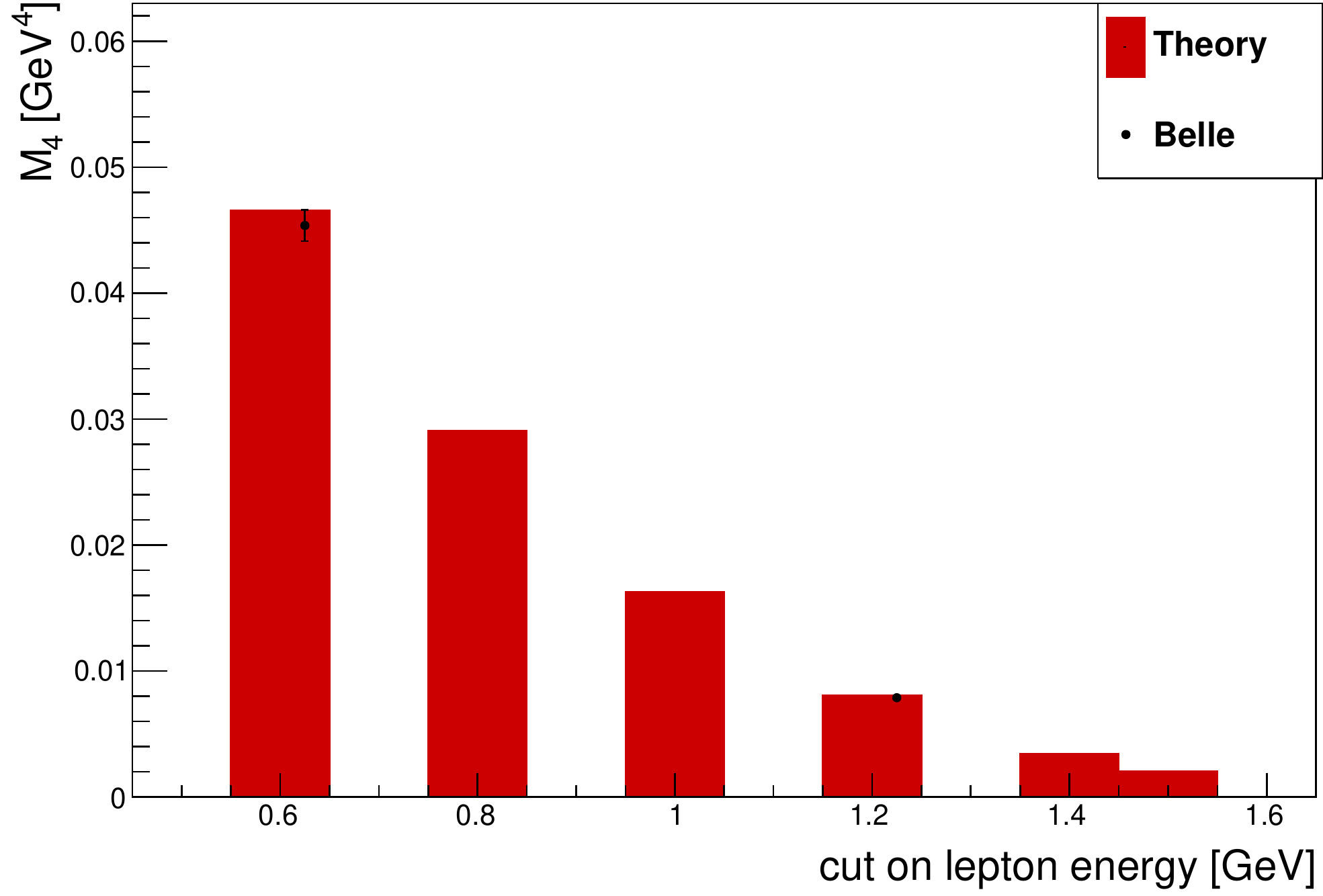}
\caption{Fourth electron moment $M_{4}$: Result of Fit 2 in Table~\ref{Results1} and the corresponding experimental data used.}
\end{figure}

\begin{figure}
	
	\includegraphics[width=\columnwidth]{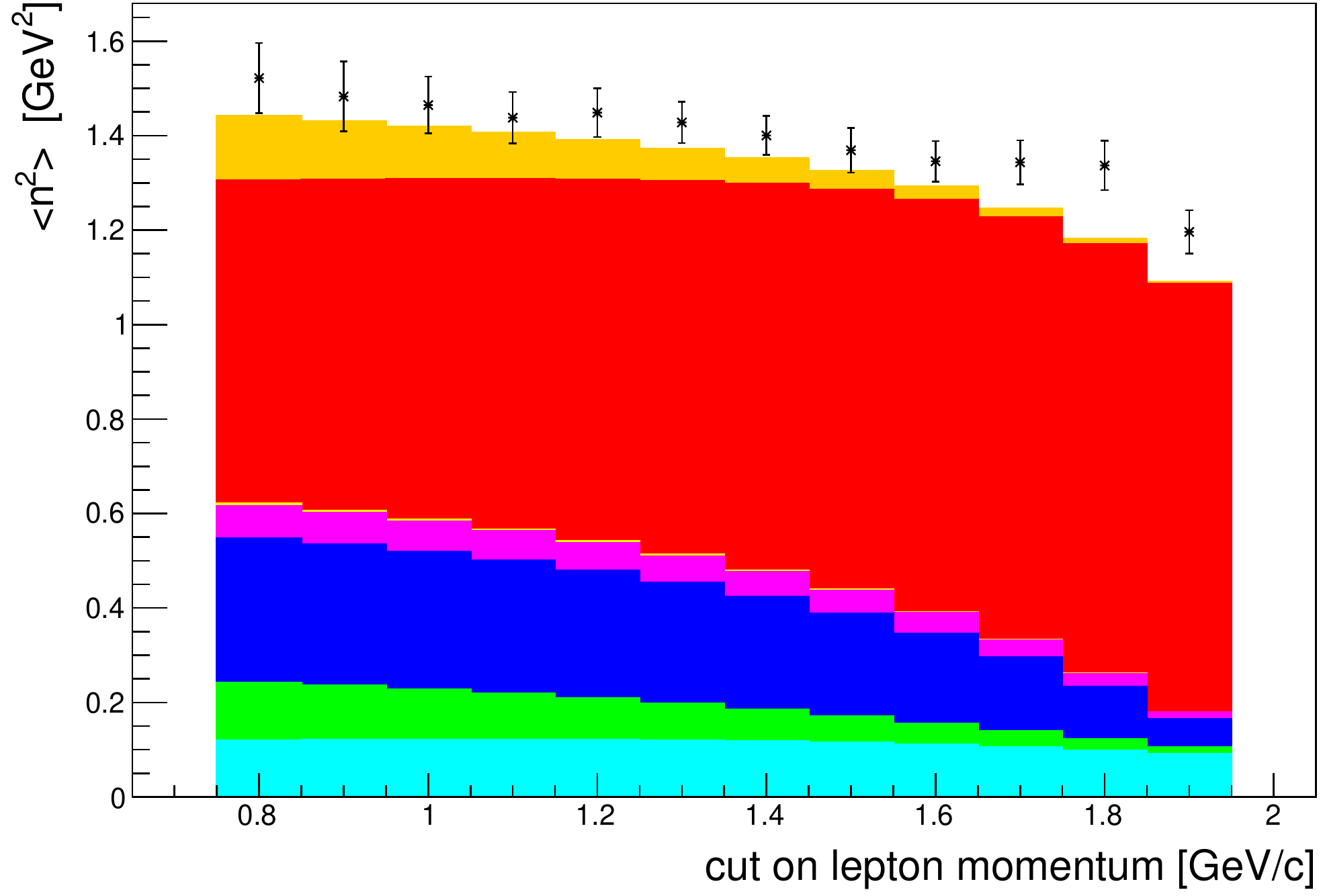}
\caption{Energy-mass moment $\left\langle n^2\right\rangle$: 
Result of Fit 2 in Table~\ref{Results1} and the corresponding experimental data used. See legend of Fig.~\ref{M1}.}
\end{figure}
\begin{figure}
	
	\includegraphics[width=\columnwidth]{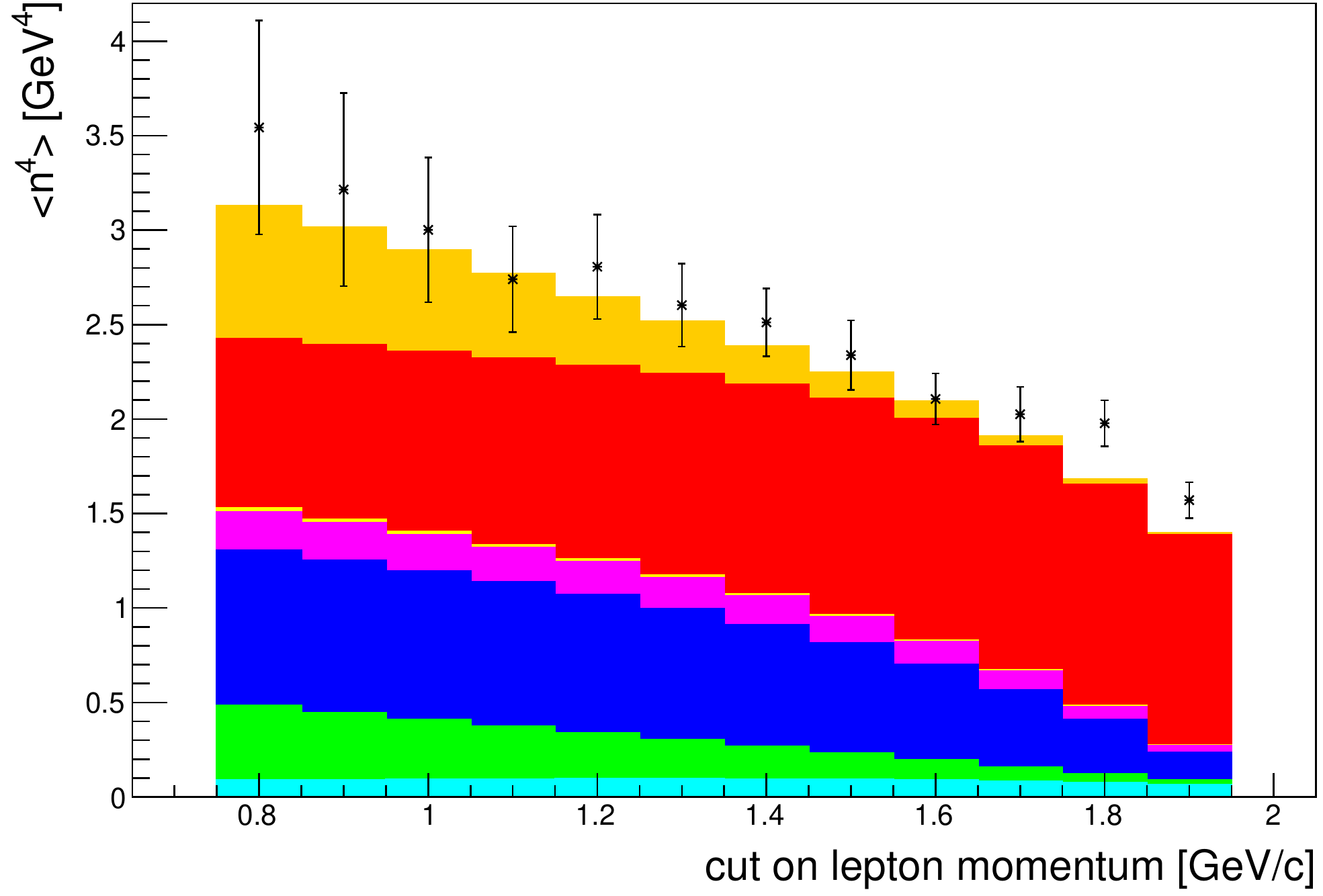}
\caption{Energy-mass moment $\left\langle n^4\right\rangle$: 
Result of Fit 2 in Table~\ref{Results1} and the corresponding experimental data used. See legend of Fig.~\ref{M1}.}
\end{figure}
\begin{figure}
	
	\includegraphics[width=\columnwidth]{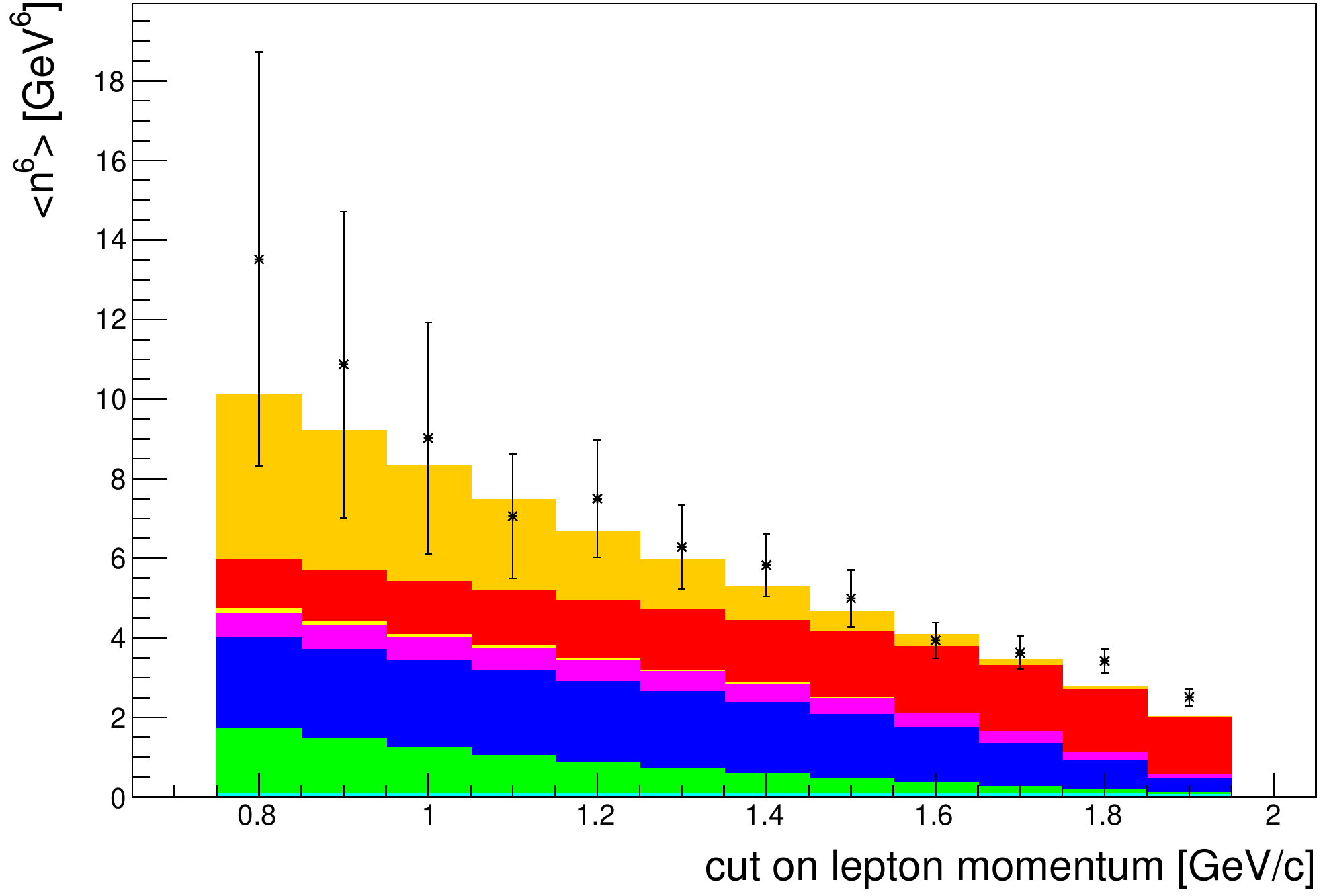}
\caption{Energy-mass moment $\left\langle n^6\right\rangle$: 
Result of Fit 2 in Table~\ref{Results1} and the corresponding experimental data used. See legend of Fig.~\ref{M1}.}
\end{figure}
\begin{figure}
	
	\includegraphics[width=\columnwidth]{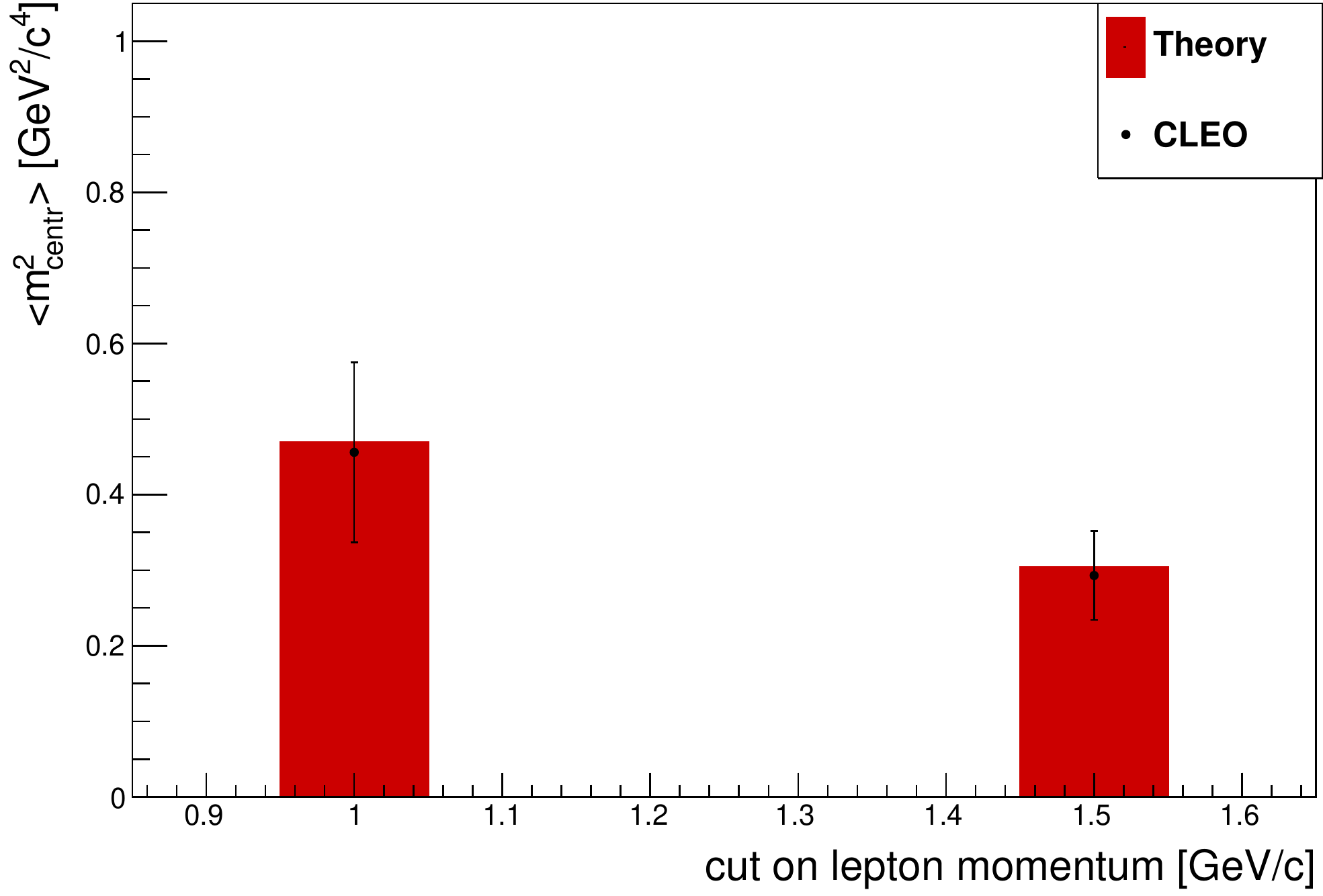}
\caption{Central moment $\left\langle m^2_{centr}\right\rangle$: Result of Fit 2 in Table~\ref{Results1} and the corresponding experimental data used.}
\end{figure}
\begin{figure}
	
	\includegraphics[width=\columnwidth]{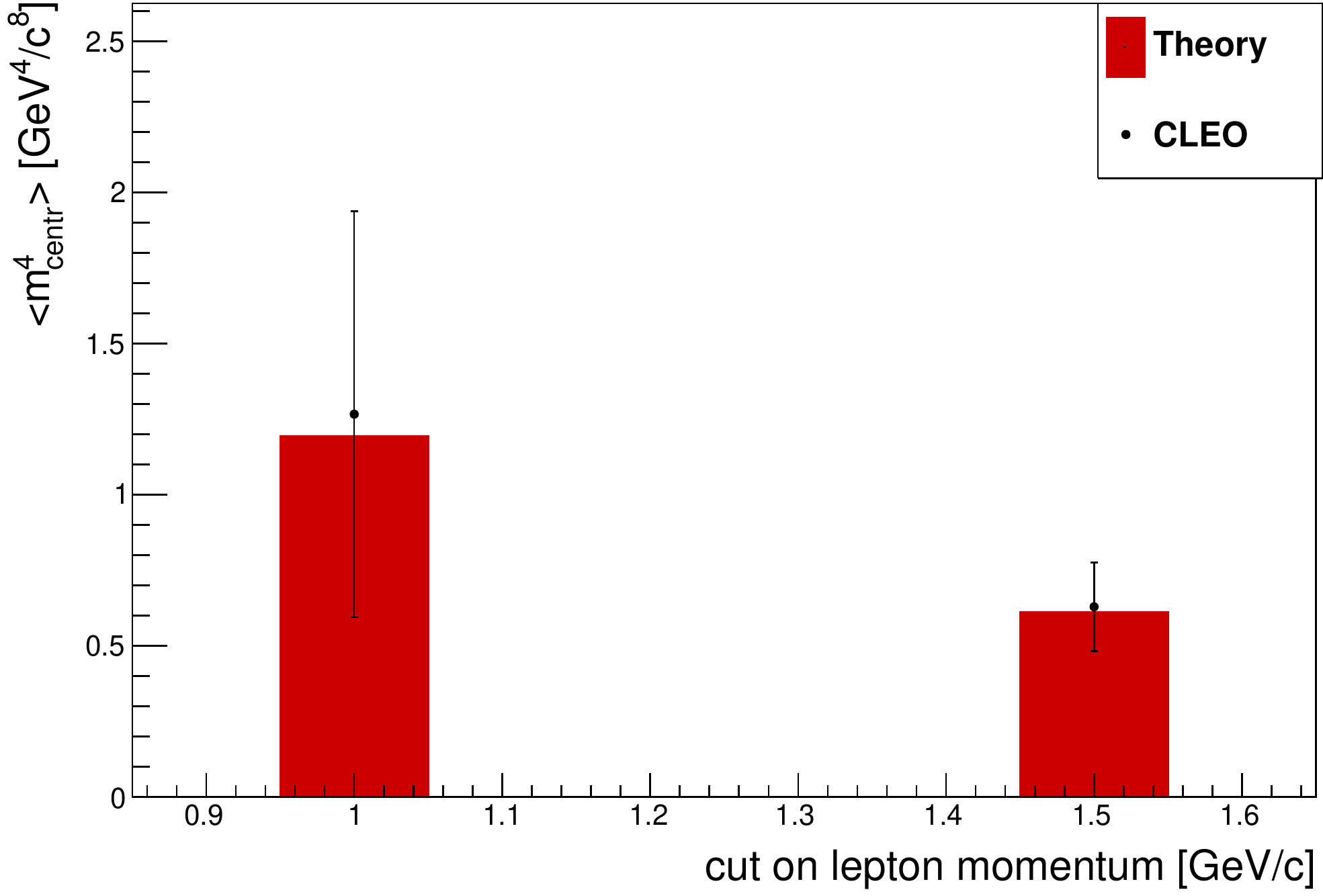}
\caption{Central moment $\left\langle m^4_{centr}\right\rangle$: Result of Fit 2 in Table~\ref{Results1} and the corresponding experimental data used.}
\end{figure}
\begin{figure}
	
	\includegraphics[width=\columnwidth]{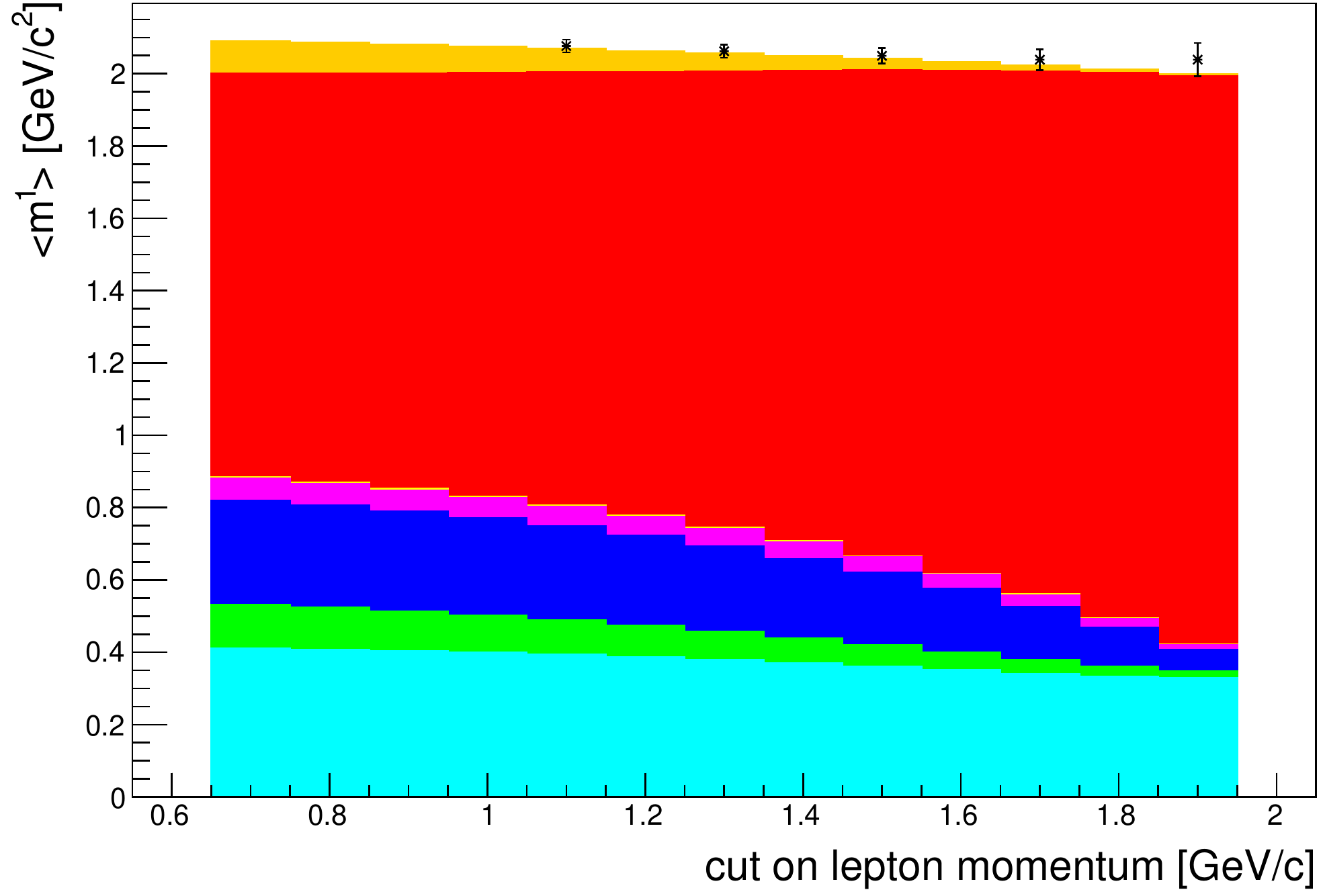}
\caption{First hadronic mass moment $\left<m^1\right>$: Result of Fit 2 in Table~\ref{Results1} and the corresponding experimental data used. See legend of Fig.~\ref{M1}.}
\end{figure}
\begin{figure}
	
	\includegraphics[width=\columnwidth]{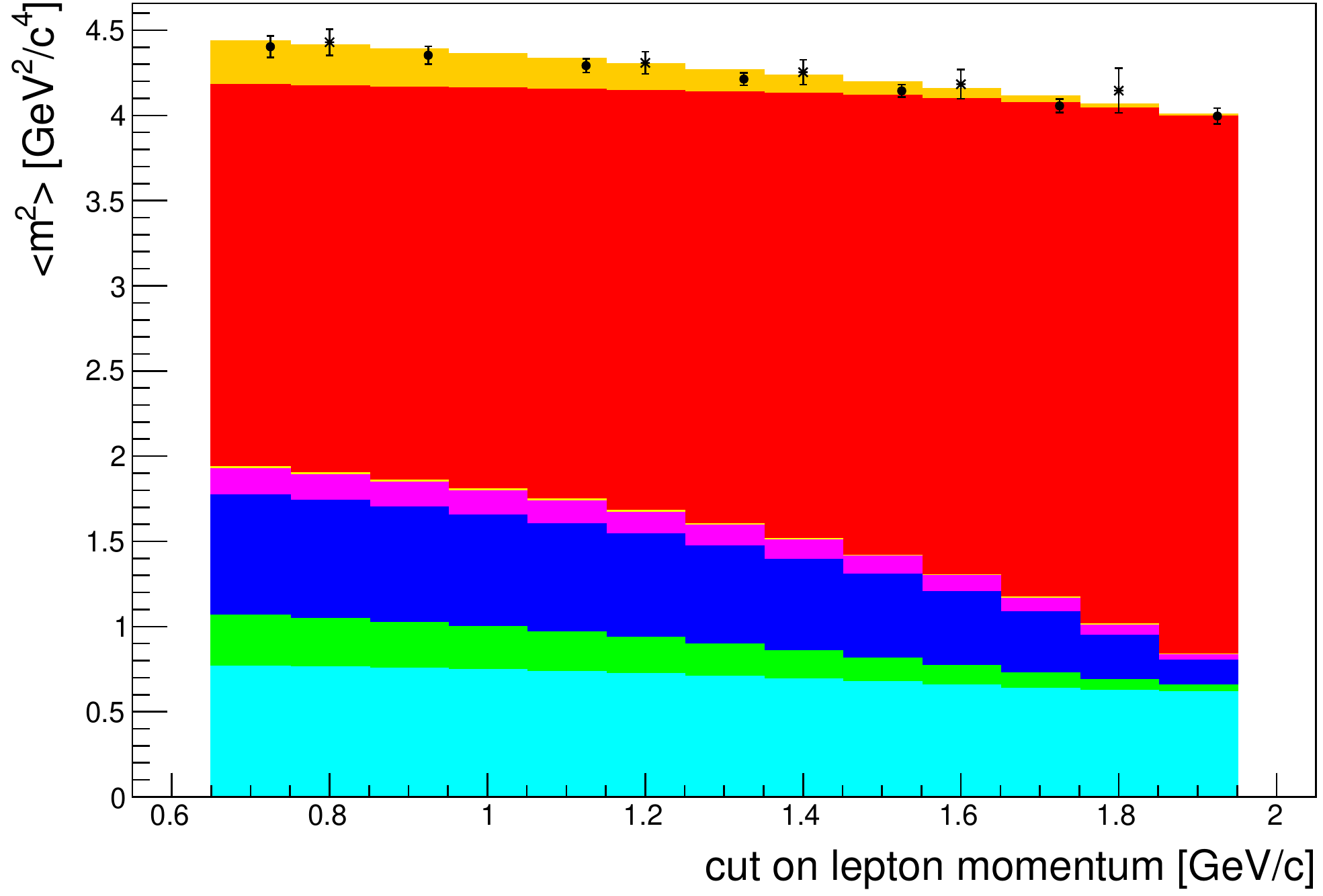}
\caption{Second hadronic mass moment $\left<m^2\right>$: Result of Fit 2 in Table~\ref{Results1} and the corresponding experimental data used. See legend of Fig.~\ref{M1}.}
\end{figure}
\begin{figure}
	
	\includegraphics[width=\columnwidth]{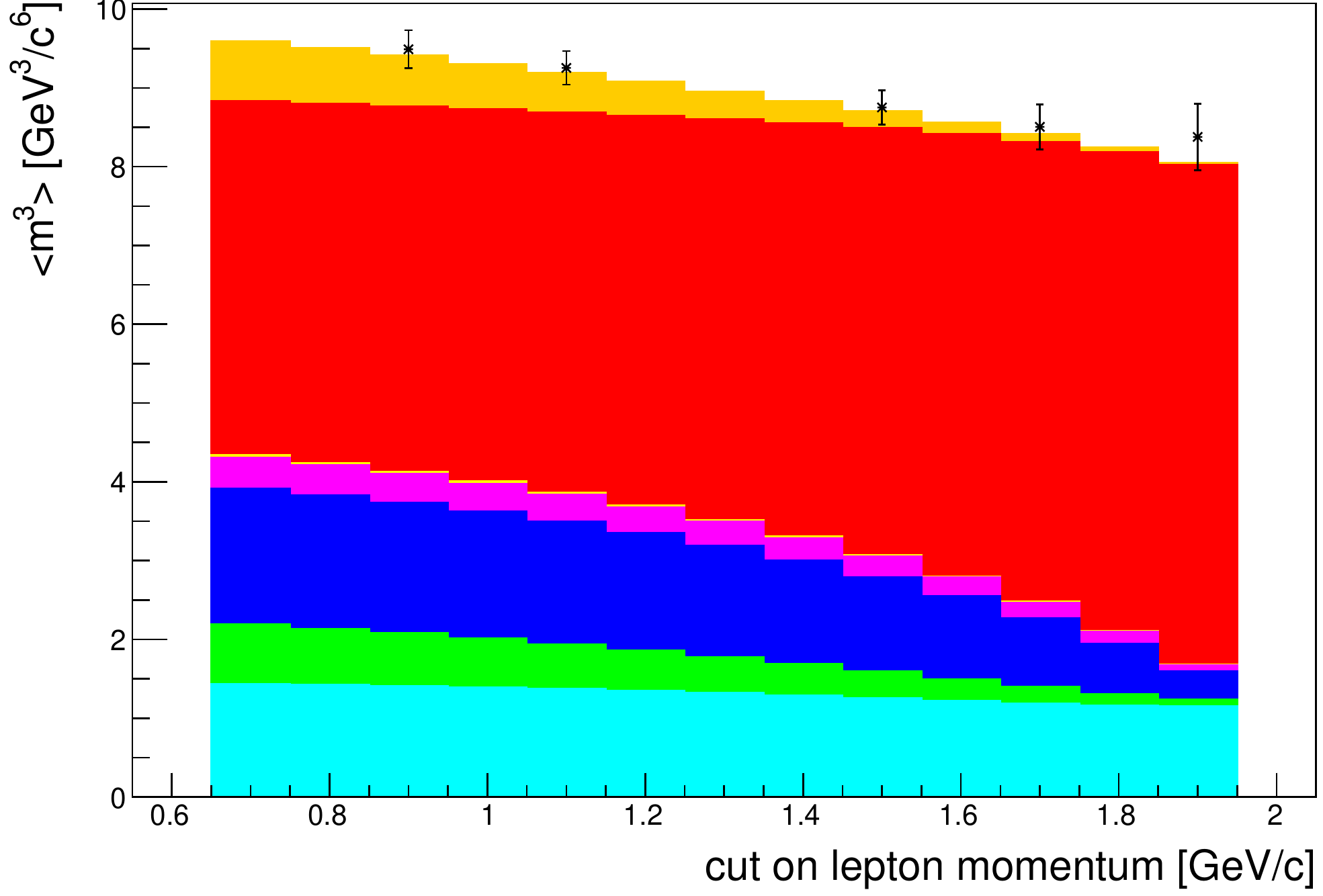}
\caption{Third hadronic mass moment $\left<m^3\right>$: Result of Fit 2 in Table~\ref{Results1} and the corresponding experimental data used. See legend of Fig.~\ref{M1}.}
\end{figure}
\begin{figure}
	
	\includegraphics[width=\columnwidth]{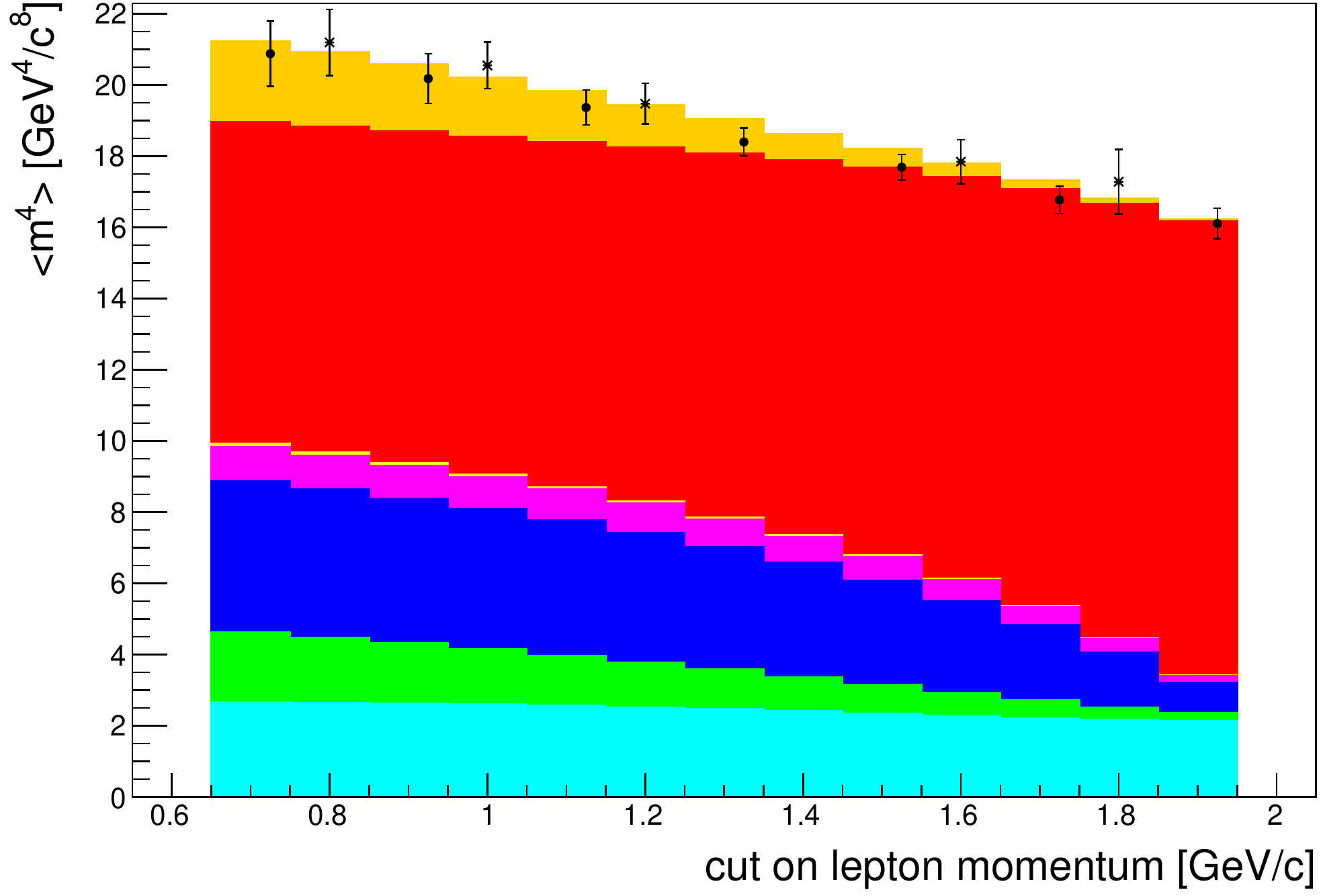}
\caption{Fourth hadronic mass moment $\left<m^4\right>$: Result of Fit 2 in Table~\ref{Results1} and the corresponding experimental data used. See legend of Fig.~\ref{M1}.}
\end{figure}
\begin{figure}
	
	\includegraphics[width=\columnwidth]{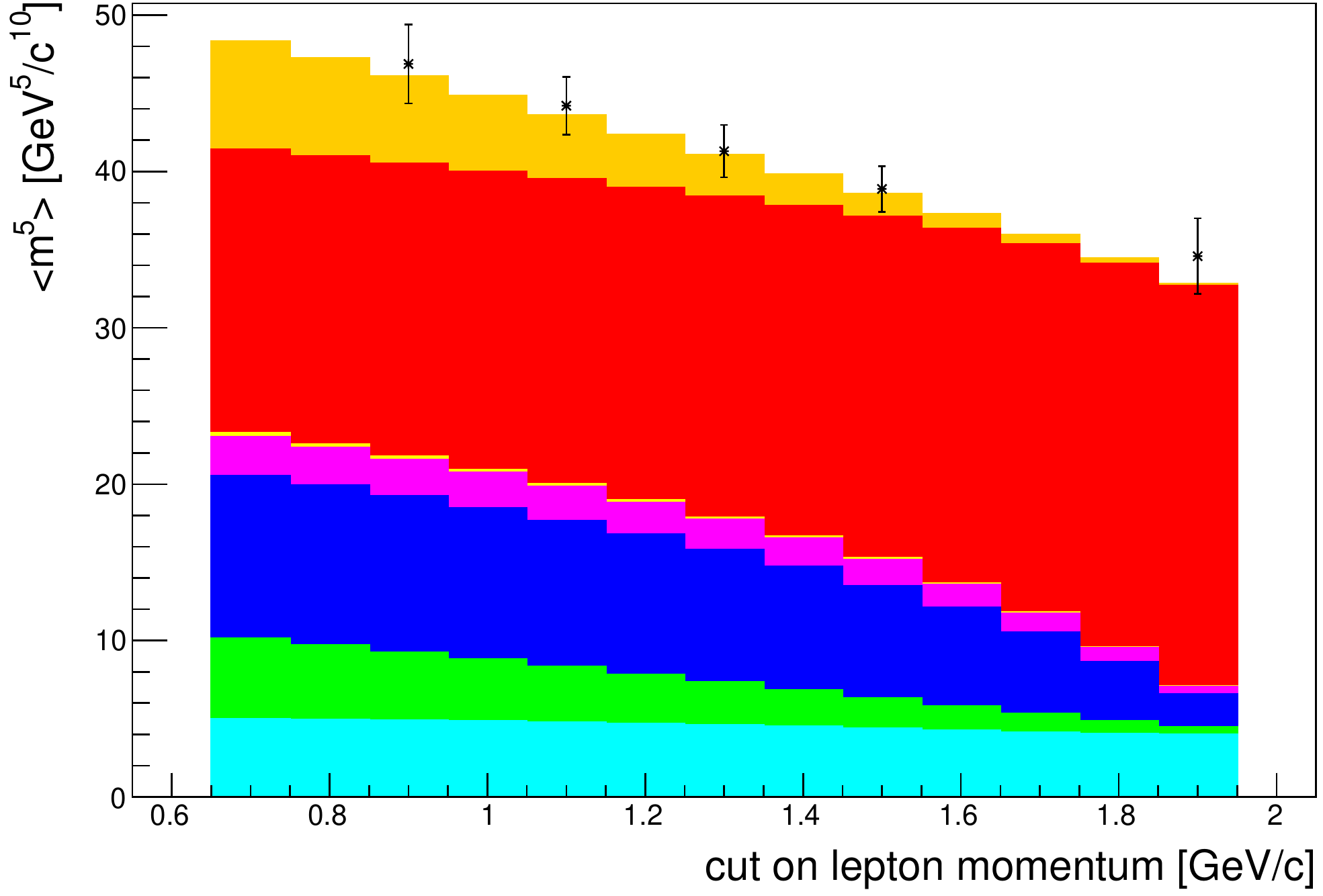}
\caption{Fifth hadronic mass moment $\left<m^5\right>$: Result of Fit 2 in Table~\ref{Results1} and the corresponding experimental data used. See legend of Fig.~\ref{M1}.}
\end{figure}
\begin{figure}[t!]
	\includegraphics[width=\columnwidth]{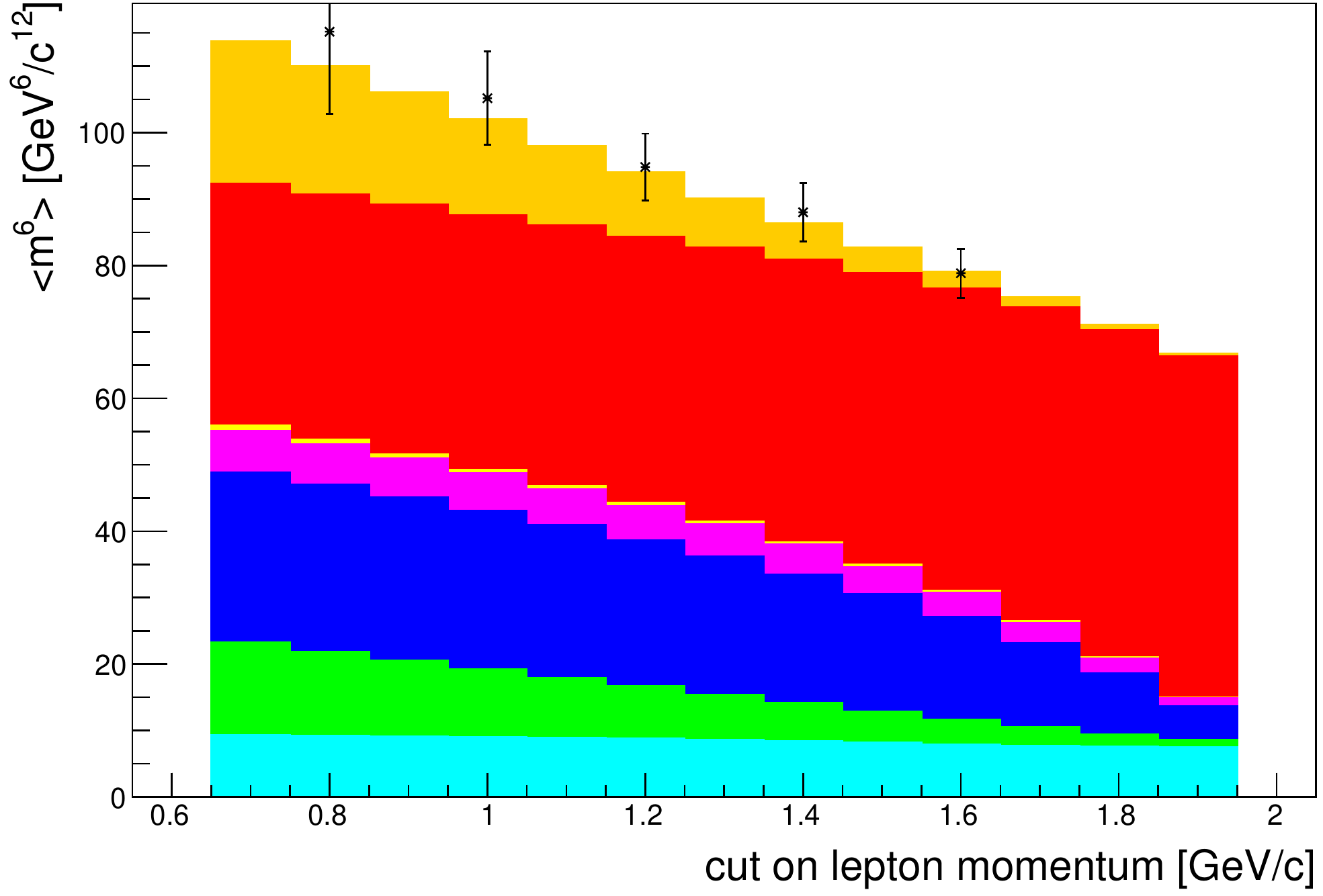}
	
\caption{Sixth hadronic mass moment $\left<m^6\right>$: Result of Fit 2 in Table~\ref{Results1} and the corresponding experimental data used. See legend of Fig.~\ref{M1}.}
\vspace{11.2cm}
\end{figure}

\onecolumn

\FloatBarrier

\section{STUDIES OF PSEUDO-DATA}\label{Toy}
\begin{figure}
	\includegraphics[width=\columnwidth]{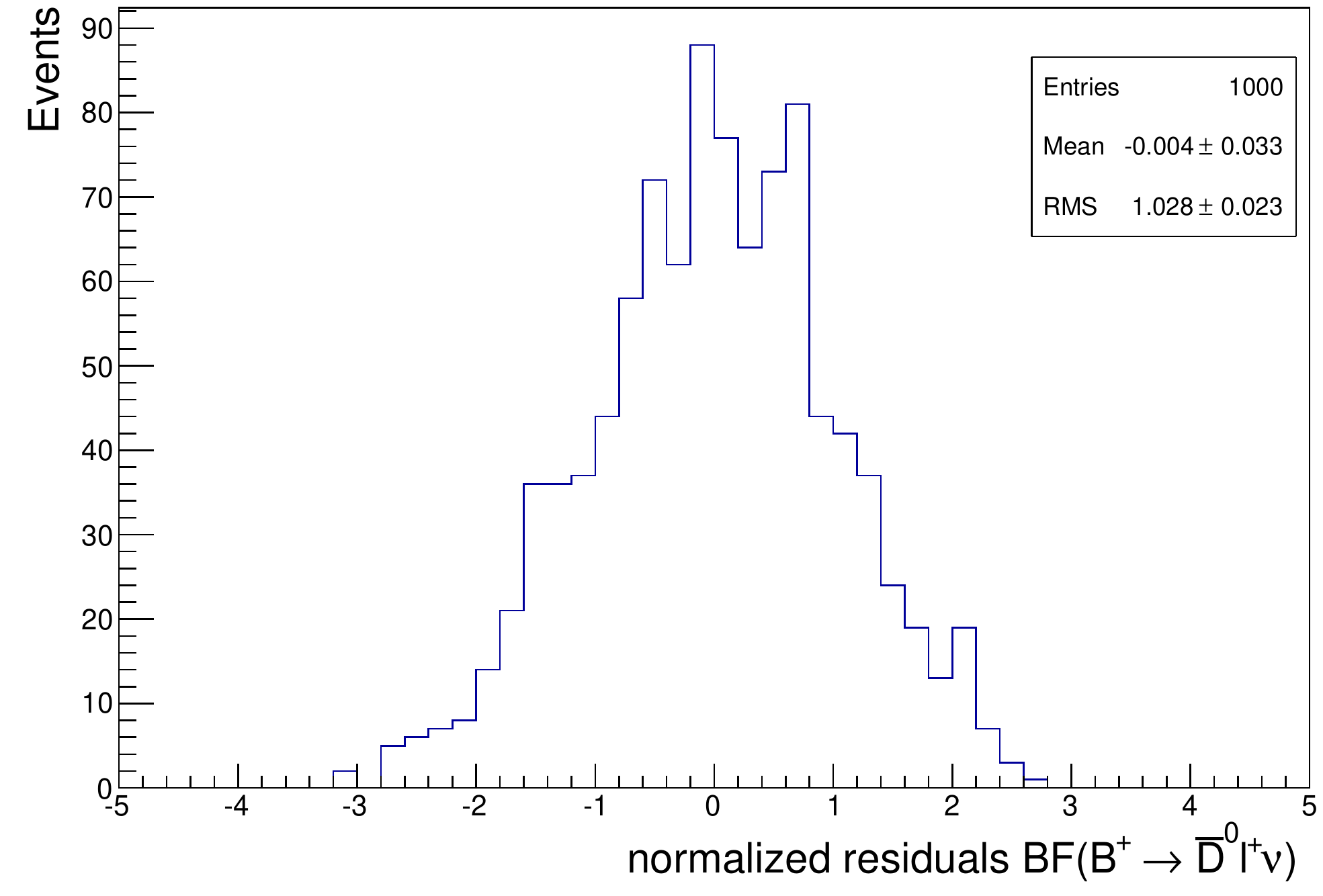}
\caption{Distribution of normalized residuals for the results of $\mathcal{B}(B^+\to \overline{D}^0l^+\nu)$ corresponding to the fit scenario of Fit 2 in Table~\ref{Results1}.}
\label{Pull1}
\end{figure}

\begin{figure}
	\includegraphics[width=\columnwidth]{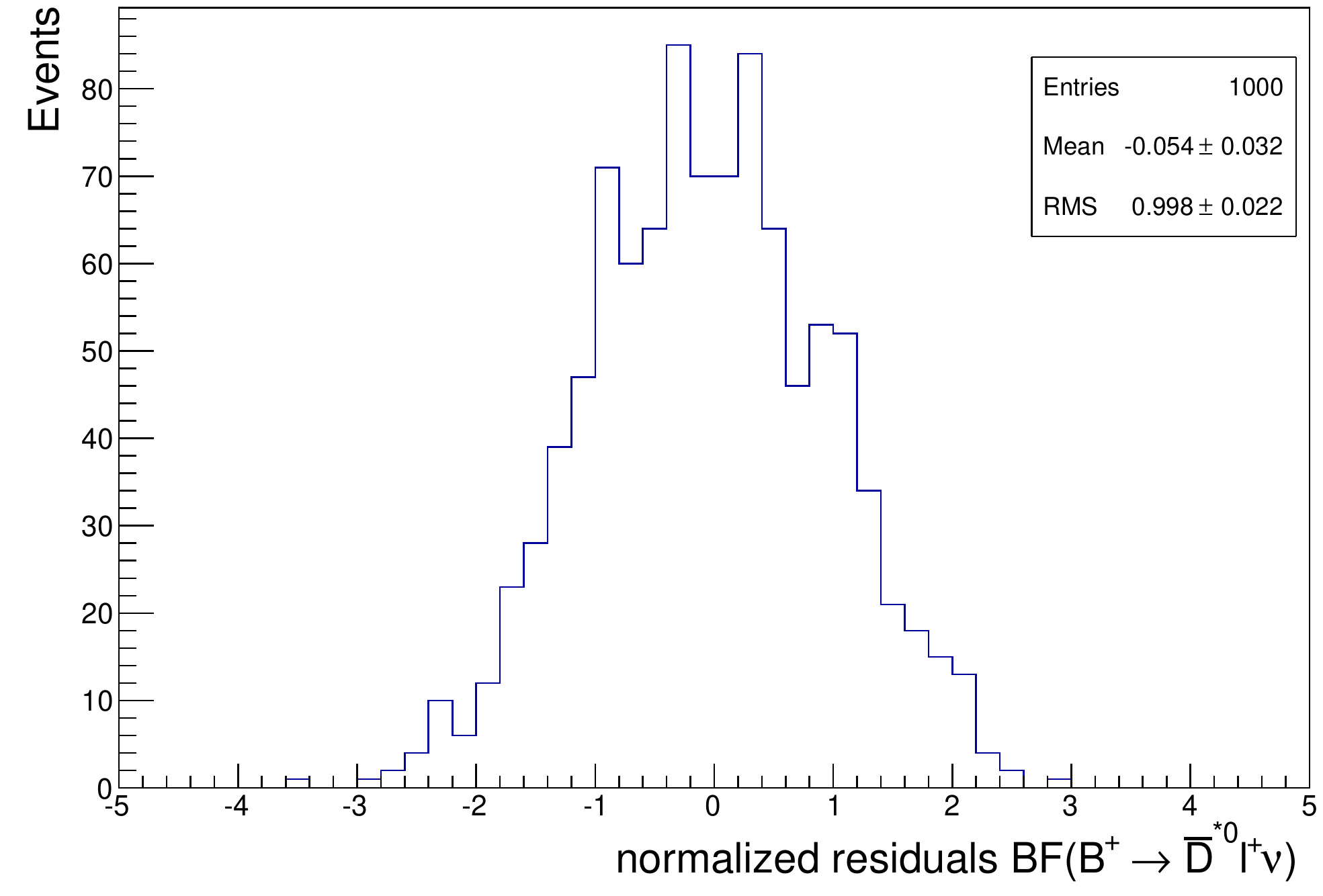}
\caption{Distribution of normalized residuals for the results of $\mathcal{B}(B^+\to \overline{D}^{*0}l^+\nu)$ corresponding to the fit scenario of Fit 2 in Table~\ref{Results1}.}
\label{Pull2}
\end{figure}

\begin{figure}
	\includegraphics[width=\columnwidth]{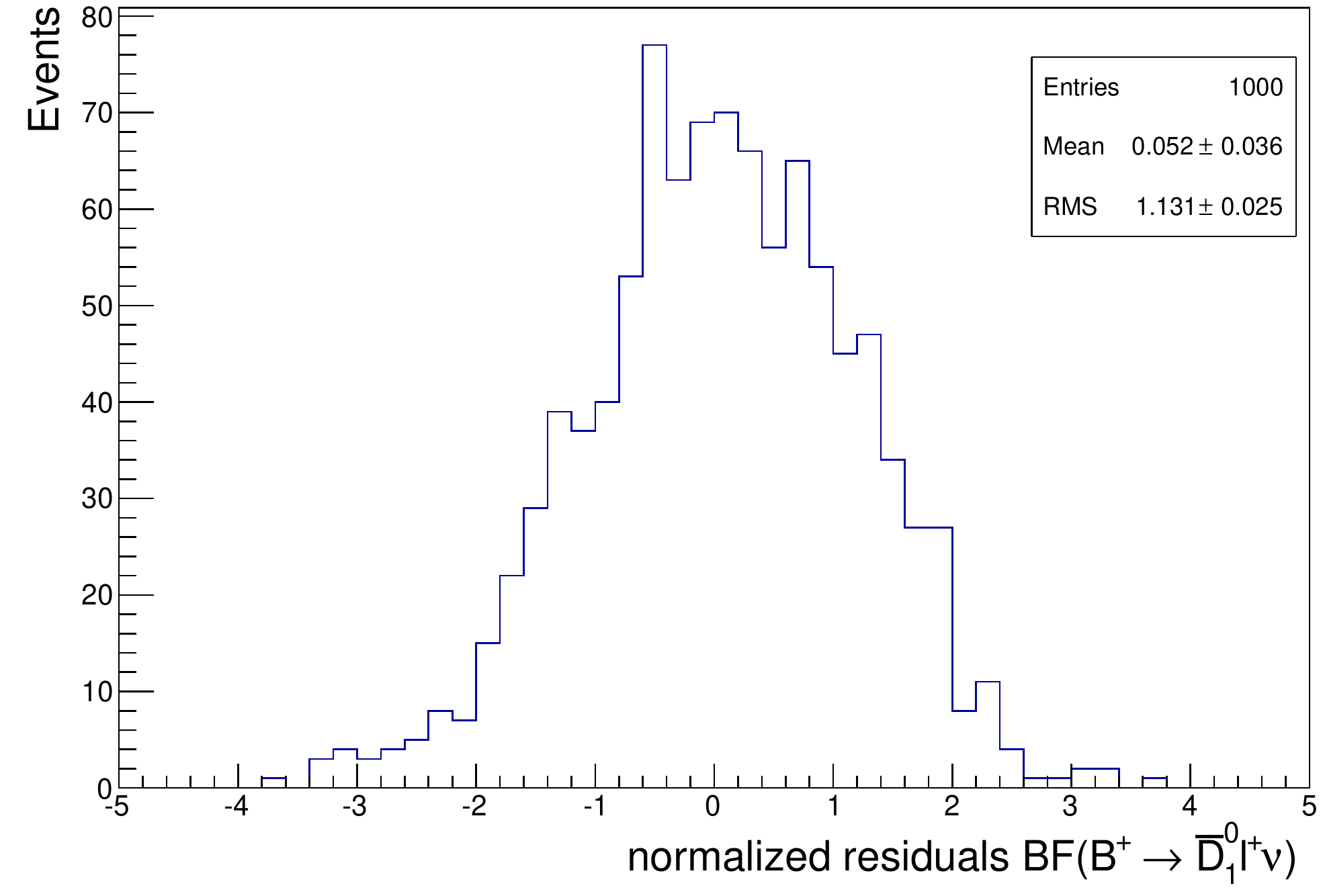}
\caption{Distribution of normalized residuals for the results of $\mathcal{B}(B^+\to \overline{D}^0_1l^+\nu)$ corresponding to the fit scenario of Fit 2 in Table~\ref{Results1}.}
\label{Pull3}
\end{figure}

\begin{figure}
	\includegraphics[width=\columnwidth]{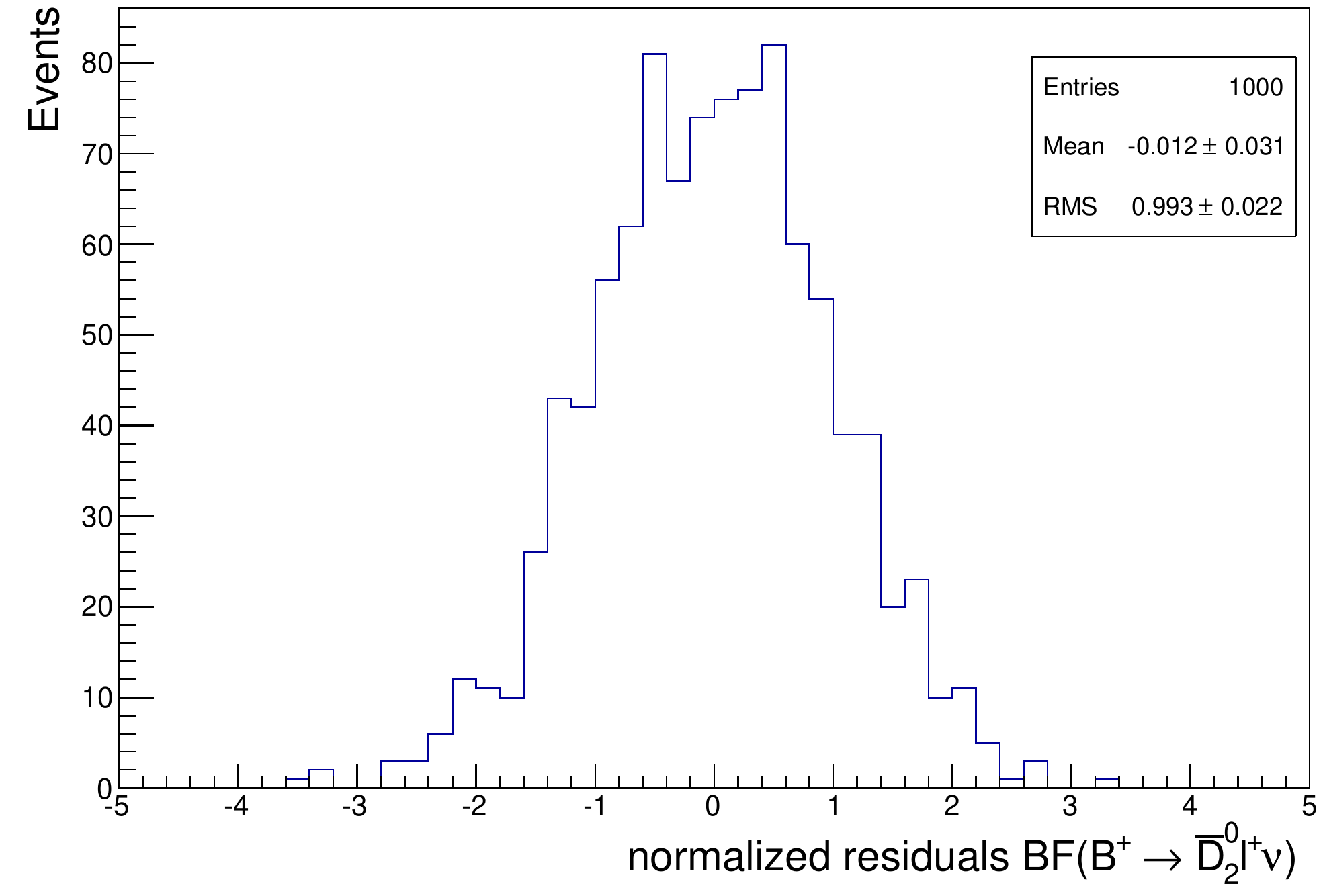}
\caption{Distribution of normalized residuals for the results of $\mathcal{B}(B^+\to \overline{D}^0_2l^+\nu)$ corresponding to the fit scenario of Fit 2 in Table~\ref{Results1}.}
\label{Pull4}
\end{figure}

\begin{figure}
	\includegraphics[width=\columnwidth]{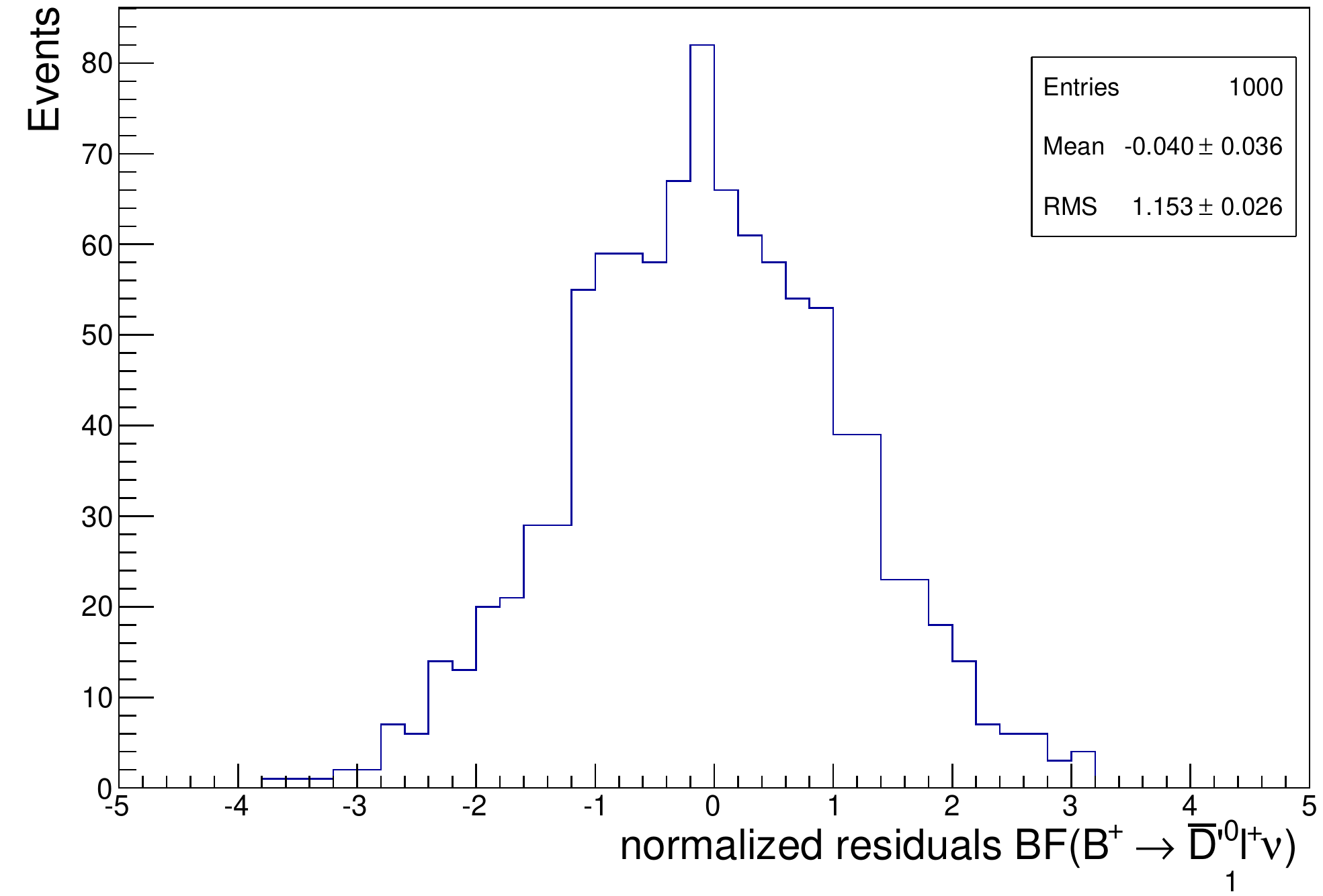}
\caption{Distribution of normalized residuals for the results of $\mathcal{B}(B^+\to \overline{D}^0_0l^+\nu)$ corresponding to the fit scenario of Fit 2 in Table~\ref{Results1}.}
\label{Pull5}
\end{figure}

\begin{figure}
	\includegraphics[width=\columnwidth]{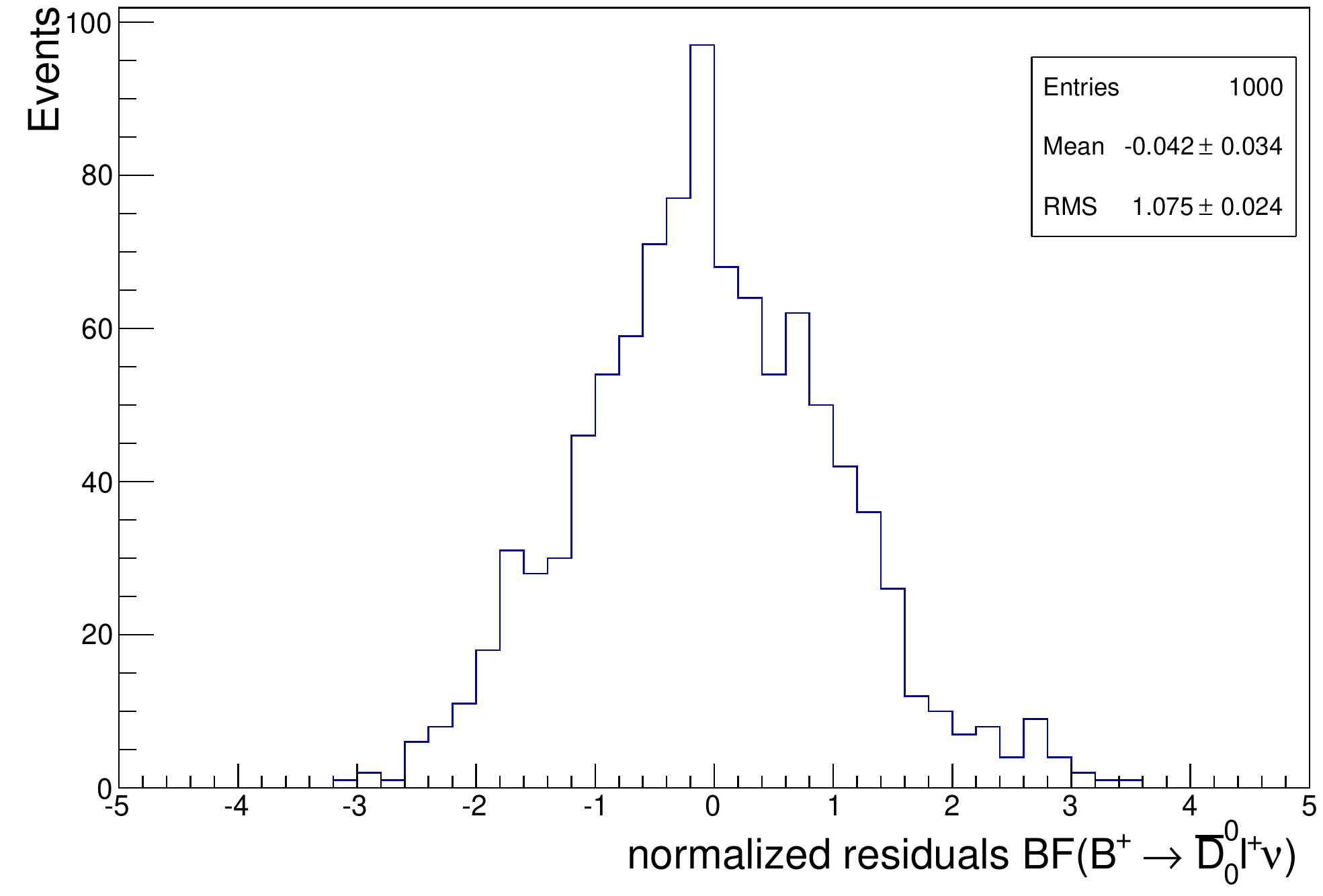}
\caption{Distribution of normalized residuals for the results of $\mathcal{B}(B^+\to \overline{D}'^0_1l^+\nu)$ corresponding to the fit scenario of Fit 2 in Table~\ref{Results1}.}
\label{Pull6}
\end{figure}

\begin{figure}
	\vspace{-0.9cm}
	\includegraphics[width=\columnwidth]{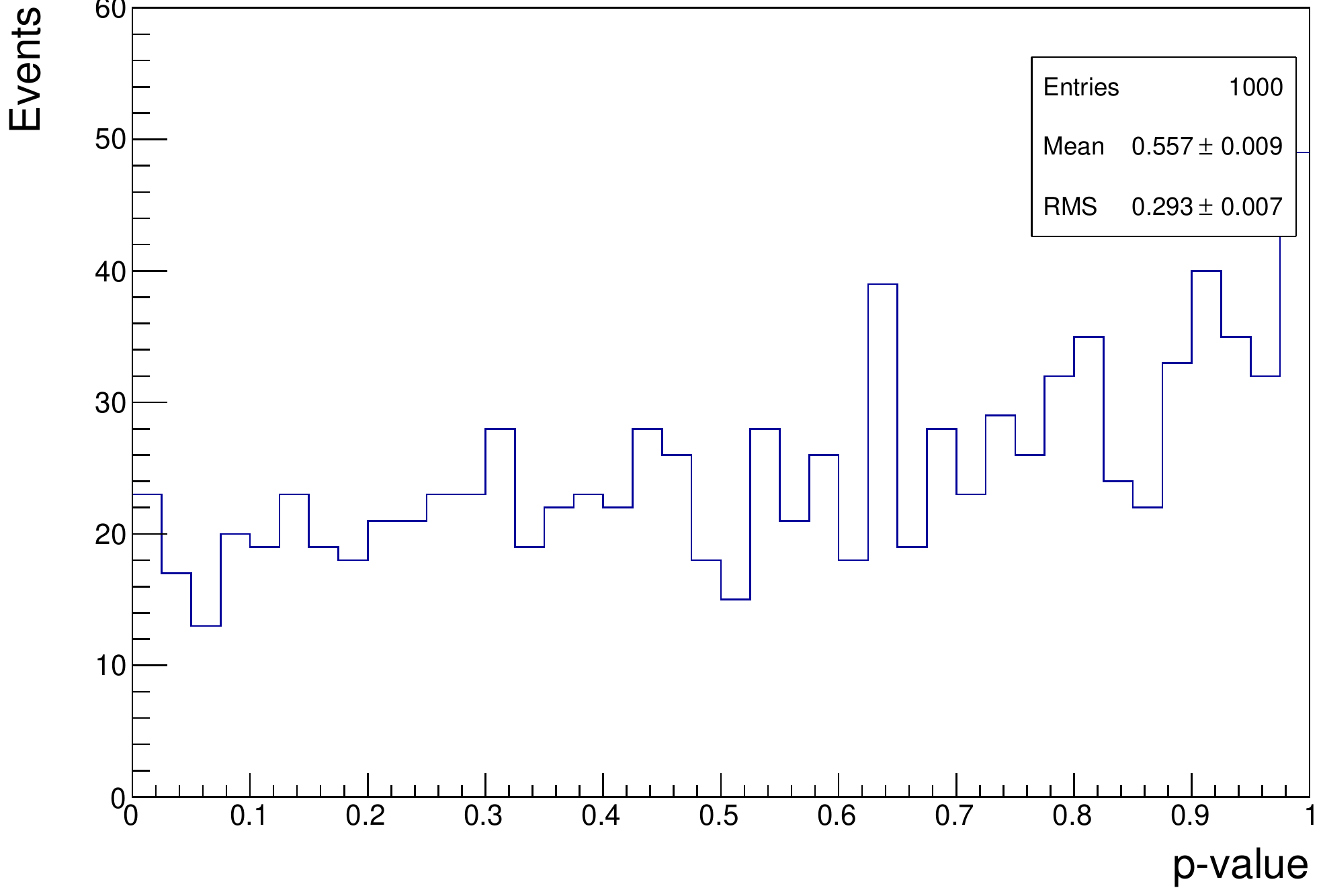}
\caption{p-value distribution of the fits to pseudo data corresponding to the fit scenario of Fit 2 in Table~\ref{Results1}.}
\label{PValue}
\end{figure}

This appendix provides an example of the results obtained from the tests 
with a statistical ensemble of 1000 pseudo-data given in Fig.\ref{Pull1}-\ref{PValue}.\\
For the fits on pseudo-datasets the distributions of the following two quantities 
are of particular interest:
\begin{itemize}
\item \bfseries The normalized residuals $p$:\normalfont\\
The normalized residual $p_{i,n}$ associated with fit scenario $n$ and branching fraction $i$ is defined as
\begin{equation}
p_{i,n}=\frac{\mathcal{B}_i^0(B\to X_c^il\nu)-\mathcal{B}_i^n(B\to X_c^il\nu)}{\sigma^n_i},
\end{equation}
where $\mathcal{B}_i^0(B\to X_c^il\nu)$ is the value of the branching fraction of 
$B\to X_c^il\nu$ decays used for the mixture of the nominal inclusive moments, 
$\mathcal{B}_i^n(B\to X_c^il\nu)$ is the fitted result of the branching fraction 
of $B\to X_c^il\nu$ decays of fit scenario $n$ and $\sigma^n_i$ is the corresponding 
calculated fit uncertainty. If the fit works properly, the expectation value of 
this quantity and its RMS (Root Mean Square) are
\begin{equation}
\left\langle p\right\rangle=0,\qquad \sigma_p=1.
\end{equation}
\item \bfseries The p-value $P$:\normalfont\\
If the fit results follow a Gaussian distribution with a standard deviation
which is well estimated by the fit uncertainty, and if the fit model correctly 
describes the data the p-value defined as 
$P=\int\limits_{\chi^2_0}^{\infty}f({\chi^2})d\chi^2$, where $f({\chi^2})$ denotes 
the $\chi^2$ probability density function and $\chi_0^2$ the result of a particular
fit, then $P$ follows a uniformly distributed random variable 
in the interval $\left[0,1\right]$.
\end{itemize}
\twocolumn
\begin{figure}
	\vspace{-0.6cm}
	\includegraphics[width=\columnwidth]{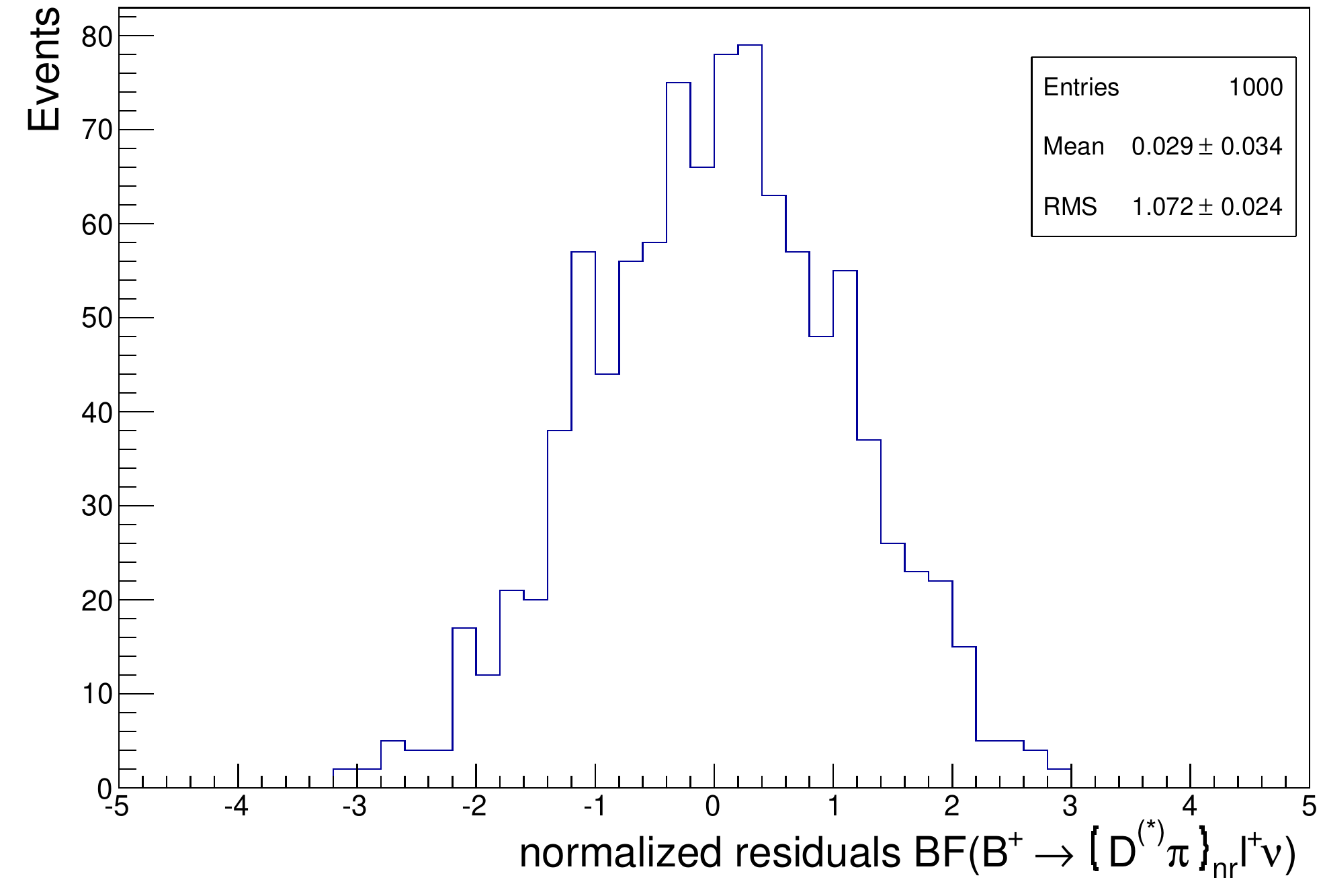}
\caption{Distribution of normalized residuals for the results of $\mathcal{B}(B^+\to (D^{(*)}\pi)_{nr} l^+\nu)$ corresponding to the fit scenario of Fit 2 in Table~\ref{Results1}.}
\end{figure}
\onecolumn

\section{ADDITIONAL RESULTS}\label{AddRes}
\FloatBarrier

\begin{figure*}
  \hspace{-0.03\textwidth}
  \includegraphics[width=1.1\textwidth]{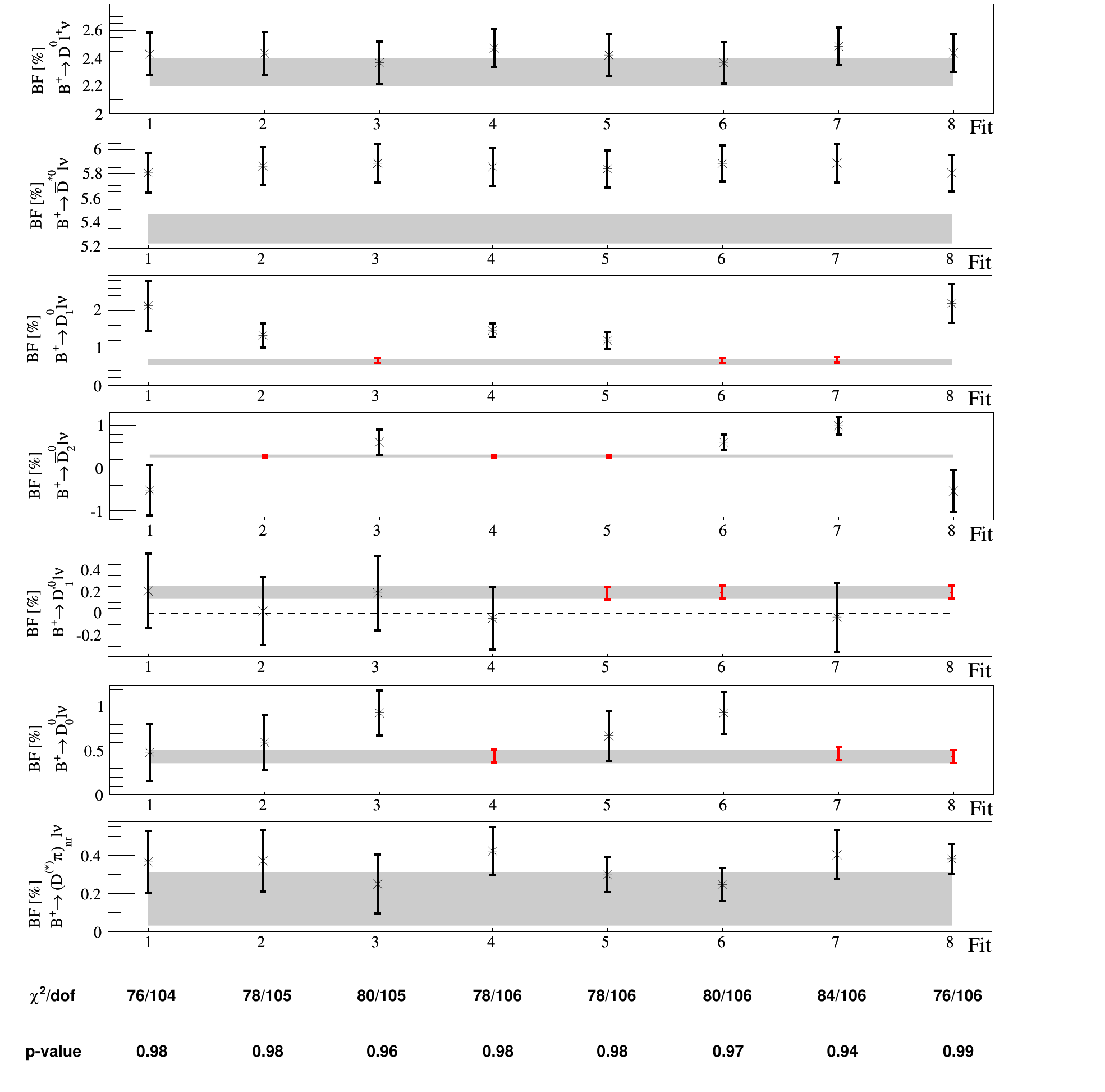}
\caption{Depiction of the fit results as quoted in Tables~\ref{Results00} and~\ref{Results01} of Appendix~\ref{AddRes}. 
In each subplot the abscissa indicates a distinctive fit scenario labeled by a number, 
whereas the ordinate represents the branching fraction. The results for a constrained 
branching fraction are depicted as red points, whereas the results for a unconstrained 
branching fraction are depicted as a black star. The grey bands correspond to the 
one-sigma error band of the corresponding direct branching-fraction measurements. 
If the ordinate includes zero, a dashed black line visualizes the zero line.
}     
\label{Plots1}
\end{figure*}

\begin{figure*}

  \hspace{-0.03\textwidth}
  \includegraphics[width=1.1\textwidth]{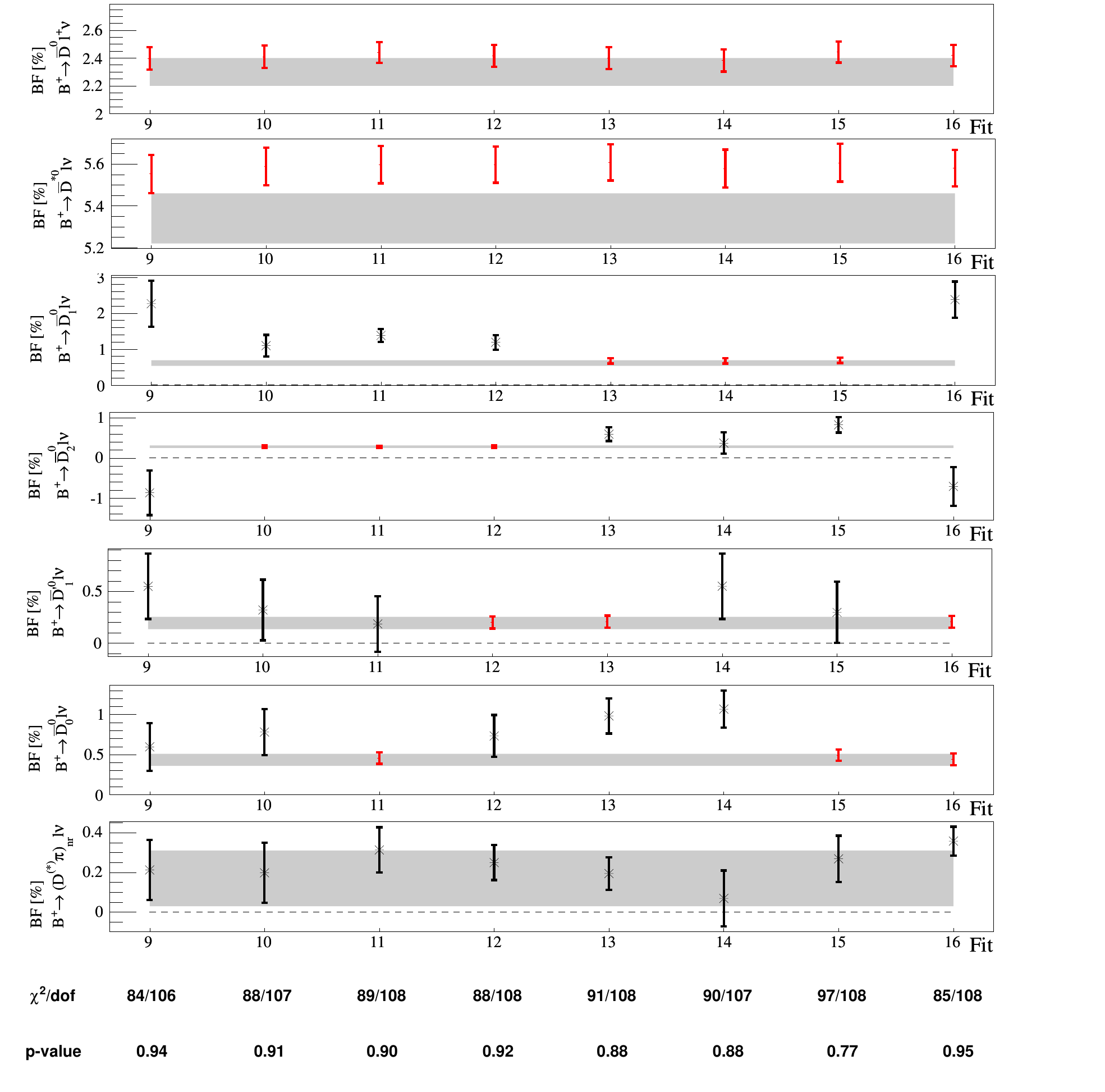}
\caption{Depiction of the fit results as quoted in Tables~\ref{Results10} and~\ref{Results11} of Appendix~\ref{AddRes}. 
In each subplot the abscissa indicates a distinctive fit scenario labeled by a number, 
whereas the ordinate represents the branching fraction. The results for a constrained 
branching fraction are depicted as red points, whereas the results for a unconstrained 
branching fraction are depicted as a black star. The grey bands correspond to the 
one-sigma error band of the corresponding direct branching-fraction measurements.  
If the ordinate includes zero, a dashed black line visualizes the zero line.
}     
\label{Plots2}
\end{figure*}

\begin{figure*}

  \hspace{-0.03\textwidth}
  \includegraphics[width=1.1\textwidth]{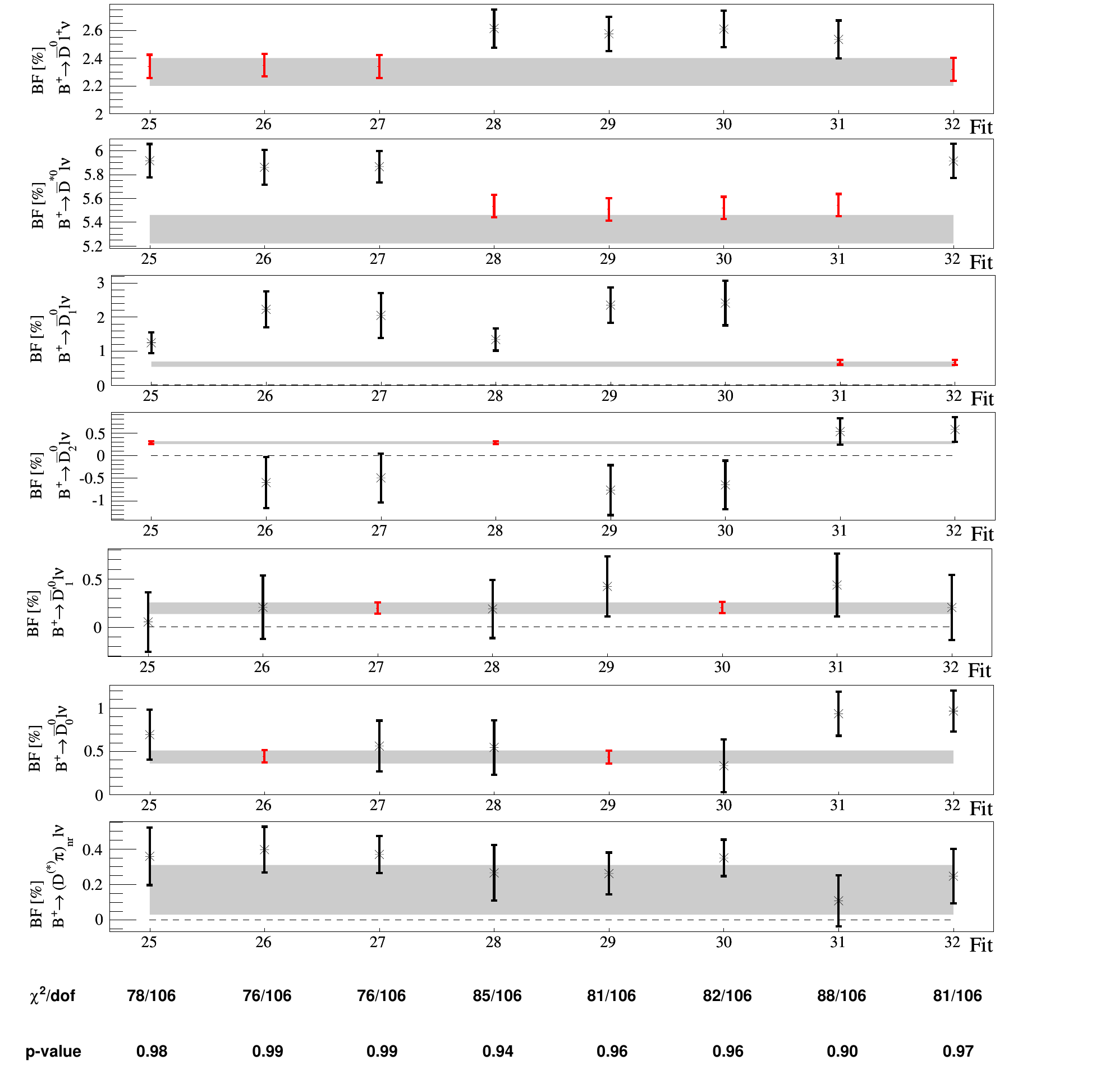}
\caption{Depiction of the fit results as quoted in Tables~\ref{Results20} and~\ref{Results21} of Appendix~\ref{AddRes}. 
In each subplot the abscissa indicates a distinctive fit scenario labeled by a number, 
whereas the ordinate represents the branching fraction. The results for a constrained 
branching fraction are depicted as red points, whereas the results for a unconstrained 
branching fraction are depicted as a black star. The grey bands correspond to the 
one-sigma error band of the corresponding direct branching-fraction measurements. 
If the ordinate includes zero, a dashed black line visualizes the zero line.
}     
\label{Plots3}
\end{figure*}

\begin{figure*}

  \hspace{-0.03\textwidth}
  \includegraphics[width=1.1\textwidth]{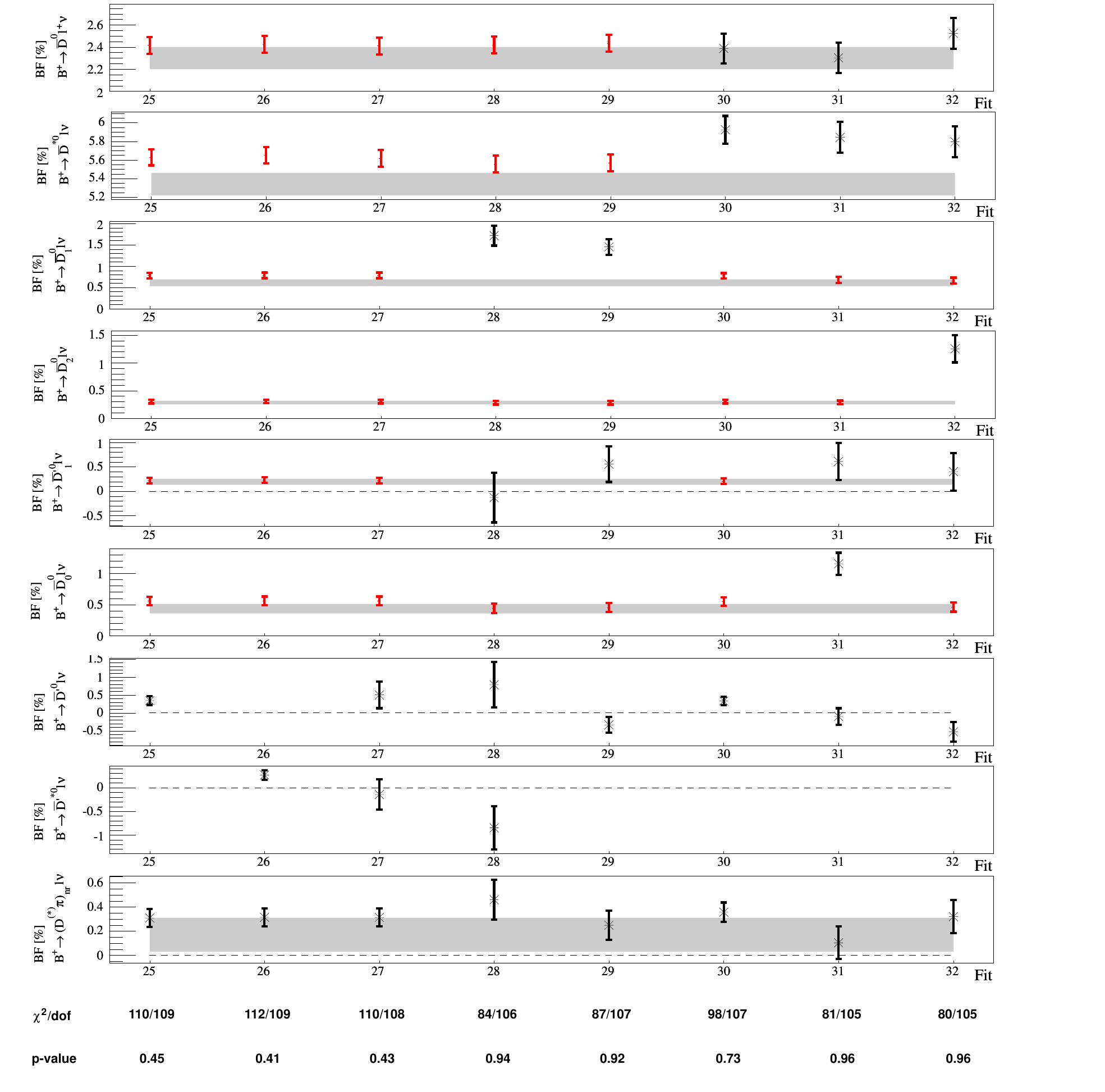}
\caption{Depiction of the fit results as quoted in Tables~\ref{Results30} and~\ref{Results31} of Appendix~\ref{AddRes}. 
In each subplot the abscissa indicates a distinctive fit scenario labeled by a number, 
whereas the ordinate represents the branching fraction. The results for a constrained 
branching fraction are depicted as red points, whereas the results for a unconstrained 
branching fraction are depicted as a black star. The grey bands correspond to the 
one-sigma error band of the corresponding direct branching-fraction measurements. 
If the ordinate includes zero, a dashed black line visualizes the zero line.
}     
\label{Plots4}
\end{figure*}

In this appendix, we quote in detail the results for all additional fit scenarios
that were studied. 
The results are grouped in eight tables (\ref{Results00}-\ref{Results31}), 
corresponding to the related plots in Fig.~\ref{Plots1}-\ref{Plots4}.

\begin{table*}[h!]
\centering
\scriptsize
\begin{tabular}{|c|c|c|c|c|c|c|c|c|c|}
 \hline
\multirow{2}{*}{$X_c$} & \multicolumn{2}{c|}{Fit 1}& \multicolumn{2}{c|}{Fit 2}& \multicolumn{2}{c|}{Fit 3}& \multicolumn{2}{c|}{Fit 4}&Measured\\ \cline{2-9}
&U/C&$\mathcal{B}[\%]$&U/C&$\mathcal{B}[\%]$&U/C&$\mathcal{B}[\%]$&U/C&$\mathcal{B}[\%]$&$\mathcal{B}[\%]$\\ \hline \hline
$ \overline{D}^{0}$&x/-&$2.43 \pm 0.15$&x/-&$2.43 \pm 0.15$&x/-&$2.37 \pm 0.15$&x/-&$2.47 \pm 0.14$&$2.30 \pm 0.10 $\\ \hline
$ \overline{D}^{*0}$&x/-&$5.81 \pm 0.16$&x/-&$5.86 \pm 0.16$&x/-&$5.89 \pm 0.16$&x/-&$5.86 \pm 0.16$&$5.34 \pm 0.12 $\\ \hline
$ \overline{D}_1^{0}$&x/-&$2.13 \pm 0.67$&x/-&$1.33 \pm 0.33$&x/x &$0.67 \pm 0.07$&x/-&$1.47 \pm 0.18$&$0.65 \pm 0.07 $\\ \hline
$ \overline{D}_2^{0}$&x/-&$-0.52 \pm 0.59$&x/x &$0.28 \pm 0.03$&x/-&$0.61 \pm 0.29$&x/x &$0.28 \pm 0.03$&$0.28 \pm 0.03 $\\ \hline
$ \overline{D}'^{0}_1$&x/-&$0.21 \pm 0.34$&x/-&$0.02 \pm 0.31$&x/-&$0.19 \pm 0.34$&x/-&$-0.04 \pm 0.28$&$0.20 \pm 0.06 $\\ \hline
$ \overline{D}_0^{0}$&x/-&$0.48 \pm 0.33$&x/-&$0.60 \pm 0.32$&x/-&$0.93 \pm 0.26$&x/x &$0.44 \pm 0.07$&$0.43 \pm 0.07 $\\ \hline
$ \overline{D}'^{0}$&-/-&-&-/-&-&-/-&-&-/-&-&-\\ \hline
$ \overline{D}'^{*0}$&-/-&-&-/-&-&-/-&-&-/-&-&-\\ \hline
$ (D^{(*)}\pi)_{nr}$&x/-&$0.36 \pm 0.16$&x/-&$0.37 \pm 0.16$&x/-&$0.25 \pm 0.15$&x/-&$0.42 \pm 0.13$&-\\ \hline
$\overline{D}^{0}_1/\overline{D}^{0}_2$& \multicolumn{2}{c|}{$1.61\pm0.33$}& \multicolumn{2}{c|}{$1.61\pm0.33$}& \multicolumn{2}{c|}{$1.27\pm0.29$}& \multicolumn{2}{c|}{$1.75\pm0.18$}&$0.94\pm0.08$\\ \hline
$\overline{D}^{0}_0/\overline{D}'^{0}_1$& \multicolumn{2}{c|}{$0.69\pm0.54$}& \multicolumn{2}{c|}{$0.62\pm0.53$}& \multicolumn{2}{c|}{$1.12\pm0.50$}& \multicolumn{2}{c|}{$0.40\pm0.30$}&$0.63\pm0.10$\\ \hline
$\overline{D}^{0}_0/\overline{D}'^{0}_1/(D^{(*)}\pi)_{nr}$& \multicolumn{2}{c|}{$1.06\pm0.40$}& \multicolumn{2}{c|}{$0.99\pm0.39$}& \multicolumn{2}{c|}{$1.37\pm0.37$}& \multicolumn{2}{c|}{$0.82\pm0.21$}&$0.63\pm0.10$\\ \hline
$\sum_iX_c^i$& \multicolumn{2}{c|}{$10.90\pm0.14$}& \multicolumn{2}{c|}{$10.90\pm0.14$}& \multicolumn{2}{c|}{$10.90\pm0.14$}& \multicolumn{2}{c|}{$10.90\pm0.14$}&$9.21\pm0.20$\\ \hline
$X_c$& \multicolumn{2}{c|}{}& \multicolumn{2}{c|}{}& \multicolumn{2}{c|}{}& \multicolumn{2}{c|}{}&$10.90 \pm 0.14$\\ \hline
\hline
$\chi^2/dof$& \multicolumn{2}{c|}{75/104 = 0.73}& \multicolumn{2}{c|}{77/105 = 0.74}& \multicolumn{2}{c|}{80/105 = 0.77}& \multicolumn{2}{c|}{77/106 = 0.73}&-\\ \hline
p-value& \multicolumn{2}{c|}{0.98}& \multicolumn{2}{c|}{0.98}& \multicolumn{2}{c|}{0.96}& \multicolumn{2}{c|}{0.98}&-\\ \hline
\end{tabular}
\caption{Results for moment fits of semileptonic decays $B^{+} \to X_{c}^{i}l^{+}\nu$ with hadronic final states $X_{c}^{i}$ 
containing $D$, $D^*$, any $D^{**}$, and $(D^{(*)}\pi)_{nr}$. Hereby, ``U/C'' stands for ``used/constrained'' and the ``x'' 
denotes ``yes'', whereas ``-'' denotes ``no'', respectively. 
The table is further described in the text of Section~\ref{sec:4}.}
\label{Results00} 
\end{table*}
\begin{table*}
\centering
\scriptsize
\begin{tabular}{|c|c|c|c|c|c|c|c|c|c|}
 \hline
\multirow{2}{*}{$X_c$} & \multicolumn{2}{c|}{Fit 5}& \multicolumn{2}{c|}{Fit 6}& \multicolumn{2}{c|}{Fit 7}& \multicolumn{2}{c|}{Fit 8}&Measured\\ \cline{2-9}
&U/C&$\mathcal{B}[\%]$&U/C&$\mathcal{B}[\%]$&U/C&$\mathcal{B}[\%]$&U/C&$\mathcal{B}[\%]$&$\mathcal{B}[\%]$\\ \hline \hline
$ \overline{D}^{0}$&x/-&$2.42 \pm 0.15$&x/-&$2.37 \pm 0.15$&x/-&$2.49 \pm 0.14$&x/-&$2.44 \pm 0.14$&$2.30 \pm 0.10 $\\ \hline
$ \overline{D}^{*0}$&x/-&$5.84 \pm 0.15$&x/-&$5.88 \pm 0.15$&x/-&$5.89 \pm 0.16$&x/-&$5.80 \pm 0.15$&$5.34 \pm 0.12 $\\ \hline
$ \overline{D}_1^{0}$&x/-&$1.20 \pm 0.23$&x/x &$0.67 \pm 0.07$&x/x &$0.68 \pm 0.07$&x/-&$2.19 \pm 0.52$&$0.65 \pm 0.07 $\\ \hline
$ \overline{D}_2^{0}$&x/x &$0.28 \pm 0.03$&x/-&$0.60 \pm 0.19$&x/-&$0.99 \pm 0.20$&x/-&$-0.54 \pm 0.50$&$0.28 \pm 0.03 $\\ \hline
$ \overline{D}'^{0}_1$&x/x &$0.19 \pm 0.06$&x/x &$0.19 \pm 0.06$&x/-&$-0.03 \pm 0.32$&x/x &$0.19 \pm 0.06$&$0.20 \pm 0.06 $\\ \hline
$ \overline{D}_0^{0}$&x/-&$0.67 \pm 0.29$&x/-&$0.94 \pm 0.24$&x/x &$0.47 \pm 0.07$&x/x &$0.44 \pm 0.07$&$0.43 \pm 0.07 $\\ \hline
$ \overline{D}'^{0}$&-/-&-&-/-&-&-/-&-&-/-&-&-\\ \hline
$ \overline{D}'^{*0}$&-/-&-&-/-&-&-/-&-&-/-&-&-\\ \hline
$ (D^{(*)}\pi)_{nr}$&x/-&$0.30 \pm 0.09$&x/-&$0.25 \pm 0.09$&x/-&$0.40 \pm 0.13$&x/-&$0.38 \pm 0.08$&-\\ \hline
$\overline{D}^{0}_1/\overline{D}^{0}_2$& \multicolumn{2}{c|}{$1.48\pm0.22$}& \multicolumn{2}{c|}{$1.27\pm0.18$}& \multicolumn{2}{c|}{$1.67\pm0.19$}& \multicolumn{2}{c|}{$1.65\pm0.11$}&$0.94\pm0.08$\\ \hline
$\overline{D}^{0}_0/\overline{D}'^{0}_1$& \multicolumn{2}{c|}{$0.86\pm0.30$}& \multicolumn{2}{c|}{$1.13\pm0.25$}& \multicolumn{2}{c|}{$0.44\pm0.33$}& \multicolumn{2}{c|}{$0.63\pm0.09$}&$0.63\pm0.10$\\ \hline
$\overline{D}^{0}_0/\overline{D}'^{0}_1/(D^{(*)}\pi)_{nr}$& \multicolumn{2}{c|}{$1.16\pm0.25$}& \multicolumn{2}{c|}{$1.38\pm0.22$}& \multicolumn{2}{c|}{$0.84\pm0.24$}& \multicolumn{2}{c|}{$1.01\pm0.10$}&$0.63\pm0.10$\\ \hline
$\sum_iX_c^i$& \multicolumn{2}{c|}{$10.90\pm0.14$}& \multicolumn{2}{c|}{$10.90\pm0.14$}& \multicolumn{2}{c|}{$10.89\pm0.14$}& \multicolumn{2}{c|}{$10.90\pm0.14$}&$9.21\pm0.20$\\ \hline
$X_c$& \multicolumn{2}{c|}{}& \multicolumn{2}{c|}{}& \multicolumn{2}{c|}{}& \multicolumn{2}{c|}{}&$10.90 \pm 0.14$\\ \hline
\hline
$\chi^2/dof$& \multicolumn{2}{c|}{77/106 = 0.74}& \multicolumn{2}{c|}{80/106 = 0.76}& \multicolumn{2}{c|}{83/106 = 0.79}& \multicolumn{2}{c|}{76/106 = 0.72}&-\\ \hline
p-value& \multicolumn{2}{c|}{0.98}& \multicolumn{2}{c|}{0.97}& \multicolumn{2}{c|}{0.94}& \multicolumn{2}{c|}{0.99}&-\\ \hline
\end{tabular}
\caption{Results for moment fits of semileptonic decays $B^{+} \to X_{c}^{i}l^{+}\nu$ with hadronic final states $X_{c}^{i}$ 
containing $D$, $D^*$, any $D^{**}$, and $(D^{(*)}\pi)_{nr}$. Hereby, ``U/C'' stands for ``used/constrained'' and the ``x'' 
denotes ``yes'', whereas ``-'' denotes ``no'', respectively. 
The table is further described in the text of Section~\ref{sec:4}.}
\label{Results01} 
\end{table*}

\begin{table*}
\centering
\scriptsize
\begin{tabular}{|c|c|c|c|c|c|c|c|c|c|}
 \hline
\multirow{2}{*}{$X_c$} & \multicolumn{2}{c|}{Fit 9}& \multicolumn{2}{c|}{Fit 10}& \multicolumn{2}{c|}{Fit 11}& \multicolumn{2}{c|}{Fit 12}&Measured\\ \cline{2-9}
&U/C&$\mathcal{B}[\%]$&U/C&$\mathcal{B}[\%]$&U/C&$\mathcal{B}[\%]$&U/C&$\mathcal{B}[\%]$&$\mathcal{B}[\%]$\\ \hline \hline
$ \overline{D}^{0}$&x/x &$2.40 \pm 0.08$&x/x &$2.41 \pm 0.08$&x/x &$2.44 \pm 0.08$&x/x &$2.42 \pm 0.08$&$2.30 \pm 0.10 $\\ \hline
$ \overline{D}^{*0}$&x/x &$5.55 \pm 0.09$&x/x &$5.59 \pm 0.09$&x/x &$5.60 \pm 0.09$&x/x &$5.60 \pm 0.09$&$5.34 \pm 0.12 $\\ \hline
$ \overline{D}_1^{0}$&x/-&$2.26 \pm 0.64$&x/-&$1.10 \pm 0.30$&x/-&$1.38 \pm 0.18$&x/-&$1.19 \pm 0.20$&$0.65 \pm 0.07 $\\ \hline
$ \overline{D}_2^{0}$&x/-&$-0.87 \pm 0.56$&x/x &$0.28 \pm 0.03$&x/x &$0.28 \pm 0.03$&x/x &$0.28 \pm 0.03$&$0.28 \pm 0.03 $\\ \hline
$ \overline{D}'^{0}_1$&x/-&$0.55 \pm 0.32$&x/-&$0.32 \pm 0.29$&x/-&$0.19 \pm 0.27$&x/x &$0.20 \pm 0.06$&$0.20 \pm 0.06 $\\ \hline
$ \overline{D}_0^{0}$&x/-&$0.60 \pm 0.30$&x/-&$0.78 \pm 0.29$&x/x &$0.46 \pm 0.07$&x/-&$0.73 \pm 0.26$&$0.43 \pm 0.07 $\\ \hline
$ \overline{D}'^{0}$&-/-&-&-/-&-&-/-&-&-/-&-&-\\ \hline
$ \overline{D}'^{*0}$&-/-&-&-/-&-&-/-&-&-/-&-&-\\ \hline
$ (D^{(*)}\pi)_{nr}$&x/-&$0.21 \pm 0.15$&x/-&$0.20 \pm 0.15$&x/-&$0.31 \pm 0.11$&x/-&$0.25 \pm 0.09$&-\\ \hline
$\overline{D}^{0}_1/\overline{D}^{0}_2$& \multicolumn{2}{c|}{$1.40\pm0.30$}& \multicolumn{2}{c|}{$1.38\pm0.30$}& \multicolumn{2}{c|}{$1.66\pm0.17$}& \multicolumn{2}{c|}{$1.47\pm0.20$}&$0.94\pm0.08$\\ \hline
$\overline{D}^{0}_0/\overline{D}'^{0}_1$& \multicolumn{2}{c|}{$1.15\pm0.49$}& \multicolumn{2}{c|}{$1.10\pm0.49$}& \multicolumn{2}{c|}{$0.64\pm0.29$}& \multicolumn{2}{c|}{$0.93\pm0.27$}&$0.63\pm0.10$\\ \hline
$\overline{D}^{0}_0/\overline{D}'^{0}_1/(D^{(*)}\pi)_{nr}$& \multicolumn{2}{c|}{$1.36\pm0.36$}& \multicolumn{2}{c|}{$1.30\pm0.36$}& \multicolumn{2}{c|}{$0.96\pm0.20$}& \multicolumn{2}{c|}{$1.18\pm0.22$}&$0.63\pm0.10$\\ \hline
$\sum_iX_c^i$& \multicolumn{2}{c|}{$10.71\pm0.12$}& \multicolumn{2}{c|}{$10.68\pm0.12$}& \multicolumn{2}{c|}{$10.65\pm0.12$}& \multicolumn{2}{c|}{$10.67\pm0.12$}&$9.21\pm0.20$\\ \hline
$X_c$& \multicolumn{2}{c|}{}& \multicolumn{2}{c|}{}& \multicolumn{2}{c|}{}& \multicolumn{2}{c|}{}&$10.90 \pm 0.14$\\ \hline
\hline
$\chi^2/dof$& \multicolumn{2}{c|}{83/106 = 0.79}& \multicolumn{2}{c|}{88/107 = 0.82}& \multicolumn{2}{c|}{89/108 = 0.83}& \multicolumn{2}{c|}{88/108 = 0.82}&-\\ \hline
p-value& \multicolumn{2}{c|}{0.94}& \multicolumn{2}{c|}{0.91}& \multicolumn{2}{c|}{0.90}& \multicolumn{2}{c|}{0.92}&-\\ \hline
\end{tabular}
\caption{Results for moment fits of semileptonic decays $B^{+} \to X_{c}^{i}l^{+}\nu$ with hadronic final states $X_{c}^{i}$ 
containing $D$, $D^*$, any $D^{**}$, and $(D^{(*)}\pi)_{nr}$. Hereby, ``U/C'' stands for ``used/constrained'' and the ``x'' 
denotes ``yes'', whereas ``-'' denotes ``no'', respectively. 
The table is further described in the text of Section~\ref{sec:4}.}
\label{Results10}
\end{table*}
\begin{table*}
\centering
\scriptsize
\begin{tabular}{|c|c|c|c|c|c|c|c|c|c|}
 \hline
\multirow{2}{*}{$X_c$} & \multicolumn{2}{c|}{Fit 13}& \multicolumn{2}{c|}{Fit 14}& \multicolumn{2}{c|}{Fit 15}& \multicolumn{2}{c|}{Fit 16}&Measured\\ \cline{2-9}
&U/C&$\mathcal{B}[\%]$&U/C&$\mathcal{B}[\%]$&U/C&$\mathcal{B}[\%]$&U/C&$\mathcal{B}[\%]$&$\mathcal{B}[\%]$\\ \hline \hline
$ \overline{D}^{0}$&x/x &$2.40 \pm 0.08$&x/x &$2.38 \pm 0.08$&x/x &$2.44 \pm 0.08$&x/x &$2.42 \pm 0.08$&$2.30 \pm 0.10 $\\ \hline
$ \overline{D}^{*0}$&x/x &$5.61 \pm 0.09$&x/x &$5.58 \pm 0.09$&x/x &$5.61 \pm 0.09$&x/x &$5.58 \pm 0.09$&$5.34 \pm 0.12 $\\ \hline
$ \overline{D}_1^{0}$&x/x &$0.67 \pm 0.07$&x/x &$0.67 \pm 0.07$&x/x &$0.69 \pm 0.07$&x/-&$2.38 \pm 0.50$&$0.65 \pm 0.07 $\\ \hline
$ \overline{D}_2^{0}$&x/-&$0.60 \pm 0.17$&x/-&$0.37 \pm 0.27$&x/-&$0.83 \pm 0.20$&x/-&$-0.71 \pm 0.48$&$0.28 \pm 0.03 $\\ \hline
$ \overline{D}'^{0}_1$&x/x &$0.21 \pm 0.06$&x/-&$0.55 \pm 0.32$&x/-&$0.30 \pm 0.30$&x/x &$0.21 \pm 0.06$&$0.20 \pm 0.06 $\\ \hline
$ \overline{D}_0^{0}$&x/-&$0.98 \pm 0.22$&x/-&$1.07 \pm 0.23$&x/x &$0.49 \pm 0.07$&x/x &$0.44 \pm 0.07$&$0.43 \pm 0.07 $\\ \hline
$ \overline{D}'^{0}$&-/-&-&-/-&-&-/-&-&-/-&-&-\\ \hline
$ \overline{D}'^{*0}$&-/-&-&-/-&-&-/-&-&-/-&-&-\\ \hline
$ (D^{(*)}\pi)_{nr}$&x/-&$0.19 \pm 0.08$&x/-&$0.07 \pm 0.14$&x/-&$0.27 \pm 0.12$&x/-&$0.36 \pm 0.07$&-\\ \hline
$\overline{D}^{0}_1/\overline{D}^{0}_2$& \multicolumn{2}{c|}{$1.27\pm0.16$}& \multicolumn{2}{c|}{$1.04\pm0.26$}& \multicolumn{2}{c|}{$1.52\pm0.18$}& \multicolumn{2}{c|}{$1.67\pm0.11$}&$0.94\pm0.08$\\ \hline
$\overline{D}^{0}_0/\overline{D}'^{0}_1$& \multicolumn{2}{c|}{$1.19\pm0.23$}& \multicolumn{2}{c|}{$1.62\pm0.45$}& \multicolumn{2}{c|}{$0.79\pm0.31$}& \multicolumn{2}{c|}{$0.65\pm0.09$}&$0.63\pm0.10$\\ \hline
$\overline{D}^{0}_0/\overline{D}'^{0}_1/(D^{(*)}\pi)_{nr}$& \multicolumn{2}{c|}{$1.38\pm0.19$}& \multicolumn{2}{c|}{$1.69\pm0.34$}& \multicolumn{2}{c|}{$1.06\pm0.23$}& \multicolumn{2}{c|}{$1.00\pm0.10$}&$0.63\pm0.10$\\ \hline
$\sum_iX_c^i$& \multicolumn{2}{c|}{$10.66\pm0.12$}& \multicolumn{2}{c|}{$10.69\pm0.12$}& \multicolumn{2}{c|}{$10.63\pm0.12$}& \multicolumn{2}{c|}{$10.67\pm0.12$}&$9.21\pm0.20$\\ \hline
$X_c$& \multicolumn{2}{c|}{}& \multicolumn{2}{c|}{}& \multicolumn{2}{c|}{}& \multicolumn{2}{c|}{}&$10.90 \pm 0.14$\\ \hline
\hline
$\chi^2/dof$& \multicolumn{2}{c|}{91/108 = 0.84}& \multicolumn{2}{c|}{90/107 = 0.84}& \multicolumn{2}{c|}{96/108 = 0.90}& \multicolumn{2}{c|}{85/108 = 0.79}&-\\ \hline
p-value& \multicolumn{2}{c|}{0.88}& \multicolumn{2}{c|}{0.88}& \multicolumn{2}{c|}{0.77}& \multicolumn{2}{c|}{0.95}&-\\ \hline
\end{tabular}
\caption{Results for moment fits of semileptonic decays $B^{+} \to X_{c}^{i}l^{+}\nu$ with hadronic final states $X_{c}^{i}$ 
containing $D$, $D^*$, any $D^{**}$, and $(D^{(*)}\pi)_{nr}$. Hereby, ``U/C'' stands for ``used/constrained'' and the ``x'' 
denotes ``yes'', whereas ``-'' denotes ``no'', respectively. 
The table is further described in the text of Section~\ref{sec:4}.}
\label{Results11}
\end{table*}

\begin{table*}
\centering
\scriptsize
\begin{tabular}{|c|c|c|c|c|c|c|c|c|c|}
 \hline
\multirow{2}{*}{$X_c$} & \multicolumn{2}{c|}{Fit 17}& \multicolumn{2}{c|}{Fit 18}& \multicolumn{2}{c|}{Fit 19}& \multicolumn{2}{c|}{Fit 20}&Measured\\ \cline{2-9}
&U/C&$\mathcal{B}[\%]$&U/C&$\mathcal{B}[\%]$&U/C&$\mathcal{B}[\%]$&U/C&$\mathcal{B}[\%]$&$\mathcal{B}[\%]$\\ \hline \hline
$ \overline{D}^{0}$&x/x &$2.34 \pm 0.08$&x/x &$2.35 \pm 0.08$&x/x &$2.34 \pm 0.08$&x/-&$2.61 \pm 0.14$&$2.30 \pm 0.10 $\\ \hline
$ \overline{D}^{*0}$&x/-&$5.92 \pm 0.14$&x/-&$5.86 \pm 0.15$&x/-&$5.87 \pm 0.13$&x/x &$5.53 \pm 0.09$&$5.34 \pm 0.12 $\\ \hline
$ \overline{D}_1^{0}$&x/-&$1.24 \pm 0.30$&x/-&$2.23 \pm 0.53$&x/-&$2.05 \pm 0.66$&x/-&$1.34 \pm 0.33$&$0.65 \pm 0.07 $\\ \hline
$ \overline{D}_2^{0}$&x/x &$0.28 \pm 0.03$&x/-&$-0.60 \pm 0.56$&x/-&$-0.50 \pm 0.54$&x/x &$0.28 \pm 0.03$&$0.28 \pm 0.03 $\\ \hline
$ \overline{D}'^{0}_1$&x/-&$0.05 \pm 0.31$&x/-&$0.20 \pm 0.33$&x/x &$0.20 \pm 0.06$&x/-&$0.19 \pm 0.30$&$0.20 \pm 0.06 $\\ \hline
$ \overline{D}_0^{0}$&x/-&$0.69 \pm 0.29$&x/x &$0.44 \pm 0.07$&x/-&$0.56 \pm 0.29$&x/-&$0.55 \pm 0.31$&$0.43 \pm 0.07 $\\ \hline
$ \overline{D}'^{0}$&-/-&-&-/-&-&-/-&-&-/-&-&-\\ \hline
$ \overline{D}'^{*0}$&-/-&-&-/-&-&-/-&-&-/-&-&-\\ \hline
$ (D^{(*)}\pi)_{nr}$&x/-&$0.36 \pm 0.16$&x/-&$0.40 \pm 0.13$&x/-&$0.37 \pm 0.10$&x/-&$0.27 \pm 0.16$&-\\ \hline
$\overline{D}^{0}_1/\overline{D}^{0}_2$& \multicolumn{2}{c|}{$1.52\pm0.30$}& \multicolumn{2}{c|}{$1.63\pm0.19$}& \multicolumn{2}{c|}{$1.55\pm0.22$}& \multicolumn{2}{c|}{$1.62\pm0.32$}&$0.94\pm0.08$\\ \hline
$\overline{D}^{0}_0/\overline{D}'^{0}_1$& \multicolumn{2}{c|}{$0.75\pm0.50$}& \multicolumn{2}{c|}{$0.65\pm0.34$}& \multicolumn{2}{c|}{$0.76\pm0.30$}& \multicolumn{2}{c|}{$0.73\pm0.52$}&$0.63\pm0.10$\\ \hline
$\overline{D}^{0}_0/\overline{D}'^{0}_1/(D^{(*)}\pi)_{nr}$& \multicolumn{2}{c|}{$1.10\pm0.36$}& \multicolumn{2}{c|}{$1.04\pm0.24$}& \multicolumn{2}{c|}{$1.13\pm0.24$}& \multicolumn{2}{c|}{$1.00\pm0.39$}&$0.63\pm0.10$\\ \hline
$\sum_iX_c^i$& \multicolumn{2}{c|}{$10.88\pm0.14$}& \multicolumn{2}{c|}{$10.88\pm0.14$}& \multicolumn{2}{c|}{$10.88\pm0.14$}& \multicolumn{2}{c|}{$10.77\pm0.13$}&$9.21\pm0.20$\\ \hline
$X_c$& \multicolumn{2}{c|}{}& \multicolumn{2}{c|}{}& \multicolumn{2}{c|}{}& \multicolumn{2}{c|}{}&$10.90 \pm 0.14$\\ \hline
\hline
$\chi^2/dof$& \multicolumn{2}{c|}{78/106 = 0.74}& \multicolumn{2}{c|}{76/106 = 0.72}& \multicolumn{2}{c|}{76/106 = 0.72}& \multicolumn{2}{c|}{84/106 = 0.80}&-\\ \hline
p-value& \multicolumn{2}{c|}{0.98}& \multicolumn{2}{c|}{0.99}& \multicolumn{2}{c|}{0.99}& \multicolumn{2}{c|}{0.94}&-\\ \hline
\end{tabular}
\caption{Results for moment fits of semileptonic decays $B^{+} \to X_{c}^{i}l^{+}\nu$ with hadronic final states $X_{c}^{i}$ 
containing $D$, $D^*$, any $D^{**}$, and $(D^{(*)}\pi)_{nr}$. Hereby, ``U/C'' stands for ``used/constrained'' and the ``x'' 
denotes ``yes'', whereas ``-'' denotes ``no'', respectively. 
The table is further described in the text of Section~\ref{sec:4}.}
\label{Results20}
\end{table*}
\begin{table*}
\centering
\scriptsize
\begin{tabular}{|c|c|c|c|c|c|c|c|c|c|}
\hline
\multirow{2}{*}{$X_c$} & \multicolumn{2}{c|}{Fit 21}& \multicolumn{2}{c|}{Fit 22}& \multicolumn{2}{c|}{Fit 23}& \multicolumn{2}{c|}{Fit 24}&Measured\\ \cline{2-9}
&U/C&$\mathcal{B}[\%]$&U/C&$\mathcal{B}[\%]$&U/C&$\mathcal{B}[\%]$&U/C&$\mathcal{B}[\%]$&$\mathcal{B}[\%]$\\ \hline \hline
$ \overline{D}^{0}$&x/-&$2.57 \pm 0.12$&x/-&$2.61 \pm 0.13$&x/-&$2.53 \pm 0.14$&x/x &$2.32 \pm 0.08$&$2.30 \pm 0.10 $\\ \hline
$ \overline{D}^{*0}$&x/x &$5.51 \pm 0.10$&x/x &$5.52 \pm 0.09$&x/x &$5.54 \pm 0.09$&x/-&$5.91 \pm 0.14$&$5.34 \pm 0.12 $\\ \hline
$ \overline{D}_1^{0}$&x/-&$2.35 \pm 0.52$&x/-&$2.41 \pm 0.66$&x/x &$0.67 \pm 0.07$&x/x &$0.67 \pm 0.07$&$0.65 \pm 0.07 $\\ \hline
$ \overline{D}_2^{0}$&x/-&$-0.76 \pm 0.55$&x/-&$-0.65 \pm 0.53$&x/-&$0.53 \pm 0.29$&x/-&$0.57 \pm 0.28$&$0.28 \pm 0.03 $\\ \hline
$ \overline{D}'^{0}_1$&x/-&$0.42 \pm 0.31$&x/x &$0.20 \pm 0.06$&x/-&$0.44 \pm 0.33$&x/-&$0.20 \pm 0.34$&$0.20 \pm 0.06 $\\ \hline
$ \overline{D}_0^{0}$&x/x &$0.43 \pm 0.07$&x/-&$0.33 \pm 0.31$&x/-&$0.93 \pm 0.25$&x/-&$0.96 \pm 0.24$&$0.43 \pm 0.07 $\\ \hline
$ \overline{D}'^{0}$&-/-&-&-/-&-&-/-&-&-/-&-&-\\ \hline
$ \overline{D}'^{*0}$&-/-&-&-/-&-&-/-&-&-/-&-&-\\ \hline
$ (D^{(*)}\pi)_{nr}$&x/-&$0.26 \pm 0.12$&x/-&$0.35 \pm 0.10$&x/-&$0.11 \pm 0.14$&x/-&$0.25 \pm 0.15$&-\\ \hline
$\overline{D}^{0}_1/\overline{D}^{0}_2$& \multicolumn{2}{c|}{$1.58\pm0.19$}& \multicolumn{2}{c|}{$1.76\pm0.23$}& \multicolumn{2}{c|}{$1.20\pm0.29$}& \multicolumn{2}{c|}{$1.24\pm0.27$}&$0.94\pm0.08$\\ \hline
$\overline{D}^{0}_0/\overline{D}'^{0}_1$& \multicolumn{2}{c|}{$0.85\pm0.33$}& \multicolumn{2}{c|}{$0.54\pm0.31$}& \multicolumn{2}{c|}{$1.37\pm0.49$}& \multicolumn{2}{c|}{$1.17\pm0.48$}&$0.63\pm0.10$\\ \hline
$\overline{D}^{0}_0/\overline{D}'^{0}_1/(D^{(*)}\pi)_{nr}$& \multicolumn{2}{c|}{$1.12\pm0.24$}& \multicolumn{2}{c|}{$0.89\pm0.25$}& \multicolumn{2}{c|}{$1.48\pm0.37$}& \multicolumn{2}{c|}{$1.41\pm0.35$}&$0.63\pm0.10$\\ \hline
$\sum_iX_c^i$& \multicolumn{2}{c|}{$10.78\pm0.13$}& \multicolumn{2}{c|}{$10.77\pm0.13$}& \multicolumn{2}{c|}{$10.75\pm0.13$}& \multicolumn{2}{c|}{$10.89\pm0.14$}&$9.21\pm0.20$\\ \hline
$X_c$& \multicolumn{2}{c|}{}& \multicolumn{2}{c|}{}& \multicolumn{2}{c|}{}& \multicolumn{2}{c|}{}&$10.90 \pm 0.14$\\ \hline
\hline
$\chi^2/dof$& \multicolumn{2}{c|}{81/106 = 0.77}& \multicolumn{2}{c|}{81/106 = 0.77}& \multicolumn{2}{c|}{88/106 = 0.83}& \multicolumn{2}{c|}{80/106 = 0.76}&-\\ \hline
p-value& \multicolumn{2}{c|}{0.96}& \multicolumn{2}{c|}{0.96}& \multicolumn{2}{c|}{0.90}& \multicolumn{2}{c|}{0.97}&-\\ \hline
\end{tabular}
\caption{Results for moment fits of semileptonic decays $B^{+} \to X_{c}^{i}l^{+}\nu$ with hadronic final states $X_{c}^{i}$ 
containing $D$, $D^*$, any $D^{**}$, and $(D^{(*)}\pi)_{nr}$. Hereby, ``U/C'' stands for ``used/constrained'' and the ``x'' 
denotes ``yes'', whereas ``-'' denotes ``no'', respectively. 
The table is further described in the text of Section~\ref{sec:4}.}
\label{Results21}
\end{table*}

\begin{table*}
\centering
\scriptsize
\begin{tabular}{|c|c|c|c|c|c|c|c|c|c|}
 \hline
\multirow{2}{*}{$X_c$} & \multicolumn{2}{c|}{Fit 25}& \multicolumn{2}{c|}{Fit 26}& \multicolumn{2}{c|}{Fit 27}& \multicolumn{2}{c|}{Fit 28}&Measured\\ \cline{2-9}
&U/C&$\mathcal{B}[\%]$&U/C&$\mathcal{B}[\%]$&U/C&$\mathcal{B}[\%]$&U/C&$\mathcal{B}[\%]$&$\mathcal{B}[\%]$\\ \hline \hline
$ \overline{D}^{0}$&x/x &$2.42 \pm 0.08$&x/x &$2.42 \pm 0.08$&x/x &$2.41 \pm 0.08$&x/x &$2.42 \pm 0.08$&$2.30 \pm 0.10 $\\ \hline
$ \overline{D}^{*0}$&x/x &$5.63 \pm 0.09$&x/x &$5.65 \pm 0.09$&x/x &$5.62 \pm 0.09$&x/x &$5.56 \pm 0.09$&$5.34 \pm 0.12 $\\ \hline
$ \overline{D}_1^{0}$&x/x &$0.78 \pm 0.07$&x/x &$0.79 \pm 0.07$&x/x &$0.79 \pm 0.07$&x/-&$1.72 \pm 0.23$&$0.65 \pm 0.07 $\\ \hline
$ \overline{D}_2^{0}$&x/x &$0.30 \pm 0.03$&x/x &$0.30 \pm 0.03$&x/x &$0.30 \pm 0.03$&x/x &$0.28 \pm 0.03$&$0.28 \pm 0.03 $\\ \hline
$ \overline{D}'^{0}_1$&x/x &$0.22 \pm 0.06$&x/x &$0.23 \pm 0.06$&x/x &$0.22 \pm 0.06$&x/-&$-0.13 \pm 0.51$&$0.20 \pm 0.06 $\\ \hline
$ \overline{D}_0^{0}$&x/x &$0.56 \pm 0.07$&x/x &$0.56 \pm 0.07$&x/x &$0.56 \pm 0.07$&x/x &$0.44 \pm 0.07$&$0.43 \pm 0.07 $\\ \hline
$ \overline{D}'^{0}$&x/-&$0.35 \pm 0.12$&-/-&-&x/-&$0.50 \pm 0.37$&x/-&$0.79 \pm 0.63$&-\\ \hline
$ \overline{D}'^{*0}$&-/-&-&x/-&$0.27 \pm 0.10$&x/-&$-0.14 \pm 0.32$&x/-&$-0.85 \pm 0.46$&-\\ \hline
$ (D^{(*)}\pi)_{nr}$&x/-&$0.31 \pm 0.07$&x/-&$0.31 \pm 0.07$&x/-&$0.31 \pm 0.07$&x/-&$0.46 \pm 0.17$&-\\ \hline
$\overline{D}^{0}_1+\overline{D}^{0}_2$& \multicolumn{2}{c|}{$1.08\pm0.07$}& \multicolumn{2}{c|}{$1.09\pm0.07$}& \multicolumn{2}{c|}{$1.09\pm0.07$}& \multicolumn{2}{c|}{$2.00\pm0.23$}&$0.94\pm0.08$\\ \hline
$\overline{D}^{0}_0+\overline{D}'^{0}_1$& \multicolumn{2}{c|}{$0.78\pm0.09$}& \multicolumn{2}{c|}{$0.79\pm0.09$}& \multicolumn{2}{c|}{$0.78\pm0.09$}& \multicolumn{2}{c|}{$0.32\pm0.52$}&$0.63\pm0.10$\\ \hline
$\overline{D}^{0}_0/\overline{D}'^{0}_1+(D^{(*)}\pi)_{nr}$& \multicolumn{2}{c|}{$1.09\pm0.10$}& \multicolumn{2}{c|}{$1.11\pm0.10$}& \multicolumn{2}{c|}{$1.09\pm0.10$}& \multicolumn{2}{c|}{$0.78\pm0.38$}&$0.63\pm0.10$\\ \hline
$\sum_iX_c^i$& \multicolumn{2}{c|}{$10.56\pm0.12$}& \multicolumn{2}{c|}{$10.54\pm0.12$}& \multicolumn{2}{c|}{$10.57\pm0.12$}& \multicolumn{2}{c|}{$10.69\pm0.12$}&$9.21\pm0.20$\\ \hline
$X_c$& \multicolumn{2}{c|}{}& \multicolumn{2}{c|}{}& \multicolumn{2}{c|}{}& \multicolumn{2}{c|}{}&$10.90 \pm 0.14$\\ \hline
\hline
$\chi^2/dof$& \multicolumn{2}{c|}{110/109 = 1.01}& \multicolumn{2}{c|}{111/109 = 1.03}& \multicolumn{2}{c|}{110/108 = 1.02}& \multicolumn{2}{c|}{84/106 = 0.79}&-\\ \hline
p-value& \multicolumn{2}{c|}{0.45}& \multicolumn{2}{c|}{0.41}& \multicolumn{2}{c|}{0.43}& \multicolumn{2}{c|}{0.94}&-\\ \hline
\end{tabular}
\caption{Results for moment fits of semileptonic decays $B^{+} \to X_{c}^{i}l^{+}\nu$ with hadronic final states $X_{c}^{i}$ 
containing $D$, $D^*$, any $D^{**}$, $D'^{(*)}$, and $(D^{(*)}\pi)_{nr}$. Hereby, ``U/C'' stands for ``used/constrained'' and 
the ``x'' denotes ``yes'', whereas ``-'' denotes ``no'', respectively. 
The table is further described in the text of Section~\ref{sec:4}.}
\label{Results30}
\end{table*}
\begin{table*}
\centering
\scriptsize
\begin{tabular}{|c|c|c|c|c|c|c|c|c|c|}
 \hline
\multirow{2}{*}{$X_c$} & \multicolumn{2}{c|}{Fit 29}& \multicolumn{2}{c|}{Fit 30}& \multicolumn{2}{c|}{Fit 31}& \multicolumn{2}{c|}{Fit 32}&Measured\\ \cline{2-9}
&U/C&$\mathcal{B}[\%]$&U/C&$\mathcal{B}[\%]$&U/C&$\mathcal{B}[\%]$&U/C&$\mathcal{B}[\%]$&$\mathcal{B}[\%]$\\ \hline \hline
$ \overline{D}^{0}$&x/x &$2.44 \pm 0.08$&x/-&$2.39 \pm 0.13$&x/-&$2.30 \pm 0.14$&x/-&$2.52 \pm 0.14$&$2.30 \pm 0.10 $\\ \hline
$ \overline{D}^{*0}$&x/x &$5.57 \pm 0.09$&x/-&$5.92 \pm 0.15$&x/-&$5.84 \pm 0.17$&x/-&$5.79 \pm 0.17$&$5.34 \pm 0.12 $\\ \hline
$ \overline{D}_1^{0}$&x/-&$1.45 \pm 0.18$&x/x &$0.77 \pm 0.07$&x/x &$0.68 \pm 0.07$&x/x &$0.66 \pm 0.07$&$0.65 \pm 0.07 $\\ \hline
$ \overline{D}_2^{0}$&x/x &$0.28 \pm 0.03$&x/x &$0.30 \pm 0.03$&x/x &$0.29 \pm 0.03$&x/-&$1.25 \pm 0.24$&$0.28 \pm 0.03 $\\ \hline
$ \overline{D}'^{0}_1$&x/-&$0.56 \pm 0.37$&x/x &$0.21 \pm 0.06$&x/-&$0.61 \pm 0.38$&x/-&$0.40 \pm 0.39$&$0.20 \pm 0.06 $\\ \hline
$ \overline{D}_0^{0}$&x/x &$0.46 \pm 0.07$&x/x &$0.55 \pm 0.07$&x/-&$1.15 \pm 0.18$&x/x &$0.46 \pm 0.07$&$0.43 \pm 0.07 $\\ \hline
$ \overline{D}'^{0}$&x/-&$-0.33 \pm 0.22$&x/-&$0.33 \pm 0.12$&x/-&$-0.10 \pm 0.23$&x/-&$-0.52 \pm 0.27$&-\\ \hline
$ \overline{D}'^{*0}$&-/-&-&-/-&-&-/-&-&-/-&-&-\\ \hline
$ (D^{(*)}\pi)_{nr}$&x/-&$0.25 \pm 0.12$&x/-&$0.36 \pm 0.08$&x/-&$0.10 \pm 0.13$&x/-&$0.32 \pm 0.14$&-\\ \hline
$\overline{D}^{0}_1/\overline{D}^{0}_2$& \multicolumn{2}{c|}{$1.74\pm0.18$}& \multicolumn{2}{c|}{$1.07\pm0.07$}& \multicolumn{2}{c|}{$0.97\pm0.08$}& \multicolumn{2}{c|}{$1.92\pm0.23$}&$0.94\pm0.08$\\ \hline
$\overline{D}^{0}_0/\overline{D}'^{0}_1$& \multicolumn{2}{c|}{$1.01\pm0.38$}& \multicolumn{2}{c|}{$0.76\pm0.09$}& \multicolumn{2}{c|}{$1.77\pm0.40$}& \multicolumn{2}{c|}{$0.86\pm0.40$}&$0.63\pm0.10$\\ \hline
$\overline{D}^{0}_0/\overline{D}'^{0}_1/(D^{(*)}\pi)_{nr}$& \multicolumn{2}{c|}{$1.26\pm0.29$}& \multicolumn{2}{c|}{$1.11\pm0.10$}& \multicolumn{2}{c|}{$1.87\pm0.30$}& \multicolumn{2}{c|}{$1.18\pm0.29$}&$0.63\pm0.10$\\ \hline
$\sum_iX_c^i$& \multicolumn{2}{c|}{$10.67\pm0.12$}& \multicolumn{2}{c|}{$10.84\pm0.14$}& \multicolumn{2}{c|}{$10.89\pm0.14$}& \multicolumn{2}{c|}{$10.89\pm0.14$}&$9.21\pm0.20$\\ \hline
$X_c$& \multicolumn{2}{c|}{}& \multicolumn{2}{c|}{}& \multicolumn{2}{c|}{}& \multicolumn{2}{c|}{}&$10.90 \pm 0.14$\\ \hline
\hline
$\chi^2/dof$& \multicolumn{2}{c|}{87/107 = 0.82}& \multicolumn{2}{c|}{97/107 = 0.91}& \multicolumn{2}{c|}{81/105 = 0.78}& \multicolumn{2}{c|}{80/105 = 0.77}&-\\ \hline
p-value& \multicolumn{2}{c|}{0.92}& \multicolumn{2}{c|}{0.73}& \multicolumn{2}{c|}{0.96}& \multicolumn{2}{c|}{0.96}&-\\ \hline
\end{tabular}
\caption{Results for moment fits of semileptonic decays $B^{+} \to X_{c}^{i}l^{+}\nu$ with hadronic final states $X_{c}^{i}$ 
containing $D$, $D^*$, any $D^{**}$, $D'^{(*)}$, and $(D^{(*)}\pi)_{nr}$. Hereby, ``U/C'' stands for ``used/constrained'' and 
the ``x'' denotes ``yes'', whereas ``-'' denotes ``no'', respectively. 
The table is further described in the text of Section~\ref{sec:4}.}
\label{Results31}
\end{table*}

\section{PLOTS OF ADDITIONAL RESULTS}\label{AddPlots}

\section{RESULTS FOR SELECTED SETS OF INPUTS}\label{DstarSys}
In this appendix, we show the results for fits were different sets of experimental 
inputs are used. This is done to check if particlar measurements drive e.g. 
$\mathcal{B}(B^+\to\overline{D}^{*0}l^+\nu)$ to the large observed values and which 
fit inputs have the strongest impact on the final fit uncertainties.
In these fits we always constrain $\mathcal{B}(B^+\to\overline{D}^{0}l^+\nu)$ and 
$\mathcal{B}(B^+\to\overline{D}_{0}^{0}l^+\nu)$ to their measured value and present 
two series of fits in which we either constrain $\mathcal{B}(B^+\to\overline{D}_{2}^{0}l^+\nu)$ 
or $\mathcal{B}(B^+\to\overline{D}_{1}^{0}l^+\nu)$.
\\
Accordingly, in Fit 1, respectively, Fit 6 only lepton energy moments are used. 
In Fit 2 and Fit 7 we use only the combined hadronic mass-energy moments,
which were only measured by the \babar experiment, whereas in Fit 3 and Fit 8 
we use only hadronic mass moments. In Fit 4 and Fit 9, we use combined hadronic 
mass-energy moments (measured only by the~{\babar}), but omit the two
data points for the lower lepton momentum cut-off equal to 1.8 GeV/c and 1.9 GeV/c. 
This check was performed since these two data points can not be described very well 
by the fitted moment distribution. Furthermore, in Fit 5 and 
Fit 10, we use only the hadronic mass moments but remove the Belle measurements
from the list of inputs.\\
We find that $\mathcal{B}(B^+\to\overline{D}^{*0}l^+\nu)$ is mainly enlarged 
due to the hadronic mass-energy moments and in particular due to the mass moments.
The fit uncertainties decrease in size when using either only combined hadronic
energy-mass moments, or only electron energy moments, or only hadronic mass moments.

\begin{table*}
\centering
{\fontsize{5}{8} \selectfont 
\begin{tabular}
{|c|c|c|c|c|c|c|c|c|c|c|c|}
\hline
\multirow{2}{*}{$X_c$} & \multicolumn{2}{c|}{Fit 1}& \multicolumn{2}{c|}{Fit 2}& \multicolumn{2}{c|}{Fit 3}& \multicolumn{2}{c|}{Fit 4}& \multicolumn{2}{c|}{Fit 5}&Measured\\ \cline{2-11}
&U/C&$\mathcal{B}[\%]$&U/C&$\mathcal{B}[\%]$&U/C&$\mathcal{B}[\%]$&U/C&$\mathcal{B}[\%]$&U/C&$\mathcal{B}[\%]$&$[\%]$\\ \hline \hline
$ \overline{D}^{0}$&x/x &$2.30 \pm 0.10$&x/x &$2.25 \pm 0.10$&x/x &$2.31 \pm 0.09$&x/x &$2.25 \pm 0.10$&x/x &$2.31 \pm 0.10$&$2.30 \pm 0.10 $\\ \hline
$ \overline{D}^{*0}$&x/-&$5.55 \pm 0.21$&x/-&$5.91 \pm 0.25$&x/-&$6.20 \pm 0.18$&x/-&$6.00 \pm 0.27$&x/-&$6.18 \pm 0.20$&$5.34 \pm 0.12 $\\ \hline
$ \overline{D}_1^{0}$&x/-&$2.01 \pm 0.42$&x/-&$1.18 \pm 0.57$&x/-&$1.29 \pm 0.26$&x/-&$1.48 \pm 0.63$&x/-&$0.85 \pm 0.39$&$0.65 \pm 0.07 $\\ \hline
$ \overline{D}_2^{0}$&x/x &$0.28 \pm 0.03$&x/x &$0.28 \pm 0.03$&x/x &$0.28 \pm 0.03$&x/x &$0.28 \pm 0.03$&x/x &$0.28 \pm 0.03$&$0.28 \pm 0.03 $\\ \hline
$ \overline{D}'^{0}_1$&x/-&$0.45 \pm 0.54$&x/-&$0.68 \pm 0.94$&x/-&$-0.14 \pm 0.39$&x/-&$0.05 \pm 1.10$&x/-&$0.51 \pm 0.58$&$0.20 \pm 0.06 $\\ \hline
$ \overline{D}_0^{0}$&x/x &$0.43 \pm 0.08$&x/x &$0.44 \pm 0.07$&x/x &$0.44 \pm 0.07$&x/x &$0.44 \pm 0.07$&x/x &$0.43 \pm 0.07$&$0.43 \pm 0.07 $\\ \hline
$ \overline{D}'^{0}$&-/-&-&-/-&-&-/-&-&-/-&-&-/-&-&-\\ \hline
$ \overline{D}'^{*0}$&-/-&-&-/-&-&-/-&-&-/-&-&-/-&-&-\\ \hline
$ (D^{(*)}\pi)_{nr}$&x/-&$-0.13 \pm 0.33$&x/-&$0.18 \pm 0.37$&x/-&$0.51 \pm 0.18$&x/-&$0.42 \pm 0.44$&x/-&$0.33 \pm 0.24$&-\\ \hline
$\overline{D}^{0}_1/\overline{D}^{0}_2$& \multicolumn{2}{c|}{$2.30\pm0.42$}& \multicolumn{2}{c|}{$1.47\pm0.57$}& \multicolumn{2}{c|}{$1.58\pm0.26$}& \multicolumn{2}{c|}{$1.77\pm0.63$}& \multicolumn{2}{c|}{$1.14\pm0.39$}&$0.94\pm0.08$\\ \hline
$\overline{D}^{0}_0/\overline{D}'^{0}_1$& \multicolumn{2}{c|}{$0.88\pm0.54$}& \multicolumn{2}{c|}{$1.11\pm0.94$}& \multicolumn{2}{c|}{$0.29\pm0.41$}& \multicolumn{2}{c|}{$0.48\pm1.11$}& \multicolumn{2}{c|}{$0.94\pm0.59$}&$0.63\pm0.10$\\ \hline
$\overline{D}^{0}_0/\overline{D}'^{0}_1/(D^{(*)}\pi)_{nr}$& \multicolumn{2}{c|}{$0.75\pm0.42$}& \multicolumn{2}{c|}{$1.29\pm0.59$}& \multicolumn{2}{c|}{$0.80\pm0.29$}& \multicolumn{2}{c|}{$0.91\pm0.69$}& \multicolumn{2}{c|}{$1.27\pm0.41$}&$0.63\pm0.10$\\ \hline
$\sum_iX_c^i$& \multicolumn{2}{c|}{$10.90\pm0.14$}& \multicolumn{2}{c|}{$10.92\pm0.14$}& \multicolumn{2}{c|}{$10.89\pm0.14$}& \multicolumn{2}{c|}{$10.92\pm0.14$}& \multicolumn{2}{c|}{$10.90\pm0.14$}&$9.21\pm0.20$\\ \hline
$X_c$& \multicolumn{2}{c|}{}& \multicolumn{2}{c|}{}& \multicolumn{2}{c|}{}& \multicolumn{2}{c|}{}& \multicolumn{2}{c|}{}&$10.90 \pm 0.14$\\ \hline
\hline
$\chi^2/dof$& \multicolumn{2}{c|}{6/23 = 0.28}& \multicolumn{2}{c|}{33/33 = 1.02}& \multicolumn{2}{c|}{25/45 = 0.56}& \multicolumn{2}{c|}{24/27 = 0.91}& \multicolumn{2}{c|}{18/31 = 0.61}&-\\ \hline
p-value& \multicolumn{2}{c|}{1.00}& \multicolumn{2}{c|}{0.44}& \multicolumn{2}{c|}{0.99}& \multicolumn{2}{c|}{0.60}& \multicolumn{2}{c|}{0.96}&-\\ \hline
\end{tabular}}
\caption{Results for fits in which only subsets of the inputs as given in Table~\ref{data}
are used. Fit 1: only with lepton energy moments; Fit 2: only with hadronic mass-energy moments; 
Fit 3: only with hadronic mass moments; 
Fit 4: only with hadronic mass-energy moments with the additional
omission of the hadronic mass-energy moments for lower lepton momentum cut-offs equal 
to 1.8 GeV/c and 1.9 GeV/c; Fit 5: with only hadronic mass moments with the
additional omission of the Belle measurements. In these fits we always constrain 
$\mathcal{B}(B^+\to\overline{D}^{0}l^+\nu)$ and $\mathcal{B}(B^+\to\overline{D}_{0}^{0}l^+\nu)$ 
and $\mathcal{B}(B^+\to\overline{D}_{2}^{0}l^+\nu)$ to their measured value.
Hereby, ``U/C'' stands for ``used/constrained'' and the ``x'' denotes ``yes'', 
whereas ``-'' denotes ``no'', respectively. The Table is further 
described in the text in Section~\ref{sec:4}.} 
\label{ResultsDstarSys1}
\end{table*}
\begin{table*}
\centering
{\fontsize{5}{8} \selectfont 
\begin{tabular}
{|c|c|c|c|c|c|c|c|c|c|c|c|}
\hline
\multirow{2}{*}{$X_c$} & \multicolumn{2}{c|}{Fit 6}& \multicolumn{2}{c|}{Fit 7}& \multicolumn{2}{c|}{Fit 8}& \multicolumn{2}{c|}{Fit 9}& \multicolumn{2}{c|}{Fit 10}&Measured\\ \cline{2-11}
&U/C&$\mathcal{B}[\%]$&U/C&$\mathcal{B}[\%]$&U/C&$\mathcal{B}[\%]$&U/C&$\mathcal{B}[\%]$&U/C&$\mathcal{B}[\%]$&$[\%]$\\ \hline \hline
$ \overline{D}^{0}$&x/x &$2.30 \pm 0.10$&x/x &$2.24 \pm 0.10$&x/x &$2.29 \pm 0.09$&x/x &$2.24 \pm 0.10$&x/x &$2.31 \pm 0.10$&$2.30 \pm 0.10 $\\ \hline
$ \overline{D}^{*0}$&x/-&$5.53 \pm 0.21$&x/-&$5.99 \pm 0.26$&x/-&$6.31 \pm 0.19$&x/-&$6.11 \pm 0.29$&x/-&$6.21 \pm 0.21$&$5.34 \pm 0.12 $\\ \hline
$ \overline{D}_1^{0}$&x/x &$0.65 \pm 0.07$&x/x &$0.65 \pm 0.07$&x/x &$0.65 \pm 0.07$&x/x &$0.65 \pm 0.07$&x/x &$0.65 \pm 0.07$&$0.65 \pm 0.07 $\\ \hline
$ \overline{D}_2^{0}$&x/-&$2.13 \pm 0.58$&x/-&$0.73 \pm 0.65$&x/-&$0.89 \pm 0.25$&x/-&$1.01 \pm 0.70$&x/-&$0.57 \pm 0.49$&$0.28 \pm 0.03 $\\ \hline
$ \overline{D}'^{0}_1$&x/-&$0.03 \pm 0.60$&x/-&$0.73 \pm 1.15$&x/-&$-0.19 \pm 0.41$&x/-&$0.12 \pm 1.31$&x/-&$0.35 \pm 0.76$&$0.20 \pm 0.06 $\\ \hline
$ \overline{D}_0^{0}$&x/x &$0.44 \pm 0.08$&x/x &$0.44 \pm 0.07$&x/x &$0.44 \pm 0.07$&x/x &$0.44 \pm 0.07$&x/x &$0.43 \pm 0.07$&$0.43 \pm 0.07 $\\ \hline
$ \overline{D}'^{0}$&-/-&-&-/-&-&-/-&-&-/-&-&-/-&-&-\\ \hline
$ \overline{D}'^{*0}$&-/-&-&-/-&-&-/-&-&-/-&-&-/-&-&-\\ \hline
$ (D^{(*)}\pi)_{nr}$&x/-&$-0.18 \pm 0.34$&x/-&$0.14 \pm 0.41$&x/-&$0.50 \pm 0.18$&x/-&$0.35 \pm 0.47$&x/-&$0.37 \pm 0.27$&-\\ \hline
$\overline{D}^{0}_1/\overline{D}^{0}_2$& \multicolumn{2}{c|}{$2.78\pm0.57$}& \multicolumn{2}{c|}{$1.38\pm0.64$}& \multicolumn{2}{c|}{$1.55\pm0.24$}& \multicolumn{2}{c|}{$1.66\pm0.70$}& \multicolumn{2}{c|}{$1.22\pm0.48$}&$0.94\pm0.08$\\ \hline
$\overline{D}^{0}_0/\overline{D}'^{0}_1$& \multicolumn{2}{c|}{$0.47\pm0.60$}& \multicolumn{2}{c|}{$1.16\pm1.16$}& \multicolumn{2}{c|}{$0.24\pm0.42$}& \multicolumn{2}{c|}{$0.56\pm1.32$}& \multicolumn{2}{c|}{$0.79\pm0.77$}&$0.63\pm0.10$\\ \hline
$\overline{D}^{0}_0/\overline{D}'^{0}_1/(D^{(*)}\pi)_{nr}$& \multicolumn{2}{c|}{$0.29\pm0.55$}& \multicolumn{2}{c|}{$1.30\pm0.76$}& \multicolumn{2}{c|}{$0.75\pm0.31$}& \multicolumn{2}{c|}{$0.91\pm0.86$}& \multicolumn{2}{c|}{$1.15\pm0.55$}&$0.63\pm0.10$\\ \hline
$\sum_iX_c^i$& \multicolumn{2}{c|}{$10.90\pm0.14$}& \multicolumn{2}{c|}{$10.92\pm0.14$}& \multicolumn{2}{c|}{$10.90\pm0.14$}& \multicolumn{2}{c|}{$10.92\pm0.14$}& \multicolumn{2}{c|}{$10.90\pm0.14$}&$9.21\pm0.20$\\ \hline
$X_c$& \multicolumn{2}{c|}{}& \multicolumn{2}{c|}{}& \multicolumn{2}{c|}{}& \multicolumn{2}{c|}{}& \multicolumn{2}{c|}{}&$10.90 \pm 0.14$\\ \hline
\hline
$\chi^2/dof$& \multicolumn{2}{c|}{6/23 = 0.29}& \multicolumn{2}{c|}{34/33 = 1.03}& \multicolumn{2}{c|}{25/45 = 0.56}& \multicolumn{2}{c|}{25/27 = 0.93}& \multicolumn{2}{c|}{18/31 = 0.60}&-\\ \hline
p-value& \multicolumn{2}{c|}{1.00}& \multicolumn{2}{c|}{0.42}& \multicolumn{2}{c|}{0.99}& \multicolumn{2}{c|}{0.57}& \multicolumn{2}{c|}{0.96}&-\\ \hline
\end{tabular}}
\caption{Results for fits in which only subsets of the inputs as given in Table~\ref{data}
are used. Fit 6: only with lepton energy moments; Fit 7: only with lepton energy 
and hadronic mass-energy moments; in Fit 8, only lepton energy and hadronic mass moments; 
Fit 9: only with lepton energy and hadronic mass-energy moments with the additional
omission of the hadronic mass-energy moments for lower lepton momentum cut-offs equal 
to 1.8 GeV/c and 1.9 GeV/c; Fit 10: leptonic energy and hadronic mass moments with the
additional omission of the Belle measurement of hadronic mass moments. 
In these fits we always constrain 
$\mathcal{B}(B^+\to\overline{D}^{0}l^+\nu)$ and $\mathcal{B}(B^+\to\overline{D}_{0}^{0}l^+\nu)$ 
and $\mathcal{B}(B^+\to\overline{D}_{1}^{0}l^+\nu)$ to their measured value.
Hereby, ``U/C'' stands for ``used/constrained'' and the ``x'' denotes ``yes'', 
whereas ``-'' denotes ``no'', respectively. The Table is further 
described in the text in Section~\ref{sec:4}.}
\label{ResultsDstarSys2}
\end{table*}

\section{RESULT TABLES WHERE $B\to (D^{(*)}\pi)_{nr}l\nu$ DECAYS ARE MODELED WITH $B\to (D^{*}\pi)_{nr}l\nu$ DECAYS}\label{ResultsDstar}
\FloatBarrier

In this appendix, we present the results for the same fit scenarios as in  Appendix~\ref{AddRes} but modeled the
$B\to (D^{(*)}\pi)_{nr}l\nu$ decays exclusively with $B\to (D^{*}\pi)_{nr}l\nu$ decays. The results are again 
grouped in eight tables (~\ref{NRResults00}-~\ref{NRResults31}).

\begin{table*}[h!]
\centering
\scriptsize
\begin{tabular}{|c|c|c|c|c|c|c|c|c|c|}
 \hline
\multirow{2}{*}{$X_c$} & \multicolumn{2}{c|}{Fit 1}& \multicolumn{2}{c|}{Fit 2}& \multicolumn{2}{c|}{Fit 3}& \multicolumn{2}{c|}{Fit 4}&Measured\\ \cline{2-9}
&U/C&$\mathcal{B}[\%]$&U/C&$\mathcal{B}[\%]$&U/C&$\mathcal{B}[\%]$&U/C&$\mathcal{B}[\%]$&$[\%]$\\ \hline \hline
$ \overline{D}^{0}$&x/-&$2.38 \pm 0.16$&x/-&$2.39 \pm 0.16$&x/-&$2.33 \pm 0.16$&x/-&$2.43 \pm 0.14$&$2.30 \pm 0.10 $\\ \hline
$ \overline{D}^{*0}$&x/-&$5.79 \pm 0.16$&x/-&$5.81 \pm 0.15$&x/-&$5.82 \pm 0.15$&x/-&$5.81 \pm 0.15$&$5.34 \pm 0.12 $\\ \hline
$ \overline{D}_1^{0}$&x/-&$2.22 \pm 0.97$&x/-&$1.62 \pm 0.57$&x/x &$0.66 \pm 0.07$&x/-&$1.87 \pm 0.27$&$0.65 \pm 0.07 $\\ \hline
$ \overline{D}_2^{0}$&x/-&$-0.45 \pm 0.95$&x/x &$0.28 \pm 0.03$&x/-&$0.82 \pm 0.57$&x/x &$0.28 \pm 0.03$&$0.28 \pm 0.03 $\\ \hline
$ \overline{D}'^{0}_1$&x/-&$-0.16 \pm 0.70$&x/-&$-0.47 \pm 0.57$&x/-&$-0.22 \pm 0.70$&x/-&$-0.67 \pm 0.40$&$0.20 \pm 0.06 $\\ \hline
$ \overline{D}_0^{0}$&x/-&$0.58 \pm 0.41$&x/-&$0.64 \pm 0.41$&x/-&$1.02 \pm 0.31$&x/x &$0.44 \pm 0.07$&$0.43 \pm 0.07 $\\ \hline
$ \overline{D}'^{0}$&-/-&-&-/-&-&-/-&-&-/-&-&-\\ \hline
$ \overline{D}'^{*0}$&-/-&-&-/-&-&-/-&-&-/-&-&-\\ \hline
$ (D^{*}\pi)_{nr}$&x/-&$0.54 \pm 0.29$&x/-&$0.62 \pm 0.27$&x/-&$0.46 \pm 0.29$&x/-&$0.73 \pm 0.17$&-\\ \hline
$\overline{D}^{0}_1/\overline{D}^{0}_2$& \multicolumn{2}{c|}{$1.77\pm0.61$}& \multicolumn{2}{c|}{$1.91\pm0.57$}& \multicolumn{2}{c|}{$1.48\pm0.57$}& \multicolumn{2}{c|}{$2.16\pm0.27$}&$0.94\pm0.08$\\ \hline
$\overline{D}^{0}_0/\overline{D}'^{0}_1$& \multicolumn{2}{c|}{$0.42\pm0.98$}& \multicolumn{2}{c|}{$0.17\pm0.91$}& \multicolumn{2}{c|}{$0.80\pm0.93$}& \multicolumn{2}{c|}{$-0.23\pm0.42$}&$0.63\pm0.10$\\ \hline
$\overline{D}^{0}_0/\overline{D}'^{0}_1/(D^{*}\pi)_{nr}$& \multicolumn{2}{c|}{$0.95\pm0.70$}& \multicolumn{2}{c|}{$0.79\pm0.66$}& \multicolumn{2}{c|}{$1.27\pm0.66$}& \multicolumn{2}{c|}{$0.50\pm0.28$}&$0.63\pm0.10$\\ \hline
$\sum_iX_c^i$& \multicolumn{2}{c|}{$10.90\pm0.14$}& \multicolumn{2}{c|}{$10.90\pm0.14$}& \multicolumn{2}{c|}{$10.90\pm0.14$}& \multicolumn{2}{c|}{$10.90\pm0.14$}&$9.21\pm0.20$\\ \hline
$X_c$& \multicolumn{2}{c|}{}& \multicolumn{2}{c|}{}& \multicolumn{2}{c|}{}& \multicolumn{2}{c|}{}&$10.90 \pm 0.14$\\ \hline
\hline
$\chi^2/dof$& \multicolumn{2}{c|}{75/104 = 0.72}& \multicolumn{2}{c|}{75/105 = 0.72}& \multicolumn{2}{c|}{77/105 = 0.74}& \multicolumn{2}{c|}{76/106 = 0.72}&-\\ \hline
p-value& \multicolumn{2}{c|}{0.98}& \multicolumn{2}{c|}{0.99}& \multicolumn{2}{c|}{0.98}& \multicolumn{2}{c|}{0.99}&-\\ \hline
\end{tabular} 
\caption{Results for moment fits of semileptonic decays $B^{+} \to X_{c}^{i}l^{+}\nu$ with hadronic final states $X_{c}^{i}$ 
containing $D$, $D^*$, any $D^{**}$, and $(D^{*}\pi)_{nr}$. Hereby, ``U/C'' stands for ``used/constrained'' and 
the ``x'' denotes ``yes'', whereas ``-'' denotes ``no'', respectively. 
The table is further described in the text of Section~\ref{sec:4}.}
\label{NRResults00} 
\end{table*}
\begin{table*}
\centering
\scriptsize
\begin{tabular}{|c|c|c|c|c|c|c|c|c|c|}
 \hline
\multirow{2}{*}{$X_c$} & \multicolumn{2}{c|}{Fit 5}& \multicolumn{2}{c|}{Fit 6}& \multicolumn{2}{c|}{Fit 7}& \multicolumn{2}{c|}{Fit 8}&Measured\\ \cline{2-9}
&U/C&$\mathcal{B}[\%]$&U/C&$\mathcal{B}[\%]$&U/C&$\mathcal{B}[\%]$&U/C&$\mathcal{B}[\%]$&$[\%]$\\ \hline \hline
$ \overline{D}^{0}$&x/-&$2.34 \pm 0.16$&x/-&$2.31 \pm 0.15$&x/-&$2.48 \pm 0.14$&x/-&$2.41 \pm 0.14$&$2.30 \pm 0.10 $\\ \hline
$ \overline{D}^{*0}$&x/-&$5.77 \pm 0.15$&x/-&$5.79 \pm 0.15$&x/-&$5.83 \pm 0.15$&x/-&$5.75 \pm 0.15$&$5.34 \pm 0.12 $\\ \hline
$ \overline{D}_1^{0}$&x/-&$1.00 \pm 0.23$&x/x &$0.66 \pm 0.07$&x/x &$0.67 \pm 0.07$&x/-&$2.71 \pm 0.64$&$0.65 \pm 0.07 $\\ \hline
$ \overline{D}_2^{0}$&x/x &$0.28 \pm 0.03$&x/-&$0.50 \pm 0.19$&x/-&$1.64 \pm 0.35$&x/-&$-1.05 \pm 0.62$&$0.28 \pm 0.03 $\\ \hline
$ \overline{D}'^{0}_1$&x/x &$0.19 \pm 0.06$&x/x &$0.19 \pm 0.06$&x/-&$-1.03 \pm 0.53$&x/x &$0.19 \pm 0.06$&$0.20 \pm 0.06 $\\ \hline
$ \overline{D}_0^{0}$&x/-&$0.99 \pm 0.29$&x/-&$1.14 \pm 0.23$&x/x &$0.47 \pm 0.07$&x/x &$0.44 \pm 0.07$&$0.43 \pm 0.07 $\\ \hline
$ \overline{D}'^{0}$&-/-&-&-/-&-&-/-&-&-/-&-&-\\ \hline
$ \overline{D}'^{*0}$&-/-&-&-/-&-&-/-&-&-/-&-&-\\ \hline
$ (D^{*}\pi)_{nr}$&x/-&$0.33 \pm 0.09$&x/-&$0.30 \pm 0.09$&x/-&$0.83 \pm 0.20$&x/-&$0.44 \pm 0.08$&-\\ \hline
$\overline{D}^{0}_1/\overline{D}^{0}_2$& \multicolumn{2}{c|}{$1.29\pm0.22$}& \multicolumn{2}{c|}{$1.16\pm0.18$}& \multicolumn{2}{c|}{$2.31\pm0.34$}& \multicolumn{2}{c|}{$1.66\pm0.11$}&$0.94\pm0.08$\\ \hline
$\overline{D}^{0}_0/\overline{D}'^{0}_1$& \multicolumn{2}{c|}{$1.18\pm0.30$}& \multicolumn{2}{c|}{$1.34\pm0.24$}& \multicolumn{2}{c|}{$-0.56\pm0.55$}& \multicolumn{2}{c|}{$0.64\pm0.10$}&$0.63\pm0.10$\\ \hline
$\overline{D}^{0}_0/\overline{D}'^{0}_1/(D^{*}\pi)_{nr}$& \multicolumn{2}{c|}{$1.50\pm0.26$}& \multicolumn{2}{c|}{$1.64\pm0.22$}& \multicolumn{2}{c|}{$0.27\pm0.37$}& \multicolumn{2}{c|}{$1.08\pm0.11$}&$0.63\pm0.10$\\ \hline
$\sum_iX_c^i$& \multicolumn{2}{c|}{$10.90\pm0.14$}& \multicolumn{2}{c|}{$10.90\pm0.14$}& \multicolumn{2}{c|}{$10.89\pm0.14$}& \multicolumn{2}{c|}{$10.90\pm0.14$}&$9.21\pm0.20$\\ \hline
$X_c$& \multicolumn{2}{c|}{}& \multicolumn{2}{c|}{}& \multicolumn{2}{c|}{}& \multicolumn{2}{c|}{}&$10.90 \pm 0.14$\\ \hline
\hline
$\chi^2/dof$& \multicolumn{2}{c|}{77/106 = 0.73}& \multicolumn{2}{c|}{78/106 = 0.74}& \multicolumn{2}{c|}{81/106 = 0.77}& \multicolumn{2}{c|}{76/106 = 0.72}&-\\ \hline
p-value& \multicolumn{2}{c|}{0.98}& \multicolumn{2}{c|}{0.98}& \multicolumn{2}{c|}{0.96}& \multicolumn{2}{c|}{0.99}&-\\ \hline
\end{tabular} 
\caption{Results for moment fits of semileptonic decays $B^{+} \to X_{c}^{i}l^{+}\nu$ with hadronic final states $X_{c}^{i}$ 
containing $D$, $D^*$, any $D^{**}$, and $(D^{*}\pi)_{nr}$. Hereby, ``U/C'' stands for ``used/constrained'' and 
the ``x'' denotes ``yes'', whereas ``-'' denotes ``no'', respectively. 
The table is further described in the text of Section~\ref{sec:4}.}
\label{NRResults01} 
\end{table*}

\begin{table*}
\centering
\scriptsize
\begin{tabular}{|c|c|c|c|c|c|c|c|c|c|}
\hline
\multirow{2}{*}{$X_c$} & \multicolumn{2}{c|}{Fit 1}& \multicolumn{2}{c|}{Fit 2}& \multicolumn{2}{c|}{Fit 3}& \multicolumn{2}{c|}{Fit 4}&Measured\\ \cline{2-9}
&U/C&$\mathcal{B}[\%]$&U/C&$\mathcal{B}[\%]$&U/C&$\mathcal{B}[\%]$&U/C&$\mathcal{B}[\%]$&$[\%]$\\ \hline \hline
$ \overline{D}^{0}$&x/x &$2.37 \pm 0.08$&x/x &$2.38 \pm 0.08$&x/x &$2.42 \pm 0.08$&x/x &$2.37 \pm 0.08$&$2.30 \pm 0.10 $\\ \hline
$ \overline{D}^{*0}$&x/x &$5.55 \pm 0.09$&x/x &$5.56 \pm 0.09$&x/x &$5.58 \pm 0.09$&x/x &$5.56 \pm 0.09$&$5.34 \pm 0.12 $\\ \hline
$ \overline{D}_1^{0}$&x/-&$2.19 \pm 0.93$&x/-&$1.16 \pm 0.52$&x/-&$1.77 \pm 0.26$&x/-&$1.01 \pm 0.20$&$0.65 \pm 0.07 $\\ \hline
$ \overline{D}_2^{0}$&x/-&$-0.93 \pm 0.93$&x/x &$0.28 \pm 0.03$&x/x &$0.28 \pm 0.03$&x/x &$0.28 \pm 0.03$&$0.28 \pm 0.03 $\\ \hline
$ \overline{D}'^{0}_1$&x/-&$0.50 \pm 0.65$&x/-&$0.03 \pm 0.53$&x/-&$-0.47 \pm 0.38$&x/x &$0.19 \pm 0.06$&$0.20 \pm 0.06 $\\ \hline
$ \overline{D}_0^{0}$&x/-&$0.80 \pm 0.36$&x/-&$0.92 \pm 0.36$&x/x &$0.46 \pm 0.07$&x/-&$1.00 \pm 0.25$&$0.43 \pm 0.07 $\\ \hline
$ \overline{D}'^{0}$&-/-&-&-/-&-&-/-&-&-/-&-&-\\ \hline
$ \overline{D}'^{*0}$&-/-&-&-/-&-&-/-&-&-/-&-&-\\ \hline
$ (D^{*}\pi)_{nr}$&x/-&$0.25 \pm 0.27$&x/-&$0.37 \pm 0.25$&x/-&$0.63 \pm 0.16$&x/-&$0.29 \pm 0.09$&-\\ \hline
$\overline{D}^{0}_1/\overline{D}^{0}_2$& \multicolumn{2}{c|}{$1.26\pm0.54$}& \multicolumn{2}{c|}{$1.44\pm0.52$}& \multicolumn{2}{c|}{$2.05\pm0.26$}& \multicolumn{2}{c|}{$1.29\pm0.20$}&$0.94\pm0.08$\\ \hline
$\overline{D}^{0}_0/\overline{D}'^{0}_1$& \multicolumn{2}{c|}{$1.30\pm0.88$}& \multicolumn{2}{c|}{$0.95\pm0.83$}& \multicolumn{2}{c|}{$-0.01\pm0.40$}& \multicolumn{2}{c|}{$1.20\pm0.26$}&$0.63\pm0.10$\\ \hline
$\overline{D}^{0}_0/\overline{D}'^{0}_1/(D^{*}\pi)_{nr}$& \multicolumn{2}{c|}{$1.55\pm0.62$}& \multicolumn{2}{c|}{$1.32\pm0.60$}& \multicolumn{2}{c|}{$0.62\pm0.27$}& \multicolumn{2}{c|}{$1.49\pm0.22$}&$0.63\pm0.10$\\ \hline
$\sum_iX_c^i$& \multicolumn{2}{c|}{$10.72\pm0.12$}& \multicolumn{2}{c|}{$10.71\pm0.12$}& \multicolumn{2}{c|}{$10.67\pm0.12$}& \multicolumn{2}{c|}{$10.72\pm0.12$}&$9.21\pm0.20$\\ \hline
$X_c$& \multicolumn{2}{c|}{}& \multicolumn{2}{c|}{}& \multicolumn{2}{c|}{}& \multicolumn{2}{c|}{}&$10.90 \pm 0.14$\\ \hline
\hline
$\chi^2/dof$& \multicolumn{2}{c|}{82/106 = 0.78}& \multicolumn{2}{c|}{83/107 = 0.79}& \multicolumn{2}{c|}{85/108 = 0.79}& \multicolumn{2}{c|}{84/108 = 0.78}&-\\ \hline
p-value& \multicolumn{2}{c|}{0.96}& \multicolumn{2}{c|}{0.95}& \multicolumn{2}{c|}{0.94}& \multicolumn{2}{c|}{0.96}&-\\ \hline
\end{tabular}
\caption{Results for moment fits of semileptonic decays $B^{+} \to X_{c}^{i}l^{+}\nu$ with hadronic final states $X_{c}^{i}$ 
containing $D$, $D^*$, any $D^{**}$, and $(D^{*}\pi)_{nr}$. Hereby, ``U/C'' stands for ``used/constrained'' and 
the ``x'' denotes ``yes'', whereas ``-'' denotes ``no'', respectively. 
The table is further described in the text of Section~\ref{sec:4}.}
\label{NRResults10}
\end{table*}
\begin{table*}
\centering
\scriptsize
\begin{tabular}{|c|c|c|c|c|c|c|c|c|c|}
\hline
\multirow{2}{*}{$X_c$} & \multicolumn{2}{c|}{Fit 5}& \multicolumn{2}{c|}{Fit 6}& \multicolumn{2}{c|}{Fit 7}& \multicolumn{2}{c|}{Fit 8}&Measured\\ \cline{2-9}
&U/C&$\mathcal{B}[\%]$&U/C&$\mathcal{B}[\%]$&U/C&$\mathcal{B}[\%]$&U/C&$\mathcal{B}[\%]$&$[\%]$\\ \hline \hline
$ \overline{D}^{0}$&x/x &$2.36 \pm 0.08$&x/x &$2.36 \pm 0.08$&x/x &$2.44 \pm 0.08$&x/x &$2.40 \pm 0.08$&$2.30 \pm 0.10 $\\ \hline
$ \overline{D}^{*0}$&x/x &$5.56 \pm 0.09$&x/x &$5.55 \pm 0.09$&x/x &$5.59 \pm 0.09$&x/x &$5.56 \pm 0.09$&$5.34 \pm 0.12 $\\ \hline
$ \overline{D}_1^{0}$&x/x &$0.66 \pm 0.07$&x/x &$0.66 \pm 0.07$&x/x &$0.67 \pm 0.07$&x/-&$2.78 \pm 0.62$&$0.65 \pm 0.07 $\\ \hline
$ \overline{D}_2^{0}$&x/-&$0.53 \pm 0.17$&x/-&$0.35 \pm 0.53$&x/-&$1.41 \pm 0.33$&x/-&$-1.11 \pm 0.60$&$0.28 \pm 0.03 $\\ \hline
$ \overline{D}'^{0}_1$&x/x &$0.20 \pm 0.06$&x/-&$0.43 \pm 0.66$&x/-&$-0.64 \pm 0.49$&x/x &$0.20 \pm 0.06$&$0.20 \pm 0.06 $\\ \hline
$ \overline{D}_0^{0}$&x/-&$1.14 \pm 0.21$&x/-&$1.20 \pm 0.27$&x/x &$0.49 \pm 0.07$&x/x &$0.45 \pm 0.07$&$0.43 \pm 0.07 $\\ \hline
$ \overline{D}'^{0}$&-/-&-&-/-&-&-/-&-&-/-&-&-\\ \hline
$ \overline{D}'^{*0}$&-/-&-&-/-&-&-/-&-&-/-&-&-\\ \hline
$ (D^{*}\pi)_{nr}$&x/-&$0.27 \pm 0.09$&x/-&$0.18 \pm 0.27$&x/-&$0.68 \pm 0.18$&x/-&$0.42 \pm 0.08$&-\\ \hline
$\overline{D}^{0}_1/\overline{D}^{0}_2$& \multicolumn{2}{c|}{$1.19\pm0.16$}& \multicolumn{2}{c|}{$1.01\pm0.53$}& \multicolumn{2}{c|}{$2.08\pm0.32$}& \multicolumn{2}{c|}{$1.67\pm0.11$}&$0.94\pm0.08$\\ \hline
$\overline{D}^{0}_0/\overline{D}'^{0}_1$& \multicolumn{2}{c|}{$1.34\pm0.22$}& \multicolumn{2}{c|}{$1.63\pm0.87$}& \multicolumn{2}{c|}{$-0.15\pm0.51$}& \multicolumn{2}{c|}{$0.64\pm0.09$}&$0.63\pm0.10$\\ \hline
$\overline{D}^{0}_0/\overline{D}'^{0}_1/(D^{*}\pi)_{nr}$& \multicolumn{2}{c|}{$1.61\pm0.19$}& \multicolumn{2}{c|}{$1.81\pm0.61$}& \multicolumn{2}{c|}{$0.53\pm0.35$}& \multicolumn{2}{c|}{$1.07\pm0.10$}&$0.63\pm0.10$\\ \hline
$\sum_iX_c^i$& \multicolumn{2}{c|}{$10.72\pm0.12$}& \multicolumn{2}{c|}{$10.73\pm0.12$}& \multicolumn{2}{c|}{$10.64\pm0.12$}& \multicolumn{2}{c|}{$10.70\pm0.12$}&$9.21\pm0.20$\\ \hline
$X_c$& \multicolumn{2}{c|}{}& \multicolumn{2}{c|}{}& \multicolumn{2}{c|}{}& \multicolumn{2}{c|}{}&$10.90 \pm 0.14$\\ \hline
\hline
$\chi^2/dof$& \multicolumn{2}{c|}{84/108 = 0.79}& \multicolumn{2}{c|}{84/107 = 0.79}& \multicolumn{2}{c|}{92/108 = 0.86}& \multicolumn{2}{c|}{83/108 = 0.77}&-\\ \hline
p-value& \multicolumn{2}{c|}{0.95}& \multicolumn{2}{c|}{0.94}& \multicolumn{2}{c|}{0.86}& \multicolumn{2}{c|}{0.96}&-\\ \hline 
\end{tabular}
\caption{Results for moment fits of semileptonic decays $B^{+} \to X_{c}^{i}l^{+}\nu$ with hadronic final states $X_{c}^{i}$ 
containing $D$, $D^*$, any $D^{**}$, and $(D^{*}\pi)_{nr}$. Hereby, ``U/C'' stands for ``used/constrained'' and 
the ``x'' denotes ``yes'', whereas ``-'' denotes ``no'', respectively. 
The table is further described in the text of Section~\ref{sec:4}.}
\label{NRResults11}
\end{table*}

\begin{table*}
\centering
\scriptsize
\begin{tabular}{|c|c|c|c|c|c|c|c|c|c|}
\hline
\multirow{2}{*}{$X_c$} & \multicolumn{2}{c|}{Fit 1}& \multicolumn{2}{c|}{Fit 2}& \multicolumn{2}{c|}{Fit 3}& \multicolumn{2}{c|}{Fit 4}&Measured\\ \cline{2-9}
&U/C&$\mathcal{B}[\%]$&U/C&$\mathcal{B}[\%]$&U/C&$\mathcal{B}[\%]$&U/C&$\mathcal{B}[\%]$&$[\%]$\\ \hline \hline
$ \overline{D}^{0}$&x/x &$2.32 \pm 0.08$&x/x &$2.34 \pm 0.08$&x/x &$2.32 \pm 0.08$&x/-&$2.55 \pm 0.15$&$2.30 \pm 0.10 $\\ \hline
$ \overline{D}^{*0}$&x/-&$5.85 \pm 0.14$&x/-&$5.84 \pm 0.14$&x/-&$5.79 \pm 0.13$&x/x &$5.53 \pm 0.09$&$5.34 \pm 0.12 $\\ \hline
$ \overline{D}_1^{0}$&x/-&$1.52 \pm 0.53$&x/-&$2.51 \pm 0.72$&x/-&$2.16 \pm 0.94$&x/-&$1.51 \pm 0.57$&$0.65 \pm 0.07 $\\ \hline
$ \overline{D}_2^{0}$&x/x &$0.28 \pm 0.03$&x/-&$-0.59 \pm 0.93$&x/-&$-0.70 \pm 0.77$&x/x &$0.28 \pm 0.03$&$0.28 \pm 0.03 $\\ \hline
$ \overline{D}'^{0}_1$&x/-&$-0.41 \pm 0.55$&x/-&$-0.27 \pm 0.62$&x/x &$0.19 \pm 0.06$&x/-&$-0.22 \pm 0.56$&$0.20 \pm 0.06 $\\ \hline
$ \overline{D}_0^{0}$&x/-&$0.73 \pm 0.36$&x/x &$0.44 \pm 0.07$&x/-&$0.73 \pm 0.33$&x/-&$0.64 \pm 0.40$&$0.43 \pm 0.07 $\\ \hline
$ \overline{D}'^{0}$&-/-&-&-/-&-&-/-&-&-/-&-&-\\ \hline
$ \overline{D}'^{*0}$&-/-&-&-/-&-&-/-&-&-/-&-&-\\ \hline
$ (D^{*}\pi)_{nr}$&x/-&$0.60 \pm 0.26$&x/-&$0.62 \pm 0.22$&x/-&$0.40 \pm 0.11$&x/-&$0.48 \pm 0.26$&-\\ \hline
$\overline{D}^{0}_1/\overline{D}^{0}_2$& \multicolumn{2}{c|}{$1.80\pm0.53$}& \multicolumn{2}{c|}{$1.92\pm0.38$}& \multicolumn{2}{c|}{$1.46\pm0.25$}& \multicolumn{2}{c|}{$1.79\pm0.57$}&$0.94\pm0.08$\\ \hline
$\overline{D}^{0}_0/\overline{D}'^{0}_1$& \multicolumn{2}{c|}{$0.32\pm0.85$}& \multicolumn{2}{c|}{$0.17\pm0.63$}& \multicolumn{2}{c|}{$0.92\pm0.34$}& \multicolumn{2}{c|}{$0.42\pm0.90$}&$0.63\pm0.10$\\ \hline
$\overline{D}^{0}_0/\overline{D}'^{0}_1/(D^{*}\pi)_{nr}$& \multicolumn{2}{c|}{$0.91\pm0.61$}& \multicolumn{2}{c|}{$0.79\pm0.43$}& \multicolumn{2}{c|}{$1.32\pm0.28$}& \multicolumn{2}{c|}{$0.90\pm0.65$}&$0.63\pm0.10$\\ \hline
$\sum_iX_c^i$& \multicolumn{2}{c|}{$10.89\pm0.14$}& \multicolumn{2}{c|}{$10.88\pm0.14$}& \multicolumn{2}{c|}{$10.89\pm0.14$}& \multicolumn{2}{c|}{$10.77\pm0.13$}&$9.21\pm0.20$\\ \hline
$X_c$& \multicolumn{2}{c|}{}& \multicolumn{2}{c|}{}& \multicolumn{2}{c|}{}& \multicolumn{2}{c|}{}&$10.90 \pm 0.14$\\ \hline
\hline
$\chi^2/dof$& \multicolumn{2}{c|}{76/106 = 0.72}& \multicolumn{2}{c|}{75/106 = 0.72}& \multicolumn{2}{c|}{75/106 = 0.71}& \multicolumn{2}{c|}{82/106 = 0.77}&-\\ \hline
p-value& \multicolumn{2}{c|}{0.99}& \multicolumn{2}{c|}{0.99}& \multicolumn{2}{c|}{0.99}& \multicolumn{2}{c|}{0.96}&-\\ \hline
\end{tabular}
\caption{Results for moment fits of semileptonic decays $B^{+} \to X_{c}^{i}l^{+}\nu$ with hadronic final states $X_{c}^{i}$ 
containing $D$, $D^*$, any $D^{**}$, and $(D^{*}\pi)_{nr}$. Hereby, ``U/C'' stands for ``used/constrained'' and 
the ``x'' denotes ``yes'', whereas ``-'' denotes ``no'', respectively. 
The table is further described in the text of Section~\ref{sec:4}.}
\label{NRResults20}
\end{table*}
\begin{table*}
\scriptsize
\centering
\begin{tabular}{|c|c|c|c|c|c|c|c|c|c|}
\hline
\multirow{2}{*}{$X_c$} & \multicolumn{2}{c|}{Fit 5}& \multicolumn{2}{c|}{Fit 6}& \multicolumn{2}{c|}{Fit 7}& \multicolumn{2}{c|}{Fit 8}&Measured\\ \cline{2-9}
&U/C&$\mathcal{B}[\%]$&U/C&$\mathcal{B}[\%]$&U/C&$\mathcal{B}[\%]$&U/C&$\mathcal{B}[\%]$&$[\%]$\\ \hline \hline
$ \overline{D}^{0}$&x/-&$2.55 \pm 0.12$&x/-&$2.53 \pm 0.14$&x/-&$2.48 \pm 0.15$&x/x &$2.31 \pm 0.08$&$2.30 \pm 0.10 $\\ \hline
$ \overline{D}^{*0}$&x/x &$5.51 \pm 0.09$&x/x &$5.51 \pm 0.09$&x/x &$5.52 \pm 0.09$&x/-&$5.83 \pm 0.14$&$5.34 \pm 0.12 $\\ \hline
$ \overline{D}_1^{0}$&x/-&$2.59 \pm 0.72$&x/-&$2.41 \pm 0.96$&x/x &$0.66 \pm 0.07$&x/x &$0.66 \pm 0.07$&$0.65 \pm 0.07 $\\ \hline
$ \overline{D}_2^{0}$&x/-&$-0.85 \pm 0.92$&x/-&$-0.79 \pm 0.78$&x/-&$0.59 \pm 0.58$&x/-&$0.79 \pm 0.54$&$0.28 \pm 0.03 $\\ \hline
$ \overline{D}'^{0}_1$&x/-&$0.12 \pm 0.59$&x/x &$0.20 \pm 0.06$&x/-&$0.19 \pm 0.69$&x/-&$-0.19 \pm 0.68$&$0.20 \pm 0.06 $\\ \hline
$ \overline{D}_0^{0}$&x/x &$0.44 \pm 0.07$&x/-&$0.55 \pm 0.35$&x/-&$1.05 \pm 0.31$&x/-&$1.04 \pm 0.28$&$0.43 \pm 0.07 $\\ \hline
$ \overline{D}'^{0}$&-/-&-&-/-&-&-/-&-&-/-&-&-\\ \hline
$ \overline{D}'^{*0}$&-/-&-&-/-&-&-/-&-&-/-&-&-\\ \hline
$ (D^{*}\pi)_{nr}$&x/-&$0.42 \pm 0.21$&x/-&$0.37 \pm 0.11$&x/-&$0.27 \pm 0.28$&x/-&$0.46 \pm 0.28$&-\\ \hline
$\overline{D}^{0}_1/\overline{D}^{0}_2$& \multicolumn{2}{c|}{$1.74\pm0.37$}& \multicolumn{2}{c|}{$1.62\pm0.26$}& \multicolumn{2}{c|}{$1.25\pm0.58$}& \multicolumn{2}{c|}{$1.45\pm0.54$}&$0.94\pm0.08$\\ \hline
$\overline{D}^{0}_0/\overline{D}'^{0}_1$& \multicolumn{2}{c|}{$0.56\pm0.60$}& \multicolumn{2}{c|}{$0.74\pm0.36$}& \multicolumn{2}{c|}{$1.25\pm0.94$}& \multicolumn{2}{c|}{$0.85\pm0.89$}&$0.63\pm0.10$\\ \hline
$\overline{D}^{0}_0/\overline{D}'^{0}_1/(D^{*}\pi)_{nr}$& \multicolumn{2}{c|}{$0.98\pm0.42$}& \multicolumn{2}{c|}{$1.12\pm0.30$}& \multicolumn{2}{c|}{$1.51\pm0.67$}& \multicolumn{2}{c|}{$1.31\pm0.62$}&$0.63\pm0.10$\\ \hline
$\sum_iX_c^i$& \multicolumn{2}{c|}{$10.78\pm0.13$}& \multicolumn{2}{c|}{$10.78\pm0.13$}& \multicolumn{2}{c|}{$10.77\pm0.13$}& \multicolumn{2}{c|}{$10.89\pm0.14$}&$9.21\pm0.20$\\ \hline
$X_c$& \multicolumn{2}{c|}{}& \multicolumn{2}{c|}{}& \multicolumn{2}{c|}{}& \multicolumn{2}{c|}{}&$10.90 \pm 0.14$\\ \hline
\hline
$\chi^2/dof$& \multicolumn{2}{c|}{80/106 = 0.76}& \multicolumn{2}{c|}{80/106 = 0.76}& \multicolumn{2}{c|}{83/106 = 0.79}& \multicolumn{2}{c|}{77/106 = 0.73}&-\\ \hline
p-value& \multicolumn{2}{c|}{0.97}& \multicolumn{2}{c|}{0.97}& \multicolumn{2}{c|}{0.94}& \multicolumn{2}{c|}{0.98}&-\\ \hline
\end{tabular}
\caption{Results for moment fits of semileptonic decays $B^{+} \to X_{c}^{i}l^{+}\nu$ with hadronic final states $X_{c}^{i}$ 
containing $D$, $D^*$, any $D^{**}$, and $(D^{*}\pi)_{nr}$. Hereby, ``U/C'' stands for ``used/constrained'' and 
the ``x'' denotes ``yes'', whereas ``-'' denotes ``no'', respectively. 
The table is further described in the text of Section~\ref{sec:4}.}
\label{NRResults21}
\end{table*}

\begin{table*}
\centering
\scriptsize
\begin{tabular}{|c|c|c|c|c|c|c|c|c|c|}
\hline
\multirow{2}{*}{$X_c$} & \multicolumn{2}{c|}{Fit 1}& \multicolumn{2}{c|}{Fit 2}& \multicolumn{2}{c|}{Fit 3}& \multicolumn{2}{c|}{Fit 4}&Measured\\ \cline{2-9}
&U/C&$\mathcal{B}[\%]$&U/C&$\mathcal{B}[\%]$&U/C&$\mathcal{B}[\%]$&U/C&$\mathcal{B}[\%]$&$[\%]$\\ \hline \hline
$ \overline{D}^{0}$&x/x &$2.41 \pm 0.08$&x/x &$2.42 \pm 0.08$&x/x &$2.41 \pm 0.08$&x/x &$2.41 \pm 0.08$&$2.30 \pm 0.10 $\\ \hline
$ \overline{D}^{*0}$&x/x &$5.58 \pm 0.09$&x/x &$5.60 \pm 0.09$&x/x &$5.59 \pm 0.09$&x/x &$5.55 \pm 0.09$&$5.34 \pm 0.12 $\\ \hline
$ \overline{D}_1^{0}$&x/x &$0.79 \pm 0.07$&x/x &$0.79 \pm 0.07$&x/x &$0.78 \pm 0.07$&x/-&$1.92 \pm 0.31$&$0.65 \pm 0.07 $\\ \hline
$ \overline{D}_2^{0}$&x/x &$0.30 \pm 0.03$&x/x &$0.30 \pm 0.03$&x/x &$0.30 \pm 0.03$&x/x &$0.28 \pm 0.03$&$0.28 \pm 0.03 $\\ \hline
$ \overline{D}'^{0}_1$&x/x &$0.22 \pm 0.06$&x/x &$0.23 \pm 0.06$&x/x &$0.22 \pm 0.06$&x/-&$-0.57 \pm 0.69$&$0.20 \pm 0.06 $\\ \hline
$ \overline{D}_0^{0}$&x/x &$0.59 \pm 0.07$&x/x &$0.59 \pm 0.07$&x/x &$0.59 \pm 0.07$&x/x &$0.45 \pm 0.07$&$0.43 \pm 0.07 $\\ \hline
$ \overline{D}'^{0}$&x/-&$0.32 \pm 0.13$&-/-&-&x/-&$0.20 \pm 0.39$&x/-&$0.47 \pm 0.57$&-\\ \hline
$ \overline{D}'^{*0}$&-/-&-&x/-&$0.26 \pm 0.11$&x/-&$0.10 \pm 0.32$&x/-&$-0.48 \pm 0.36$&-\\ \hline
$ (D^{*}\pi)_{nr}$&x/-&$0.39 \pm 0.09$&x/-&$0.40 \pm 0.09$&x/-&$0.39 \pm 0.09$&x/-&$0.68 \pm 0.20$&-\\ \hline
$\overline{D}^{0}_1/\overline{D}^{0}_2$& \multicolumn{2}{c|}{$1.09\pm0.07$}& \multicolumn{2}{c|}{$1.09\pm0.07$}& \multicolumn{2}{c|}{$1.09\pm0.07$}& \multicolumn{2}{c|}{$2.20\pm0.31$}&$0.94\pm0.08$\\ \hline
$\overline{D}^{0}_0/\overline{D}'^{0}_1$& \multicolumn{2}{c|}{$0.82\pm0.09$}& \multicolumn{2}{c|}{$0.82\pm0.09$}& \multicolumn{2}{c|}{$0.81\pm0.09$}& \multicolumn{2}{c|}{$-0.12\pm0.70$}&$0.63\pm0.10$\\ \hline
$\overline{D}^{0}_0/\overline{D}'^{0}_1/(D^{*}\pi)_{nr}$& \multicolumn{2}{c|}{$1.20\pm0.11$}& \multicolumn{2}{c|}{$1.21\pm0.11$}& \multicolumn{2}{c|}{$1.20\pm0.11$}& \multicolumn{2}{c|}{$0.56\pm0.53$}&$0.63\pm0.10$\\ \hline
$\sum_iX_c^i$& \multicolumn{2}{c|}{$10.60\pm0.12$}& \multicolumn{2}{c|}{$10.58\pm0.12$}& \multicolumn{2}{c|}{$10.59\pm0.12$}& \multicolumn{2}{c|}{$10.70\pm0.12$}&$9.21\pm0.20$\\ \hline
$X_c$& \multicolumn{2}{c|}{}& \multicolumn{2}{c|}{}& \multicolumn{2}{c|}{}& \multicolumn{2}{c|}{}&$10.90 \pm 0.14$\\ \hline
\hline
$\chi^2/dof$& \multicolumn{2}{c|}{110/109 = 1.02}& \multicolumn{2}{c|}{110/109 = 1.02}& \multicolumn{2}{c|}{110/108 = 1.02}& \multicolumn{2}{c|}{83/106 = 0.79}&-\\ \hline
p-value& \multicolumn{2}{c|}{0.44}& \multicolumn{2}{c|}{0.43}& \multicolumn{2}{c|}{0.41}& \multicolumn{2}{c|}{0.95}&-\\ \hline
\end{tabular}
\caption{Results for moment fits of semileptonic decays $B^{+} \to X_{c}^{i}l^{+}\nu$ with hadronic final states $X_{c}^{i}$ 
containing $D$, $D^*$, any $D^{**}$, $D'^{(*)}$, and $(D^{*}\pi)_{nr}$. Hereby, ``U/C'' stands for ``used/constrained'' and 
the ``x'' denotes ``yes'', whereas ``-'' denotes ``no'', respectively. 
The table is further described in the text of Section~\ref{sec:4}.}
\label{NRResults30}
\end{table*}
\begin{table*}
\centering
\scriptsize
\begin{tabular}{|c|c|c|c|c|c|c|c|c|c|}
 \hline
\multirow{2}{*}{$X_c$} & \multicolumn{2}{c|}{Fit 5}& \multicolumn{2}{c|}{Fit 6}& \multicolumn{2}{c|}{Fit 7}& \multicolumn{2}{c|}{Fit 8}&Measured\\ \cline{2-9}
&U/C&$\mathcal{B}[\%]$&U/C&$\mathcal{B}[\%]$&U/C&$\mathcal{B}[\%]$&U/C&$\mathcal{B}[\%]$&$[\%]$\\ \hline \hline
$ \overline{D}^{0}$&x/x &$2.42 \pm 0.08$&x/-&$2.42 \pm 0.13$&x/-&$2.29 \pm 0.15$&x/-&$2.51 \pm 0.14$&$2.30 \pm 0.10 $\\ \hline
$ \overline{D}^{*0}$&x/x &$5.57 \pm 0.09$&x/-&$5.79 \pm 0.14$&x/-&$5.76 \pm 0.16$&x/-&$5.76 \pm 0.16$&$5.34 \pm 0.12 $\\ \hline
$ \overline{D}_1^{0}$&x/-&$1.72 \pm 0.27$&x/x &$0.78 \pm 0.07$&x/x &$0.66 \pm 0.07$&x/x &$0.66 \pm 0.07$&$0.65 \pm 0.07 $\\ \hline
$ \overline{D}_2^{0}$&x/x &$0.28 \pm 0.03$&x/x &$0.30 \pm 0.03$&x/x &$0.29 \pm 0.03$&x/-&$1.63 \pm 0.35$&$0.28 \pm 0.03 $\\ \hline
$ \overline{D}'^{0}_1$&x/-&$-0.18 \pm 0.62$&x/x &$0.22 \pm 0.06$&x/-&$0.64 \pm 0.50$&x/-&$-0.46 \pm 0.68$&$0.20 \pm 0.06 $\\ \hline
$ \overline{D}_0^{0}$&x/x &$0.46 \pm 0.07$&x/x &$0.58 \pm 0.07$&x/-&$1.23 \pm 0.19$&x/x &$0.46 \pm 0.07$&$0.43 \pm 0.07 $\\ \hline
$ \overline{D}'^{0}$&x/-&$-0.17 \pm 0.30$&x/-&$0.31 \pm 0.13$&x/-&$-0.16 \pm 0.31$&x/-&$-0.39 \pm 0.30$&-\\ \hline
$ \overline{D}'^{*0}$&-/-&-&-/-&-&-/-&-&-/-&-&-\\ \hline
$ (D^{*}\pi)_{nr}$&x/-&$0.57 \pm 0.18$&x/-&$0.41 \pm 0.09$&x/-&$0.18 \pm 0.15$&x/-&$0.72 \pm 0.22$&-\\ \hline
$\overline{D}^{0}_1/\overline{D}^{0}_2$& \multicolumn{2}{c|}{$2.00\pm0.27$}& \multicolumn{2}{c|}{$1.09\pm0.07$}& \multicolumn{2}{c|}{$0.95\pm0.08$}& \multicolumn{2}{c|}{$2.29\pm0.34$}&$0.94\pm0.08$\\ \hline
$\overline{D}^{0}_0/\overline{D}'^{0}_1$& \multicolumn{2}{c|}{$0.28\pm0.64$}& \multicolumn{2}{c|}{$0.80\pm0.09$}& \multicolumn{2}{c|}{$1.87\pm0.45$}& \multicolumn{2}{c|}{$-0.00\pm0.69$}&$0.63\pm0.10$\\ \hline
$\overline{D}^{0}_0/\overline{D}'^{0}_1/(D^{*}\pi)_{nr}$& \multicolumn{2}{c|}{$0.85\pm0.49$}& \multicolumn{2}{c|}{$1.21\pm0.12$}& \multicolumn{2}{c|}{$2.05\pm0.35$}& \multicolumn{2}{c|}{$0.72\pm0.50$}&$0.63\pm0.10$\\ \hline
$\sum_iX_c^i$& \multicolumn{2}{c|}{$10.68\pm0.12$}& \multicolumn{2}{c|}{$10.82\pm0.14$}& \multicolumn{2}{c|}{$10.90\pm0.14$}& \multicolumn{2}{c|}{$10.89\pm0.14$}&$9.21\pm0.20$\\ \hline
$X_c$& \multicolumn{2}{c|}{}& \multicolumn{2}{c|}{}& \multicolumn{2}{c|}{}& \multicolumn{2}{c|}{}&$10.90 \pm 0.14$\\ \hline
\hline
$\chi^2/dof$& \multicolumn{2}{c|}{85/107 = 0.80}& \multicolumn{2}{c|}{101/107 = 0.95}& \multicolumn{2}{c|}{78/105 = 0.75}& \multicolumn{2}{c|}{79/105 = 0.76}&-\\ \hline
p-value& \multicolumn{2}{c|}{0.94}& \multicolumn{2}{c|}{0.62}& \multicolumn{2}{c|}{0.98}& \multicolumn{2}{c|}{0.97}&-\\ \hline
\end{tabular}
\caption{Results for moment fits of semileptonic decays $B^{+} \to X_{c}^{i}l^{+}\nu$ with hadronic final states $X_{c}^{i}$ 
containing $D$, $D^*$, any $D^{**}$, $D'^{(*)}$, and $(D^{*}\pi)_{nr}$. Hereby, ``U/C'' stands for ``used/constrained'' and 
the ``x'' denotes ``yes'', whereas ``-'' denotes ``no'', respectively. 
The table is further described in the text of Section~\ref{sec:4}.}
\label{NRResults31}
\end{table*}

\end{document}